%% file: main.tex
\documentclass[twocolumn, dvipsnames]{aastex62}
\usepackage{amsmath}
\usepackage{graphicx}
\usepackage{graphbox}
\usepackage{tikz}
\usepackage{float}
\usepackage{multirow}
\usepackage{color}
\usepackage{makecell}
\usepackage{wrapfig}
\usepackage{comment}

\usepackage{lineno}

\usepackage[colorinlistoftodos]{todonotes}

\accepted{in The Astrophysical Journal Supplement Series - February 2022}

\shorttitle{Five Key Exoplanet Questions Answered}
\shortauthors{Changeat, Edwards et al. 2022}

\begin{document}
	\title{Five key exoplanet questions answered via the analysis of 25 hot Jupiter atmospheres in eclipse.}
	
	\correspondingauthor{Q. Changeat}
	\email{quentin.changeat.18@ucl.ac.uk}
	\author[0000-0001-6516-4493]{Q. Changeat}\thanks{These authors contributed equally to this work}
	\affil{Department of Physics and Astronomy,
		University College London,
		Gower Street,WC1E 6BT London, United Kingdom}
	\author[0000-0002-5494-3237]{B. Edwards}\thanks{These authors contributed equally to this work}
	\affil{Department of Physics and Astronomy,
		University College London,
		Gower Street,WC1E 6BT London, United Kingdom}
	\affil{AIM, CEA, CNRS, Universit\'e Paris-Saclay, Universit\'e de Paris, F-91191 Gif-sur-Yvette, France}
	\author[0000-0003-2241-5330]{A.F. Al-Refaie}
	\affil{Department of Physics and Astronomy,
		University College London,
		Gower Street,WC1E 6BT London, United Kingdom}
	\author[0000-0003-3840-1793]{A. Tsiaras}
	\affil{Department of Physics and Astronomy,
		University College London,
		Gower Street,WC1E 6BT London, United Kingdom}
	\affil{INAF,
		Osservatorio Astrofisico di Arcetri,
		Largo E. Fermi 5, I-50125 Firenze, Italy}
	\author[0000-0002-5263-385X]{J. W. Skinner}
	\affil{School of Physics and Astronomy,
		Queen Mary University of London,
		Mile End Road, London, E1 4NS, United Kingdom}
	\author[0000-0002-4525-5651]{J. Y-K. Cho}
	\affil{Center for Computational Astrophysics,
		Flatiron Institute,
		162 Fifth Avenue, New York, NY 10010, USA}
	\author[0000-0002-9616-1524]{K. H. Yip}
	\affil{Department of Physics and Astronomy,
		University College London,
		Gower Street,WC1E 6BT London, United Kingdom}
	\author[0000-0002-7771-6432]{L. Anisman}
	\affil{Department of Physics and Astronomy,
		University College London,
		Gower Street,WC1E 6BT London, United Kingdom}
	\author[0000-0002-5658-5971]{M. Ikoma}
	\affil{Division of Science,
		National Astronomical Observatory of Japan,
		2-21-1 Osawa, Mitaka, Tokyo 181-8588, Japan}
	\affil{Department of Astronomical Science,
		The Graduate University for Advanced Studies (SOKENDAI),
		2-21-1 Osawa, Mitaka, Tokyo 181-8588, Japan}
	\author[0000-0001-9166-3042]{M. F. Bieger}
	\affil{College of Engineering, Mathematics and Physical Sciences,
		University of Exeter,
		North Park Road, Exeter, UK}
	\author[0000-0003-2854-765X]{O. Venot}
	\affil{Universit\'{e} de Paris and Univ Paris Est Creteil,
		CNRS, LISA,
		F-75013 Paris, France}
	\author[0000-0002-5418-6336]{S. Shibata}
	\affil{Institute for Computational Science, Center for Theoretical Astrophysics \& Cosmology,
		University of Zurich,
		Winterthurerstr. 190, 8057 Zurich, Switzerland}
	\author[0000-0002-4205-5267]{I. P. Waldmann}
	\affil{Department of Physics and Astronomy,
		University College London,
		Gower Street,WC1E 6BT London, United Kingdom}
	\author[0000-0001-6058-6654]{G. Tinetti}
	\affil{Department of Physics and Astronomy,
		University College London,
		Gower Street,WC1E 6BT London, United Kingdom}
	
\begin{abstract}

Population studies of exoplanets are key to unlocking their statistical properties. So far the inferred properties have been mostly limited to planetary, orbital and stellar parameters extracted from, e.g.,  Kepler, radial velocity, and GAIA data. More recently an increasing number of exoplanet atmospheres have been observed in detail from space and the ground. Generally, however, these atmospheric studies have focused on individual planets, with the exception of a couple of works which have detected the presence of water vapor and clouds in populations of gaseous planets via transmission spectroscopy. Here, using a suite of retrieval tools, we analyse spectroscopic and photometric data of 25 hot Jupiters, obtained with the Hubble and Spitzer Space Telescopes via the eclipse technique. By applying the tools uniformly across the entire set of 25 planets, we extract robust trends in the thermal structure and chemical properties of hot Jupiters not obtained in past studies. With the recent launch of JWST and the upcoming missions Twinkle, and Ariel, population based studies of exoplanet atmospheres, such as the one presented here, will be a key approach to understanding planet characteristics, formation, and evolution in our galaxy. \\

\end{abstract}


\section{Introduction}

More than 4700 exoplanets are currently known and for about 80 of these we have atmospheric data recorded with the Hubble Space Telescope (HST), Spitzer Space Telescope and ground-based instruments. To date, most data-oriented atmospheric analyses have focused on individual targets and only a very limited number of studies are available on populations of exo-atmospheres observed with the transit or eclipse technique \citep{sing_pop, tsiaras_30planets, pinhas, mansfield_2021_pop}. In this article, we analyse  eclipse spectra and photometric data, observed with the HST G141 grism and the Spitzer Space Telescope, for 25 planets. Eclipse spectra recorded in the infrared are sensitive to vertical thermal profiles and can provide strong constraints on the chemistry and thermal properties of the day-sides of exoplanets.
These data, taken collectively and interpreted with a suite of retrieval tools uniformly applied, are used to answer five key open questions in exoatmospheric chemistry, circulation and planet formation. The questions are: \\

\begin{figure*}
    \includegraphics[width = 1\textwidth]{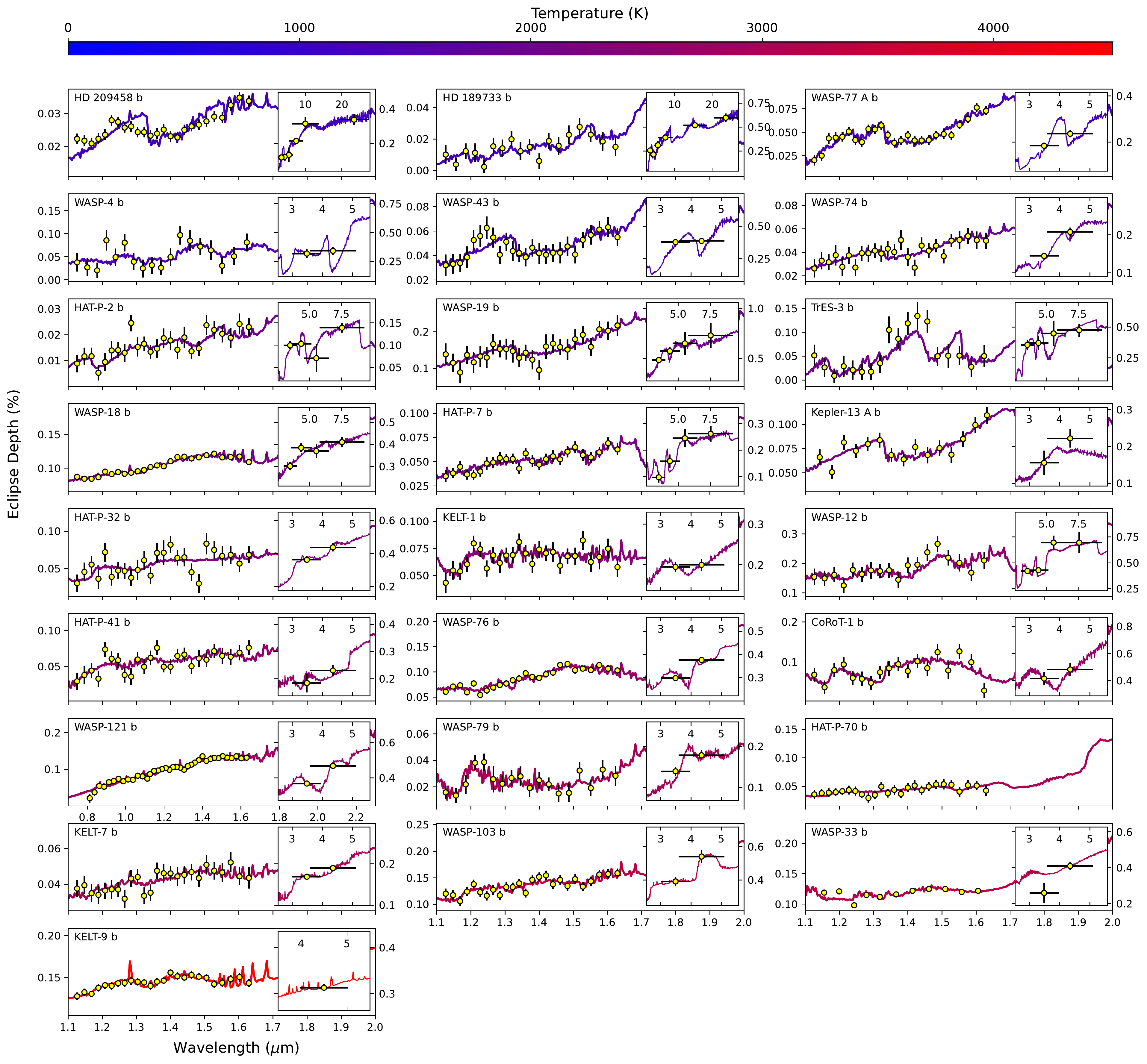}
    \caption{Best fit spectra of the 25 planets considered in this study for the HST+Spitzer observations in eclipse. Individual analyses and additional retrievals, including a simpler Blackbody fit, can be found in the Appendix. The planets are ordered in increasing atmospheric temperatures, which are traced by the colours (from blue to red). The retrieved atmospheric temperature is obtained by weighting the retrieved thermal profile by the contribution function. Unlabeled x-axes range from 1.1\,$\mu$m to 2\,$\mu$m.}
    \label{fig:pop_spectra}
\end{figure*}

1)	{\bf Do metal oxides and hydrides cause thermal inversions in exoplanet atmospheres?} Early theoretical and observational studies of hot Jupiters have highlighted the potential impact of metal hydrides and metal oxides in regulating the thermal properties of the day-side of hot Jupiters \citep{Burrows_spectra_windows, Spiegel_TiO_invertion, Hubeny_2009, Nugroho_W33b}.   Theoretical predictions  \citep{Lodders_2002, Fortney_2008} have suggested that, in analogy to Brown dwarfs, planets hotter than $\sim 1700$\,K might have metal hydrides and metal oxides in the gaseous form: these molecular species are excellent absorbers in the optical /near infrared and might cause a hot layer in the atmosphere, and therefore a thermal inversion. By contrast, planets colder than $\sim 1700$\,K might not exhibit such thermal inversions, as metal hydrides and metal oxides have been sequestered into condensates.
 \\

2)	{\bf Are the eclipse spectra of the hottest atmospheres consistent with blackbody curves?} Recent theoretical studies \citep{Bell_2018, Lothringer_2018, Parmentier_2018_w121photodiss} have suggested H$^-$ being an important opacity source, generated by the thermal dissociation of H$_2$ and H$_2$O at very hot temperatures (T $> 2500\,K$). Eclipse spectra of very hot atmospheres would therefore resemble blackbody curves due to the continuous shape of H$^-$ emission, and the absence of other molecules which have been dissociated. Observational evidence in favor of those predictions were obtained by studies of individual planets with HST and Spitzer \citep{mansfield_hatp7, arcangeli_w18_phase, Kreidberg_w103}, and more recently in the HST population analysis from \cite{mansfield_2021_pop}. However, many other works \citep{Haynes_Wasp33b_spectrum_em,Evans_wasp121_e3,Edwards_2020_ares,Changeat_2021_k9} have found spectral signatures in similarly hot atmospheres.   \\

3)	{\bf What is the dayside-terminator contrast in exoplanet atmospheres?} Planetary atmospheres present vertical and horizontal inhomogeneities: this is particularly true for tidally locked planets, for which large day-night thermal gradients, sometimes in excess of 1000\,K, and asymmetric circulation patterns have been predicted \citep{ Cho_2003, Showman_2010, Cowan_2011, Roth_2021}. Recent simulations using  pseudo-spectral methods \citep{Skinner_2021, Skinner_2021_modons, Cho_2021} have revealed the complex and turbulent nature of these atmospheres, displaying highly dynamic small and large scale storms that develop and evolve in time. Some studies \citep{tan, Roth_2021} have hypothesised that optical absorbers and thermal dissociation/recombination processes would impact the atmospheric dynamics. Namely, if H$_2$ dissociates at the day-side and recombines at the night-side, this would increase the energy transport, thus reducing the day-night temperature contrast. \cite{mansfield_k9} provided observational evidence of these effects in the Spitzer phase-curve of KELT-9\,b. On the other hand, optical absorbers such as TiO, VO and FeH are more thermally stable, so they would not contribute much to this effect. The comparison between eclipse and transit observations might constrain these dynamical processes. \\

4)	{\bf Are metallicity and C/O viable  observables to understand planet formation?} Hot Jupiters are thought to form via a three step process \citep{Mizuno_1980, Bodenheimer_1986, Ikoma_2000}: solid core accretion, runaway gas accretion and migration. The core accretion is believed to occur in the outer regions of a protoplanetary disk, where the abundance of solid materials leads to a rapid growth of a planetary core before disk gas dispersal. If that is the case, the composition of giant exoplanets, which is a direct outcome of these planetary formation processes, is predicted to be sub-stellar in heavy elements such as C, O, and refractory elements, because most of the heavy elements would be sequestered in the cores. The inferred super-stellar bulk metallicities of Jupiter, Saturn and warm exo-Jupiters \citep{Saumon_2004, Miller_2011, Thorngren_2016, Welbanks_2019}, however, suggest a more complicated picture, where heavy elements are also captured through some other processes, probably, during gas accretion and/or migration \citep{Hasegawa_2018, Shibata_2019, Shibata_2020, Turrini_2021}. Also, while difficult to determine today, the C/O ratio of exoplanets may place constraints on where and how the giant planets collect gas and solids in evolving protoplanetary disks \citep{madhu_formation, Mordasini_2016, Brewer_2017, Booth_2017, Eistrup_2018, Cridland_2019}. Constraints on those two parameters would significantly improve our understanding of planetary formation. \\

5)	{\bf Can refractory elements help us understand exoplanet formation?} In addition to the metallicity and C/O ratio, other elemental ratios, such as N/O, S/O \citep{Turrini_2021} or even  refractory elements \citep{Lothringer_2020}, may help constrain planet formation scenarios. Their potential, however, remains unexplored by observational studies, as their tracers are more difficult to detect. While HST is not particularly sensitive to N and S bearing species, refractory elements such as TiO, VO and FeH have been detected previously in eclipse spectra. 

\section{Methodology}

\begin{figure*}
    \includegraphics[width = 0.95\textwidth]{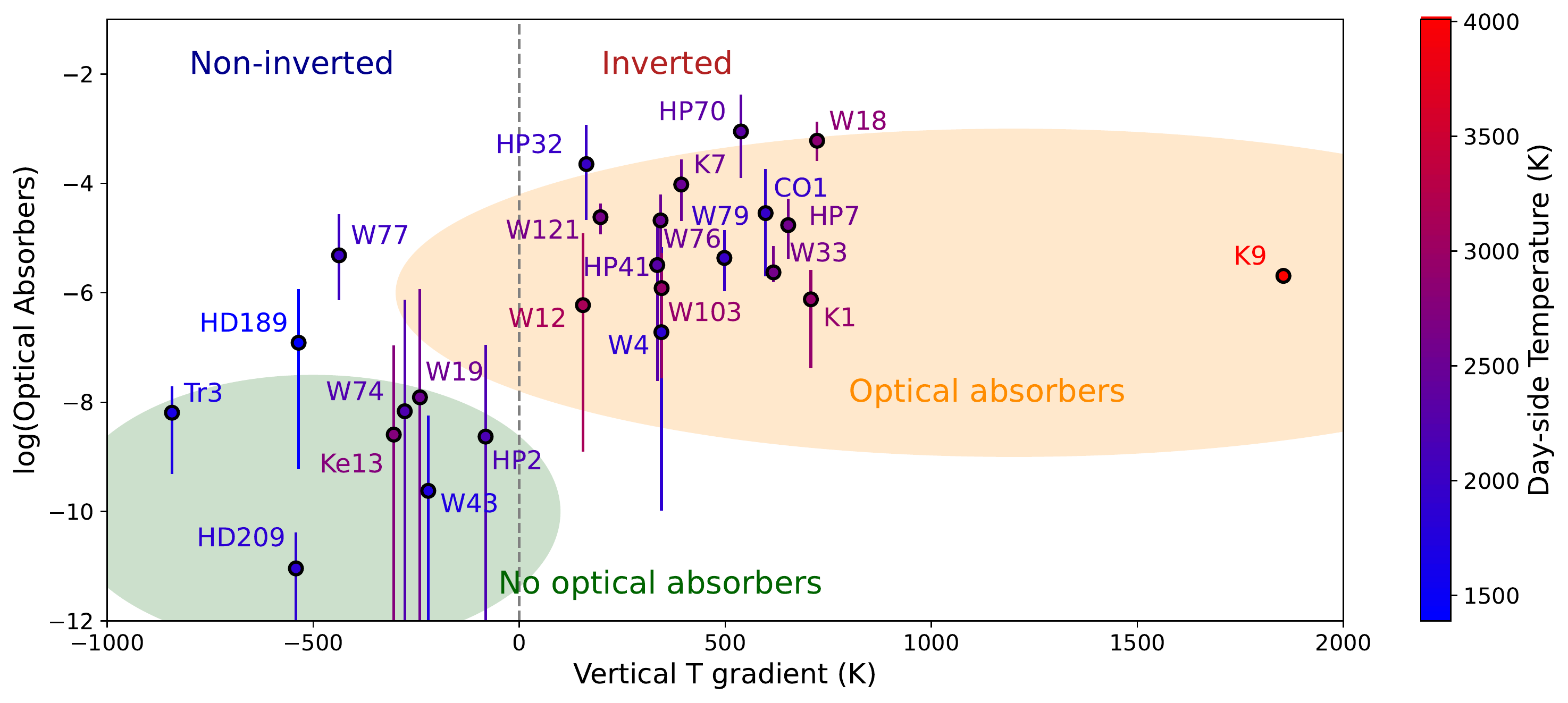}
    \caption{Correlations between the retrieved abundances of optical absorbers  (TiO, VO, FeH and H$^-$) and the thermal gradient in the atmosphere. The optical absorber abundances are estimated using the weighted average of the individually retrieved abundances for the detected molecules. If no optical absorber is detected, only a upper limit is available. The thermal gradient is the difference between atmospheric temperatures at 5\% and 95\% of the atmospheric contribution function. Colors on the datapoints represent averaged retrieved temperatures. Overall, Planets with inverted thermal profiles have temperatures above 2000K and possess optical absorbers (shaded orange region). Planets without thermal inversions have temperatures below 2500K and do not possess optical absorbers (shaded green region).}
    \label{fig:correlate_opt_grad}
\end{figure*}

Our study encompasses data for 25 hot Jupiters observed in eclipse with the HST WFC3 G141 Grism and Spitzer: CoRoT-1\,b (CO1), HAT-P-2\,b (HP2), HAT-P-7\,b (HP7), HAT-P-32\,b (HP32), HAT-P-41\,b (HP41), HAT-P-70\,b (HP70), HD\,189733\,b (HD189), HD\,209458\,b (HD209), KELT-1\,b (K1), KELT-7\,b (K7), KELT-9\,b (K9), Kepler-13\,A\,b (Ke13), TrES-3\,b (Tr3), WASP-4\,b (W4), WASP-12\,b (W12), WASP-18\,b (W18), WASP-19\,b (W19), WASP-33\,b (W33), WASP-43\,b (W43), WASP-74\,b (W74), WASP-76\,b (W76), WASP-77\,A\,b (W77), WASP-79\,b (W79), WASP-103\,b (W103) and WASP-121\,b (W121). For WASP-121\,b, we also add the available G102 Grism. While some of these datasets have already been published individually in previous works, to ensure the consistency of our analysis, we have re-analysed the raw HST data with our open-source pipeline, \emph{Iraclis} \citep{tsiaras_hd209}. Below, we also refer to the data analysis leading from raw observational data to spectra as `data reduction', to distinguish this process from the analysis/interpretation of the atmospheric spectrum.
Only two exoplanets, Kepler-13A\,b and WASP-33\,b, were taken as is from the literature. WASP-33\,b orbits a pulsating $\delta$-Scuti star and Kepler-13A\,b is part of a triple star system. Reduction of the data for these particular targets requires a particularly careful modelling of the host stars, which is not explored in this study. For our retrieval analyses we have used \emph{Alfnoor} \citep{Changeat_2020_alfnoor}, a tool that extends the atmospheric retrieval capabilities of \emph{TauREx3} \citep{2019_al-refaie_taurex3} to populations of atmospheres. More technical details on our tools and methods are reported in Materials and Methods. 

For each planet, we tested predefined scenarios, varying the molecular species included, the model assumptions, and whether Spitzer data are included.
As potential biases in retrieval studies can arise from combining HST and Spitzer observations \citep{Yip_2020_LC,Changeat_2020_K11}, we assessed the robustness of our results against possible biases by artificially modifying the Spitzer data and performing additional retrievals. More specifically, we repeated our analysis with Spitzer data offset by +100ppm, -100ppm, and doubled uncertainties. 

The free retrieval runs assumed constant with altitude abundances and included molecular species such as H$_2$O, CO, CO$_2$, CH$_4$ in a `reduced' run. Refractory species (TiO, VO and FeH) and H$^-$ were added in a `full' run. Please note that H$^-$ absorbing properties are traced by the e$^-$ abundance \citep{John_1988_hmin}. Atomic and ionic species are not considered here as they do not absorb in the HST and Spitzer wavelength regions. In the rest of the paper, we use the term `refractory' to refer to TiO, VO and FeH, and we use the term `optical absorber' to refer to TiO, VO, FeH and H$^-$.   
We have also attempted equilibrium chemistry retrievals, using the GGChem code \citep{Woitke_2018_GG} and including all the supported species, on our entire population. We used the Bayesian Evidence to compare the results with the free runs. In all models, the temperature profile is determined using a heuristic N-point profile. Since the thermal profile varies with altitude, the reported value is the mean atmospheric temperature obtained in the retrievals, weighted by the atmospheric contribution function. It characterizes the thermal conditions of the atmosphere in the region probed by the observations, Note that this is not equal to the Blackbody temperature or the equilibrium temperature. For reference, best-fit Blackbody temperatures are provided in the Appendix D, which analyses the planets individually.

For a subset of 17 planets we have also analysed the G141 transit observations  using \emph{Iraclis}, to include additional information about the terminator region of those planets in our study.


\section{Results}

The HST eclipse observations reduced with our \emph{Iraclis} pipeline are shown in Figure \ref{fig:pop_spectra}, alongside  the Spitzer photometric data.  A summary of the observations considered in this paper and relative references are given in Table \ref{tab:planet_params} of the Appendix. 
We also show in Figure \ref{fig:pop_spectra} the best-fit spectra obtained with \emph{Alfnoor}. The cooler planets in the sample, such as WASP-43\,b, may exhibit H$_2$O features  in absorption (around 1.4\,$\mu$m) and a steep increase in the spectrum towards longer wavelengths. Hotter planets, such as WASP-121\,b, show emission features from optical absorbers (from 1.1 to 1.4\,$\mu$m) and their near-Infrared spectra are more leveled. 

While a detailed analysis of each planet is provided in Appendix D, we summarise our key findings in Table \ref{tab:summary_detections}. 
Overall, after inspecting the retrieval results of our selected sample of 25 planets, we find that the HST-only spectra are difficult to interpret using free retrievals, as these may also converge to unphysical solutions. This behaviour can be explained by the narrow wavelength range covered by HST (see Appendix C), which leads to large degeneracies and does not support the complexity of an emission model where both the thermal profile and atmospheric chemistry have to be disentangled. When Spitzer IRAC observations at 3.6\,$\mu$m and 4.5\,$\mu$m are included, the thermal profiles can be retrieved more easily. In this case, a free chemical retrieval which includes only H$_2$O, CO, CO$_2$, CH$_4$ (reduced model) does not fit well the entire population of exoplanets. To improve our fit, we have to include other plausible absorbers such as refractory species (TiO, VO and FeH) and H$^-$ (full model). These species are referred as optical absorbers throughout this work.
When Spitzer data is considered, the full model obtains the highest Bayesian Evidence \citep{jeffreys1961theory}, ln(E), compared to a simple blackbody fit or a reduced model. The full model is rejected in only one case, i.e. HD 209458\,b (this case is developed further in Appendix B and in the planet's section). Equilibrium chemistry is expected to be a relatively valid assumption to describe  planets in the 1000 to 2000\,K range. However,  atmospheric retrievals following equilibrium chemistry, while not strongly rejected by the Bayesian Evidence (e.g $\Delta$ ln(E) $<$ 3), are almost never favoured as best solutions. Possibly, subtle departures from equilibrium chemistry  may be better captured by the free retrievals. Another possibility could be that the elemental abundances of refractory elements (Ti, V and Fe), for which their ratios (Ti/O, V/O and Fe/O) remain solar in our equilibrium tests, are enhanced (see key question 5 bellow). 

Combining HST and Spitzer observations should be done carefully as it can introduce biases in retrievals \citep{Yip_2021_W96}. We tested the robustness of our conclusions by including offsets in the Spitzer data, or artificially increasing the photometric uncertainties. We find that while planets can be affected individually, our results on the entire population remain unchanged. 

The transit observations and their associated fits are reported in Figure\,\ref{fig:pop_spectra_tr} of Appendix A. 
At the terminator, water vapour is recovered most of the times, independently from the planet's atmospheric temperature. There, atmospheric clouds are also detected in many planets. Concerning the key open questions listed in the Introduction, focusing more on the results of the full model with HST+Spitzer, we find: \\

{\bf 1) Refractory molecules and H$^-$ correlate with thermal inversions.}

We show in Figure\,\ref{fig:correlate_opt_grad} the weighted mean abundances of optical absorbers (TiO, VO, FeH and H$^-$) as a function of the retrieved thermal gradient in the atmospheres of all the planets. Maps displaying the retrieved abundances for each individual species as a function of the atmospheric temperature are available in the Appendix (Figures \ref{fig:correlatee-_fullspz} to \ref{fig:correlateCH4_fullspz}).
We see in Figure\,\ref{fig:correlate_opt_grad} a clear correlation between the retrieved abundances of optical absorbers and the retrieved vertical thermal gradients:  planets with positive thermal gradients show features associated with optical absorbers.  While correlation does not imply causation, refractory elements (here TiO, VO and FeH) and H$^-$ are efficient absorbers of stellar light. In the hottest atmospheres (T $>$ 2000\,K), refractory elements are thermally stable and can remain in the gas phase. They could therefore provide a natural explanation for thermal inversions. At even higher temperatures, H$_2$ is thermally dissociated, leading to abundant H$^-$ opacity. H$^-$ also absorbs in the visible and could be another contributor to the retrieved thermally inverted profiles. \\

{\bf 2a) Spectra of the hottest exoplanets do not resemble blackbodies. }

\begin{figure*}
    \includegraphics[width = 0.95\textwidth]{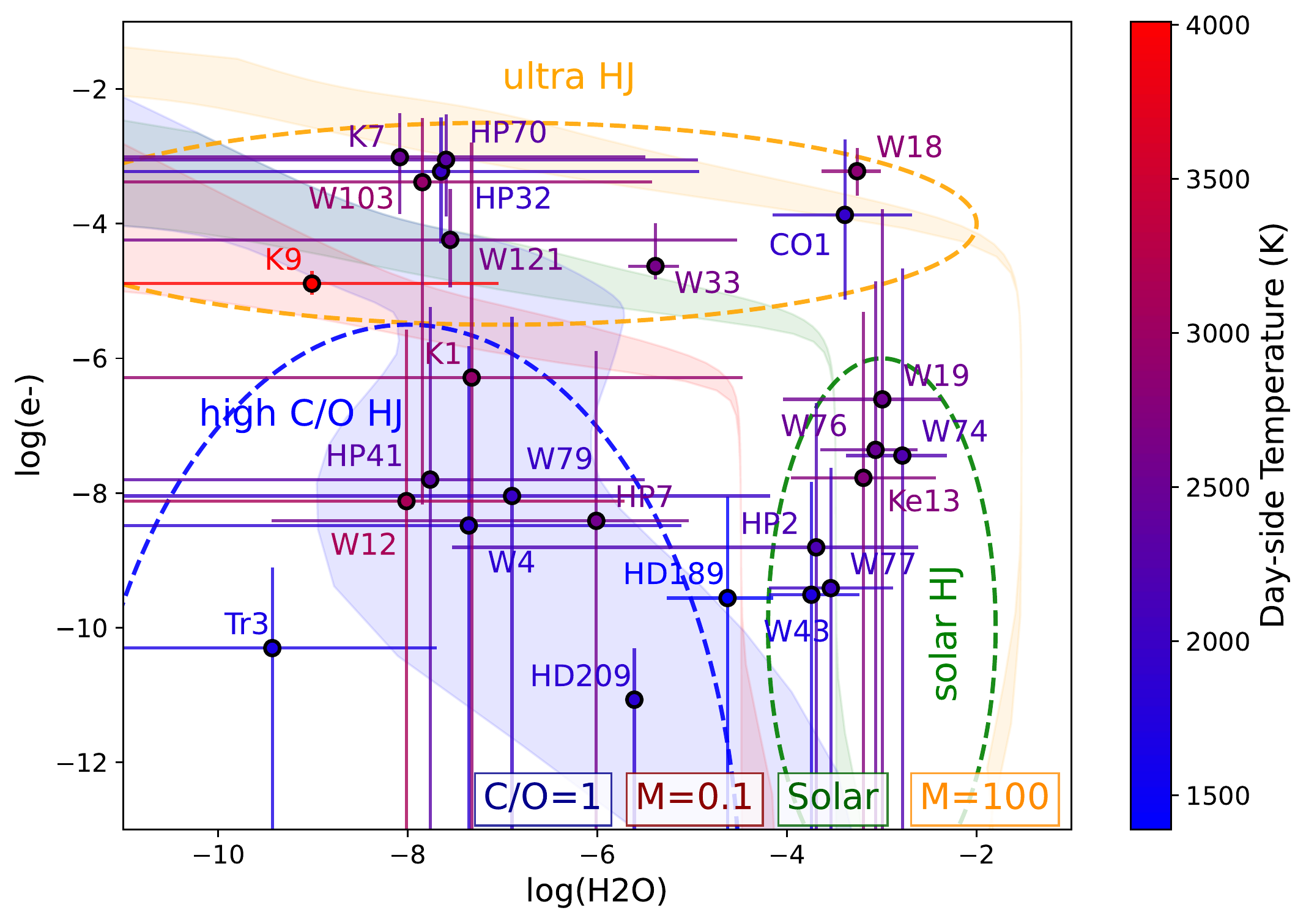}
    \caption{Retrieved abundances of e$^-$ versus H$_2$O at the day-side in our sample of 25 hot Jupiters recovered in the HST+Spitzer full runs. The colours on the datapoints indicate the retrieved atmospheric temperatures weighted by the contribution functions with legend in the colobar. The shaded green region indicates the predicted abundances from equilibrium chemistry at solar metallicity (M=1) and C/O (C/O=0.55)  between 1 bar and 0.01 bar, which is around the region probed by the observations. In orange, the metallicity is increased to 100 times solar. In red, the metallicity is decreased to 0.1 times solar. In blue, the metallicity is solar and the C/O  is increased to 1. The planets separate in three regimes: A solar hot Jupiter regime where water vapour is detected in moderately hot atmospheres associated with a decreasing thermal profile (dashed green); An ultra-hot Jupiter regime where thermal inversions and H$^-$ emission are detected in very hot atmospheres (dashed orange); A high C/O hot Jupiter regime where temperatures remain moderate, but where water is not detected (dashed blue).  }
    \label{fig:correlateH2O_e-}
\end{figure*}

Figure \ref{fig:pop_spectra} demonstrates that the hottest planets in our sample show features associated with molecular species. This is also confirmed by the Bayesian Evidence of retrievals including absorbing species as opposed to featureless blackbody curves as shown in the complementary figure \ref{fig:Bayesian_evidence} of the Appendix). The presence of refractory elements lead to spectral features that make the spectra deviate from the pure blackbody. \\

{\bf 2b)  Dissociation processes do occur in the hottest atmospheres. }

\begin{figure*}
    \includegraphics[width = 1\textwidth]{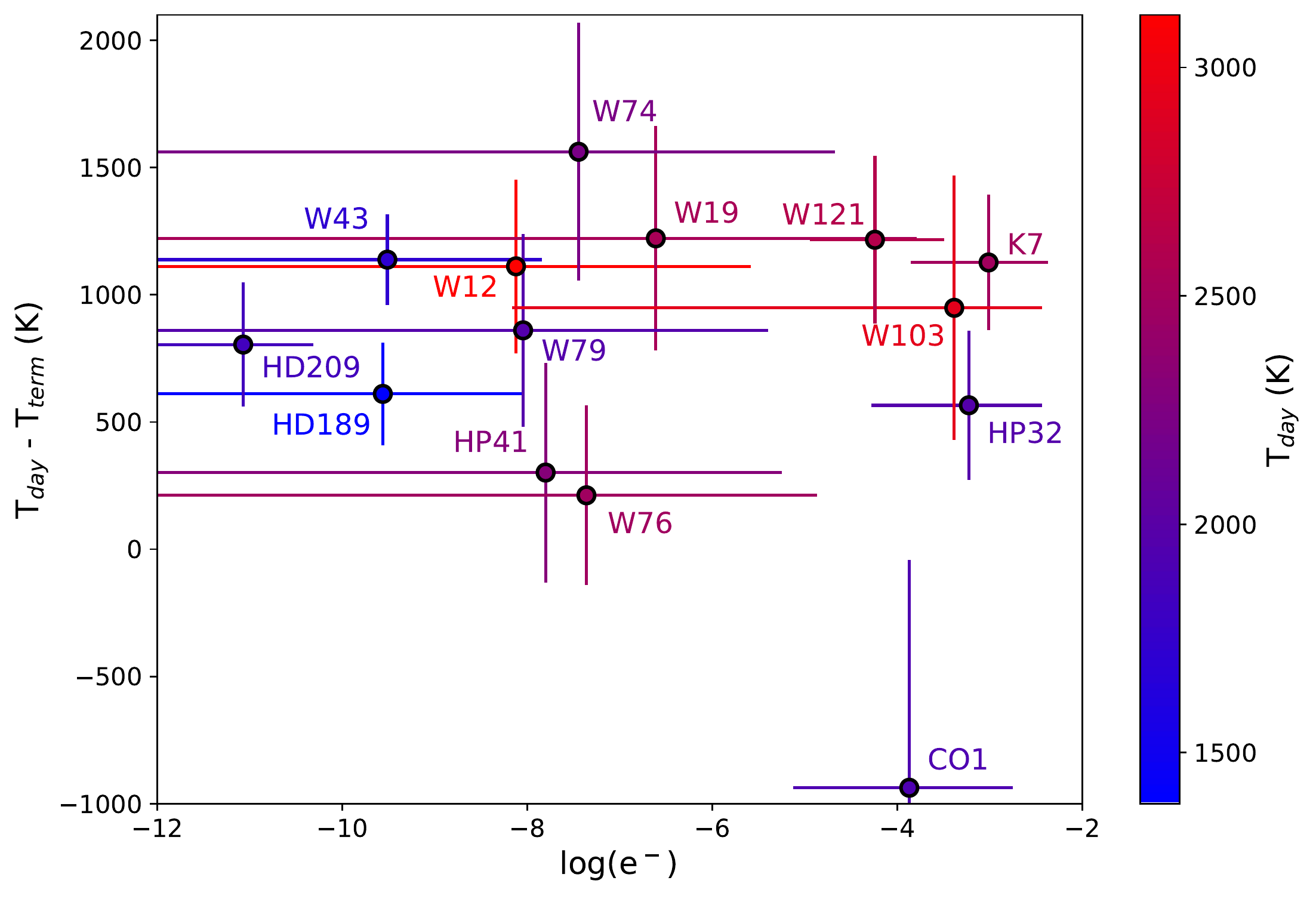}
    \caption{Day-terminator thermal gradient as a function of the e$^-$ abundance recovered using the free retrievals. Only planets that have both eclipse and transit spectra and for which the terminator temperature is constrained are shown in this plot.}
    \label{fig:dynamics_day_term_e-}
\end{figure*}

Our results also confirm the presence of dissociation processes in the hottest atmospheres:  Figure\,\ref{fig:correlateH2O_e-} indicates the apparition of H$^-$ opacity from the dissociation of H$_2$ and H$_2$O. The trend appears to be driven by the retrieved atmospheric temperature: H$_2$O is detected in the spectra of cooler atmospheres (high C/O and solar HJ) and H$^-$ becomes more dominant for atmospheres hotter than 2500\,K (ultra HJ). As demonstrated by the different shaded regions of this figure, the metallicity and C/O ratio impacts the transition from non-dissociated to thermally dissociated regimes. Additional figures are available in the Appendix A, showing the temperature dependence for each of the investigated molecules.  \\

{\bf 3) The day-terminator contrast of exoplanets does not appear to correlate with temperature and/or dissociation.}

In Figure\,\ref{fig:dynamics_day_term_e-_indiv} we compare the atmospheric temperature on the day-side, as retrieved from eclipse spectra, and the atmospheric temperature at the terminator, as retrieved from transit spectra, to the estimated equilibrium temperature. Overall, we find that the day-side is about 20\% hotter than the equilibrium temperature, while the terminator is about 30\% cooler as also expected by theoretical studies \citep{Cho_2003, tan, parmentier}. We observe some scatter across the population in the recovered temperature (see Figure \ref{fig:dynamics_day_term_e-_indiv} in the Appendix), which could be explained by uncertainties in the observation, intrinsic variability in the population or atmospheric temporal variability \citep{Cho_2003, cho_2015, Komacek_2020, Cho_2021, Skinner_2021, Skinner_2021_modons}.  Our retrieved temperature at the terminator is compatible with the predicted temperatures from General Circulation Models (GCMs) at the night-side. This confirms potential biases of transit observations towards lower retrieved temperatures, most likely due to 3D effects or the presence of clouds \citep{Caldas_2019,  MacDonald_2020,  skaf_aresII, Pluriel_2020}.

The presence of H$^-$ on the day-side does not appear to be correlated with an increased of the day-terminator thermal gradient, as seen in Figure\,\ref{fig:dynamics_day_term_e-}. For instance, for CoRoT-1\,b, the terminator is found hotter than the day-side. While our results have large uncertainties, our study cannot confirm that the dissociation and recombination of H$_2$ have large effects on day-terminator contrast \citep{tan, Roth_2021}, as demonstrated in other observational studies. To our knowledge, the only planet for which such effect has been reported is KELT-9\,b \citep{mansfield_k9}, but given the fact that we do not have transit observations for this planet, we are unable to verify this claim. \\

{\bf 4) Planets have solar to sub-solar water vapour abundances, which suggest two reservoirs for planet formation.}

\begin{figure*}
    \includegraphics[width = 0.95\textwidth]{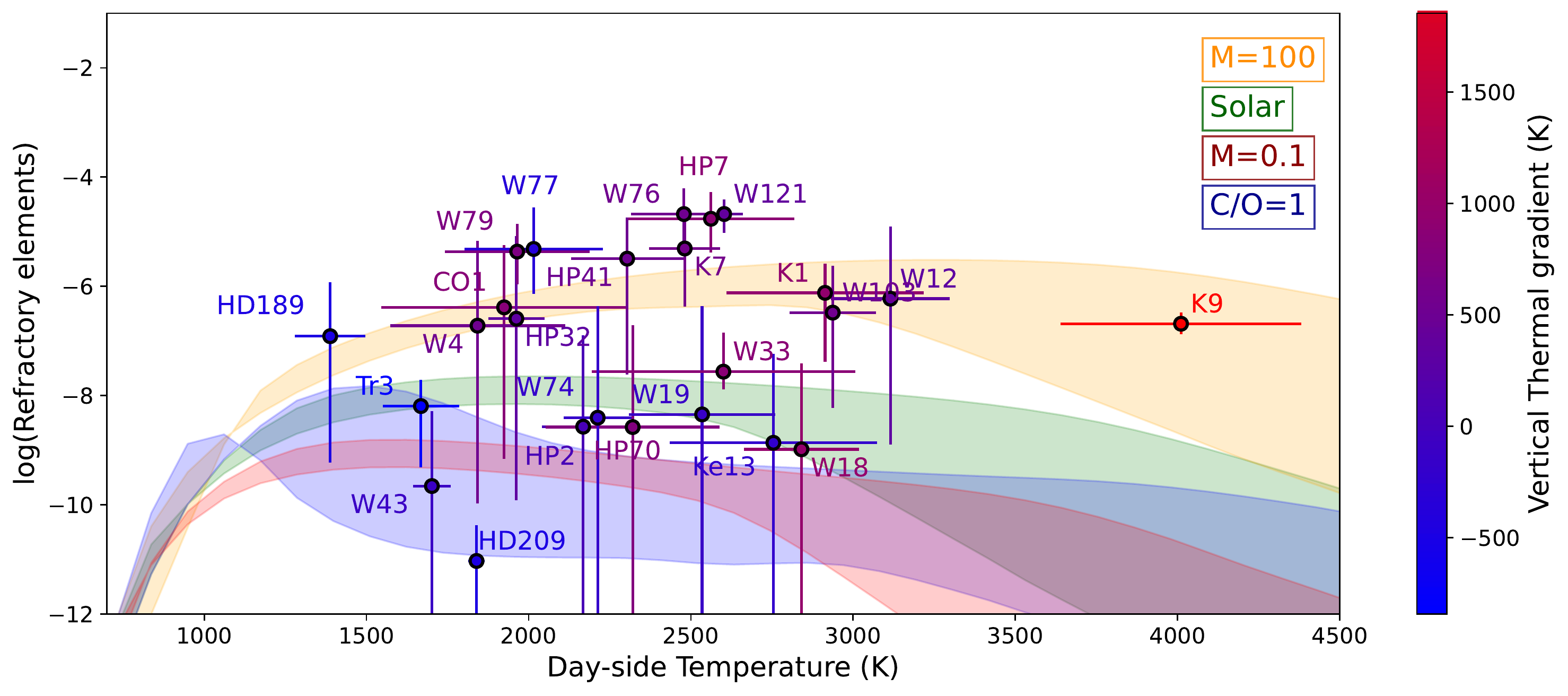}
    \caption{Averaged refractory element (TiO, VO and FeH) abundances recovered in the HST+Spitzer full runs, weighted by their standard deviation, and plotted against the retrieved atmospheric temperature on the day-side. Green area: predictions from a solar composition and equilibrium chemistry between 1 bar and 0.01 bar. Orange area: the metallicity is assumed to be 100 times solar. Red area: the metallicity is assumed to be 0.1 times solar. Blue area: The metallicity is solar and C/O is assumed equal 1.}
    \label{fig:coorelate_opticals.pdf}
\end{figure*}


In most studies, metallicity and C/O ratio of exoplanets are inferred from the detection of water vapour and C-bearing species, or from chemical equilibrium. Here, we demonstrate that for the coolest planets, water vapour -- if detected -- can be used as a proxy for metallicity, allowing us to estimate the O/H ratio. This is shown in Figure\,\ref{fig:met_co_full_spz}, where the estimated metallicity from O/H is consistent with the equilibrium retrievals (see the inset). For these planets, the chemistry of H$_2$O is relatively well understood and should be well described by chemical equilibrium. For the hotter planets (T $>$ 2500\,K ), water vapour dissociates, as shown in point 2), and  H$_2$O  becomes a poor indicator of the planet metallicity. In many cases, inferring metallicity and C/O ratio from water only is degenerate. We attempted to constrain the abundances of the carbon-bearing species and estimate C/O from our free retrievals. At the equilibrium, CH$_4$ is predicted to be scarce in hot atmospheres, while CO should be the main carbon-carrier. We detected evidence of CO only in HD\,209458\,b. Instead, in a few instances, such as in WASP-43\,b or HAT-P-7\,b, we found evidence for high abundances of CO$_2$. These detections are likely driven by the stronger absorption of CO$_2$ in the 4.5\,$\mu$m Spitzer channels and its additional absorption at 1.5\,$\mu$m. We note that, except for HAT-P-7\,b, performing retrievals without CO$_2$ in the cases where it is detected leads to detection of CO instead, albeit with a slightly lower Bayesian evidence ($\Delta$ ln(E) $<$ 5). As such, we believe the recovered abundance for CO$_2$ are not reliable and might here be overestimated.


While directly inferring the C/O from C-bearing species remains difficult, the abundance of water vapour also affects this parameter, which is shown in Figure\,\ref{fig:correlateH2O_e-} by the blue shaded region. In this figure, three regimes are identified. The first regime (orange dashed line) includes the hottest planets for which H$^-$ is detected. We label those as ultra HJ. For cooler planets two reservoirs exist. The first reservoir (green dashed line) contains planets for which water vapour is detected. The recovered abundances are roughly consistent with solar predictions (see also Figure \ref{fig:correlatee-_fullspz}). There are exceptions, such as HD\,189733\,b and HD\,209458\,b, for which the data allows us to detect much lower water abundances. Note also that WASP-18\,b and CoRoT-1\,b have water abundances that are consistent with solar, but those planets also present a particularly high dissociation.
The third reservoir (blue dashed line) encompasses planets which do not show the spectral feature of H$_2$O and H$^-$. The non-detection of water in those planets could be due to either a depletion of the molecule, or other molecules/clouds masking the water signal. If these atmospheres are indeed depleted of H$_2$O, as also suggested in previous studies \citep{Madhu_H2O_HJ, Welbanks_2019, pinhas}, this provides strong constraints on metallicity and C/O ratio. The depletion could be explained by either an overall sub-solar metallicity, or alternatively a high C/O (see blue region in Figure \ref{fig:correlateH2O_e-}). When this information is combined with point 5), we argue that the latter is more probable. At present, the bulk metallicities and C/O ratios of exoplanets are difficult to constrain from the HST and Spitzer data only, partly due to the limited amount of tracers that our observations are sensitive to. Future telescopes will have the potential to characterise the metallicity and C/O for exoplanets very accurately and confirm these predictions. \\

{\bf 5) A contradiction exists between the abundances of volatiles and refractory elements.}

For the hotter planets, we can constrain refractory elements from the abundance of metal oxides and hydrides. As their individual abundances could be unreliable due to their  overlapping features, we show in Figure\,\ref{fig:coorelate_opticals.pdf} their averaged retrieved abundances weighted by standard deviation. Shaded regions demonstrate that refractory species (TiO, VO and FeH) are expected to condensate at the cooler temperatures (T $<$ 1300K) and thermally dissociate for temperatures higher than 3500K. For intermediate temperatures, they can exist in a gaseous form and be detectable. For planets hotter than 2500\,K, at least one metal oxide/hydride is present and, interestingly, the recovered weighted abundance is higher than the one predicted from equilibrium chemistry for solar metallicity values. However, it is compatible with 100 times solar. While we cannot entirely exclude systematic errors from  missing molecules in our retrievals or other biases arising from model assumptions, this result contradicts our estimates of the metallicity from water vapour, which suggested solar to sub-solar water abundances (see Figure\,\ref{fig:correlateH2O_e-}). If confirmed, our result suggests that the water depletion identified here and found in previous studies, would come from a depletion in oxygen rather than an overall sub-solar metallicity, which would have strong implications for planetary formation. We note that a recent study \citep{Welbanks_2019} reached the same conclusion when using Alkali as secondary tracers of metallicity, which were also found to have super-solar abundances.



\section{Discussion and Conclusion} \label{sec:conc}

We have presented here the first retrieval population study of exoplanet atmospheres observed in eclipse with the Hubble and Spitzer Space Telescopes. Our sample includes 25 hot gaseous planets. When combining the HST data with the available Spitzer data,  the trends in our population are stable to changes of $\pm$100\,ppm in the Spitzer data, and an increase of observed noise by a factor two.  Overall, we have found that:

$\bullet$ the coolest planets in our sample (T $<$ 2000\,K) have non-inverted thermal profiles with signatures of water absorption.

$\bullet$ The hottest planets in our sample (T $>$ 2000\,K) have  inverted thermal profiles with signatures from thermal dissociation (H$^-$) and refractory species (TiO, VO or FeH). Their spectra in the HST wavelength range are not consistent with a simple blackbody emission.

$\bullet$ The dayside-terminator thermal gradient is not found to be correlated with equilibrium temperature or H$^-$ opacity.

$\bullet$ Metallicity and C/O  are difficult quantities to constrain from free retrievals of current data. Our results suggest  water abundance being solar to sub-solar in the sample analyses. If confirmed, this result would have important implications on planet formation.

$\bullet$ Metal oxides and hydrides are found in excess of solar abundances in the hotter planets, thus contrasting with our results on the water abundances. If confirmed, this result would inform our current understanding of giant planets' formation.

$\bullet$ For a number of planets in our sample, equilibrium chemistry retrievals are not the preferred solution. While we cannot strongly reject equilibrium chemistry, this suggests that disequilibrium mechanisms might be important and highlight the importance of carrying unbiased free retrieval approaches. Evidence for dis-equilibrium processes have also been found in previous transiting studies \citep{Roudier_2021, Kawashima_2021}. 

Population studies, such as the one presented here, pave the way to future studies based on the next generation of space telescopes. In the next decade, JWST \citep{gardner_jwst}, Twinkle \citep{Edwards_twinkle} and Ariel \citep{Tinetti_ariel,Tinetti_2021_redbook} will provide atmospheric data for thousands of diverse worlds, enabling the study of chemical regimes, circulation patterns and formation mechanisms well beyond the  parameter space explored here.   \\




\begin{table*} 
\centering
\begin{tabular}{|c|cccccc|}\hline\hline
Planet & Term detections & Term clouds & Term T(K) & Day detections & Day T(K) & Day profile \\ \hline\hline 
CoRoT-1\,b & { \color{Orange} VO} & No & 2862$^{+537}_{-1105}$ & { \color{ForestGreen} H$_2$O, VO, H$^-$} & 1924$\pm$374 & Inverted \\
HAT-P-2\,b &  - & - & - & { \color{Red} H$_2$O, VO} & 2168$\pm$123 & Not Inverted \\
HAT-P-7\,b & None & Featureless & Unknown & { \color{ForestGreen} $^*$H$_2$O, $^*$FeH, $^*$CO$_2$} & 2562$\pm$253 & Inverted \\
HAT-P-32\,b & { \color{ForestGreen} H$_2$O} & Yes & 2396$^{+261}_{-314}$ & { \color{Red} H$^-$} & 1962$\pm$83 & Inverted \\
HAT-P-41\,b & { \color{ForestGreen} H$_2$O} & Yes & 2002$^{+364}_{-446}$ & None & 2304$\pm$169 & Not Inverted \\
HAT-P-70\,b (HST) &  - & - & - & { \color{Orange} H$^-$} & 2321$\pm$263 & Inverted \\
HD 189733\,b & { \color{ForestGreen} H$_2$O} & Yes & 778$^{+179}_{-116}$ & { \color{ForestGreen} $^*$H$_2$O, $^*$CO$_2$,} { \color{Red} $^*$FeH} & 1388$\pm$104 & Not Inverted \\
HD 209458\,b & { \color{ForestGreen} H$_2$O} & Yes & 1035$^{+285}_{-182}$ & { \color{ForestGreen} $^*$H$_2$O, $^*$CO, $^*$CH$_4$} & 1839$\pm$21 & Not Inverted \\
KELT-1\,b & None & Featureless & Unknown & { \color{ForestGreen} FeH, VO} & 2913$\pm$300 & Inverted \\
KELT-7\,b & { \color{ForestGreen} H$_2$O, H$^-$} & No & 1354$^{+243}_{-263}$ & { \color{Orange} $^*$CH$_4$, $^*$TiO, $^*$VO, H$^-$} & 2480$\pm$105 & Inverted \\
KELT-9\,b & - & - & - & { \color{ForestGreen} TiO, VO, FeH, $^*$H$^-$} & 4011$\pm$367 & Inverted \\
Kepler-13A\,b & - & - & - & { \color{ForestGreen} H$_2$O} & 2754$\pm$316 & Not Inverted \\
TrES-3\,b & - & - & - & None & 1668$\pm$114 & Inverted \\
WASP-4\,b & - & - & - & None & 1842$\pm$265 & Not Inverted \\
WASP-12\,b & { \color{ForestGreen} H$_2$O} & Yes & 2003$^{+266}_{-289}$ & { \color{ForestGreen} $^*$CO$_2$} & 3114$\pm$179 & Inverted  \\
WASP-18\,b & None & Featureless & Unknown & { \color{ForestGreen} H$_2$O, H$^-$} & 2841$\pm$174 & Inverted  \\
WASP-19\,b & { \color{ForestGreen} H$_2$O} & No & 1313$^{+418}_{-284}$ & { \color{ForestGreen} H$_2$O} & 2535$\pm$221 & Not Inverted \\
WASP-33\,b & - & - & - & { \color{ForestGreen} H$_2$O, TiO, H$^-$} & 2600$\pm$402 & Inverted \\
WASP-43\,b & { \color{ForestGreen} H$_2$O} & No &  564$^{+147}_{-140}$ & { \color{ForestGreen} $^*$H$_2$O, $^*$CO$_2$} & 1701$\pm$54 & Not Inverted \\
WASP-74\,b & { \color{Red} H$_2$O, CH$_4$} & Yes &  650$^{+521}_{-329}$ &  { \color{ForestGreen} $^*$H$_2$O, $^*$CH$_4$} & 2212$\pm$101 & Not Inverted \\
WASP-76\,b &  { \color{ForestGreen} H$_2$O} & Yes & 2267$^{+267}_{-309}$ & { \color{ForestGreen} H$_2$O, TiO} & 2478$\pm$159 & Inverted \\
WASP-77\,A\,b &  - & - & - & { \color{ForestGreen} H$_2$O, TiO, $^*$CH$_4$, $^*$CO$_2$} & 2015$\pm$210 & Not Inverted \\
WASP-79\,b &  { \color{ForestGreen} H$_2$O, H$^-$} & No & 1105$^{+410}_{-275}$ & { \color{Orange} VO, FeH} & 1965$\pm$219 & Inverted \\
WASP-103\,b &  { \color{ForestGreen} VO, TiO} & No & 1988$^{+499}_{-399}$ & { \color{Red} VO, FeH, $^*$CO$_2$, $^*$H$^-$} & 2937$\pm$129 & Inverted \\
WASP-121\,b &  { \color{ForestGreen} H$_2$O, H$^-$} & No & 1386$^{+340}_{-366}$ & { \color{ForestGreen}  $^*$TiO, VO, H$^-$} & 2602$\pm$53 & Inverted \\
\hline\hline
\end{tabular}%
\caption{Summary of our `full' retrievals on HST+Spitzer eclipse data (Day) and the HST transit data (Term). The colours reflect how trustworthy molecular detections are by comparing the Bayesian Evidence to the simpler models, $\Delta$ln(E). Green: Decisive evidence (e.g: $\Delta$ln(E) $>$ 5); Orange: Strong evidence (e.g: $\Delta$ln(E) $>$ 3); Red: Tentative evidence (e.g: $\Delta$ln(E) $>$ 1). Molecules that are only detected when Spitzer is added are marked with a star symbol ($^*$). We also report if clouds were found at the planet's terminator and we indicate the retrieved temperature. The stated temperature is the retrieved atmospheric temperature, weighted by the contribution function. This temperature is not equal to the temperature obtained by the simpler Blackbody fit, which we provide in the individual planet analyses in Appendix D. }
\label{tab:summary_detections}
\end{table*}

\section*{Acknowledgements}

This work utilised the OzSTAR national facility at Swinburne University of Technology. The OzSTAR program receives funding in part from the Astronomy National Collaborative Research Infrastructure Strategy (NCRIS) allocation provided by the Australian Government. This work utilised the Cambridge Service for Data Driven Discovery (CSD3), part of which is operated by the University of Cambridge Research Computing on behalf of the STFC DiRAC HPC Facility (www.dirac.ac.uk). The DiRAC component of CSD3 was funded by BEIS capital funding via STFC capital grants ST/P002307/1 and ST/R002452/1 and STFC operations grant ST/R00689X/1. DiRAC is part of the National e-Infrastructure.

The authors wish to thank an anonymous reviewer for providing constructive suggestions that greatly improved the manuscript. Also, the authors wish to thank the AAS Editors for advising and helping with the presentation of our data and results.

\section*{Funding sources}

We acknowledge the funding bodies that provided support to this research: \\
- European Research Council (ERC), ExoAI 758892: QC, BE, KY, IW. \\
- Science and Technology Funding Council (STFC), ST/K502406/1: QC, BE, GT. \\
- Science and Technology Funding Council (STFC), ST/P000282/1: QC, BE, GT. \\
- Science and Technology Funding Council (STFC), ST/P002153/1: QC, BE, GT. \\
- Science and Technology Funding Council (STFC), ST/S002634/1: QC, BE, GT. \\
- Science and Technology Funding Council (STFC), ST/T001836/1: QC, BE, GT. \\
- Science and Technology Funding Council (STFC), ST/P000592/1: JS. \\
- UK Space Agency (UKSA), ST/W00254X/1: QC. \\
- Paris Region Fellowship Programme under the Marie Sklodowska-Curie grant agreement no. 945298: BE. \\
- University College London, UCL Cities Partnership London-Paris: QC, BE, IW, GT, OV. \\
- Centre National d'\'{E}tudes Spatiales (CNES): OV. \\
- CNRS/INSU Programme National de Plan\'etologie (PNP): OV. \\
- Agence Nationale de la Recherche (ANR), EXACT ANR-21-CE49-0008-01: OV. \\
- Japan Society for the Promotion of Science (JSPS), KAKENHI JP18H05439: MI, SS. \\
- Japan Society for the Promotion of Science (JSPS), KAKENHI JP17H01153: MI. \\
- Japan Society for the Promotion of Science (JSPS), Core-to-Core Program, Planet$^2$: MI, SS. \\
- UK Research and Innovation-Science and Technology Funding Council (UKRI-STFC) studentship: MFB.

\section*{Author contributions}

QC and BE contributed equally to this work.\\
- Data reduction: BE, QC, AT. \\
- Retrieval analysis: QC, BE, AA, IW. \\
- Code development (\emph{Iraclis}, \emph{TauREx}, \emph{Alfnoor}): QC, AT, AA, IW. \\
- Scientific interpretation: all authors contributed to the scientific interpretation of the results. \\
- Paper redaction: all authors contributed to the redaction of the manuscript. 

\section*{Competing interests}

Authors declare that they have no competing interests.

\section*{Data and materials availability}

This work is based upon observations with the NASA/ESA Hubble Space Telescope, obtained at the Space Telescope Science Institute (STScI) operated by AURA, Inc. The raw data used in this work are available from the Hubble Archive which is part of the Mikulski Archive for Space Telescopes. This work is also based in part on observations made with the Spitzer Space Telescope, which is operated by the Jet Propulsion Laboratory, California Institute of Technology, under a contract with NASA. We are thankful to those who operate these telescopes and their corresponding archives, the public nature of which increases scientific productivity and accessibility \citep{Peek_2019}. 

Our HST reduction pipeline, \emph{Iraclis}\footnote{https://github.com/ucl-exoplanets/Iraclis}, and our atmospheric retrieval code \emph{TauREx3}\footnote{https://github.com/ucl-exoplanets/TauREx3\_public}, are publicly available on Github and can be installed from pypi. 
The chemical code \emph{GGChem} is also publicly available. Its plugin version, compatible with the \emph{TauREx3} framework will be released upon acceptance of this manuscript. 
To perform our retrievals on the entire population of 25 exoplanets, we utilised the \emph{Alfnoor} code. While a public version of \emph{Alfnoor} is not yet available, it uses the \emph{TauREx} retrieval code, ensuring that our results on individual planets can be reproduced. The data products of this manuscript are available at: \url{https://github.com/QuentChangeat/HST_WFC3_Population}.

\bibliographystyle{aasjournal}
\bibliography{main}

\clearpage
\section*{\centering Appendix for: ``Five key exoplanet questions answered via the analysis of 25 hot Jupiter atmospheres in eclipse''.}

{\centering
{\bf Quentin Changeat, Billy Edwards}, \\ 
Ahmed F. Al-Refaie, Angelos Tsiaras, Jack Skinner, James Cho,  \\ 
Kai Hou Yip, Lara Anisman, Masahiro Ikoma, Michelle Bieger, \\
Olivia Venot, Sho Shibata, Ingo P. Waldmann and Giovanna Tinetti.\\}

\bigskip
\bigskip
This Appendix file contains A - Complementary Figures to the Main Article including Figures A1-A8, B - Materials and Methods including Figures B1-B5 and Tables B1-B5, C - Supplementary Text including Figures C1-C2 and D - Individual Planet Analyses including Figures D1-D3 and Tables D1.

\clearpage

\renewcommand\thesection{\Alph{section}}
\renewcommand\thesubsection{\thesection.\arabic{subsection}}
\value{section} = 0

\section{Complementary Figures to the Main Article}

This appendix contains the complementary figures to the main article, Figure \ref{fig:pop_spectra_tr} to Figure \ref{fig:met_co_full_spz}.

\renewcommand{\thefigure}{A\arabic{figure}}
\setcounter{figure}{0}

\begin{figure}[H]
    \includegraphics[width = \textwidth]{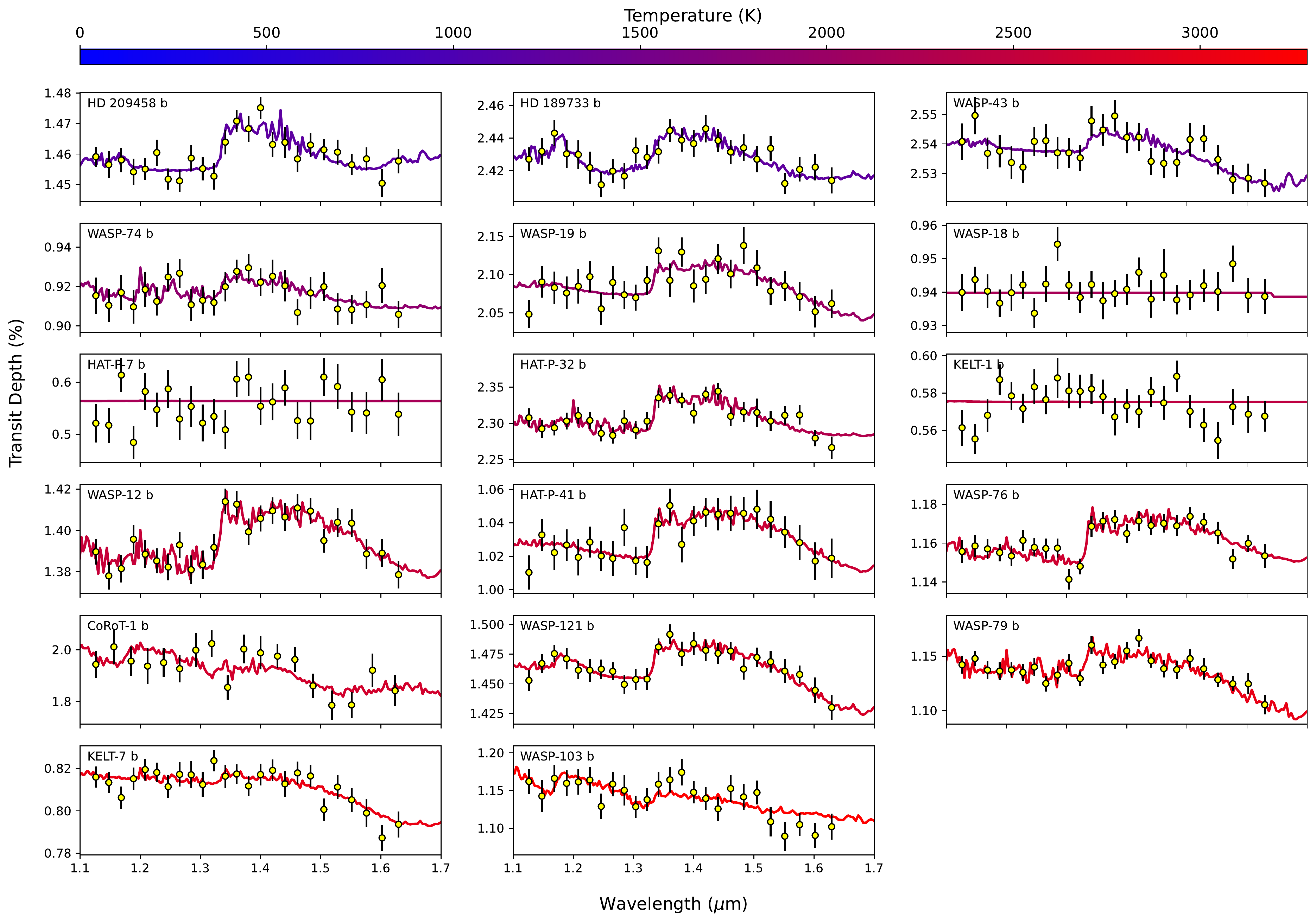}
    \caption{Best fit spectra of the full runs for the planets observed in transit. The planets are ordered in increasing atmospheric temperatures, which are traced by the colours (from blue to red). Unlabeled x-axes range from 1.1\,$\mu$m to 2\,$\mu$m. }
    \label{fig:pop_spectra_tr}
\end{figure}

\begin{figure}[H]
\centering
    \includegraphics[width = 0.74\textwidth]{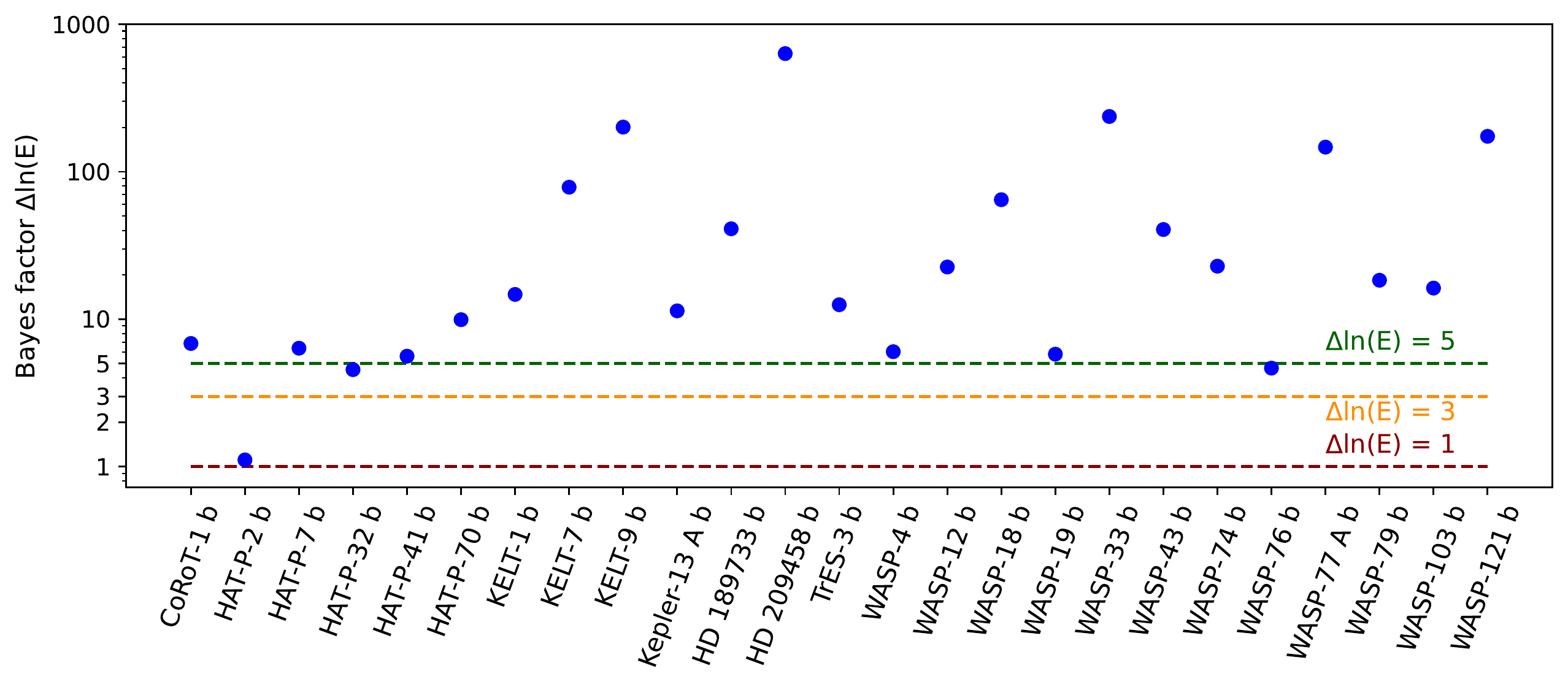}
    \caption{Bayes factor between the full HST+Sptizer retrieval and the Blackbody fit for the eclipses. The figure shows that a Blackbody is rejected for all planets in the combined dataset, except HAT-P-2\,b. }
    \label{fig:Bayesian_evidence}
\end{figure}

\begin{figure}[H]
    \includegraphics[width = 0.95\textwidth]{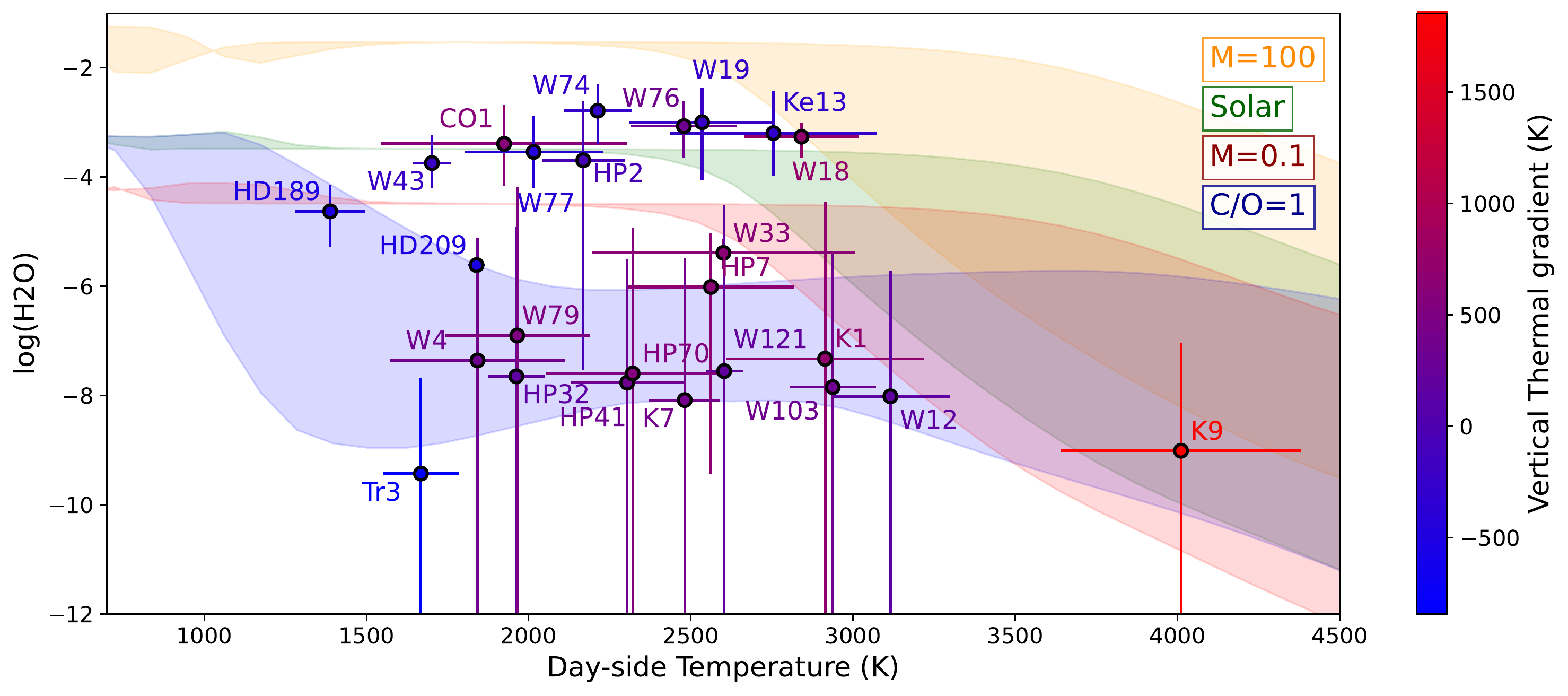}
     \includegraphics[width = 0.95\textwidth]{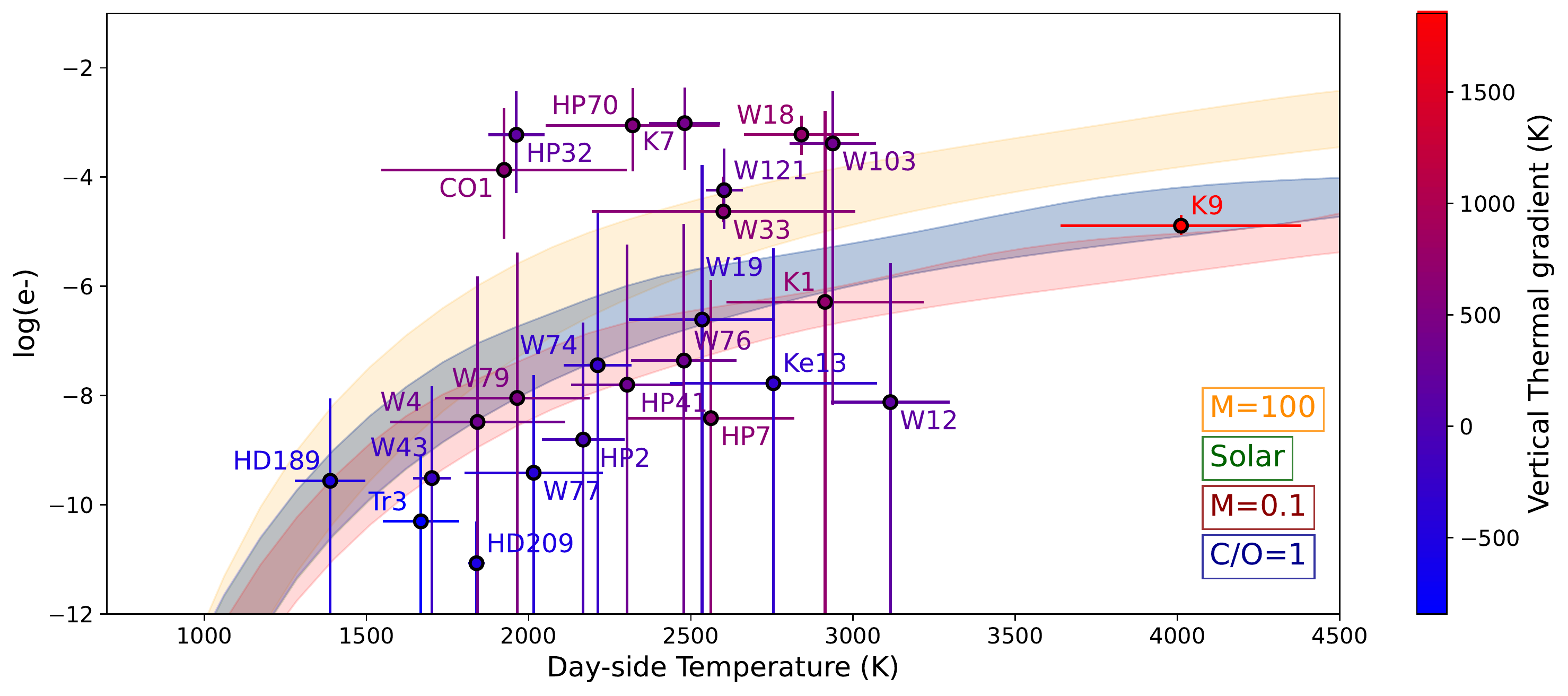}
    \caption{Correlations at the day-side between the H$_2$O (top) and e$^-$ (bottom) abundances and the mean retrieved temperature weighted by the contribution function. The colours indicate the retrieved thermal gradient, defined as the temperature differences between the upper and lower quantiles of the contribution function. The shaded green region is the predicted abundances from equilibrium chemistry at solar metallity and C/O ratio between 1 bar and 0.01 bar. In orange, the metallicity is increased to 100 times solar. In red, the metallicity is decreased to 0.1 times solar. In blue, the metallicity is solar and the C/O ratio is increased to 1. Note that in the e$^-$ plot (bottom), the Solar and C/O=1 cases overlap.}
    \label{fig:correlatee-_fullspz}
\end{figure}

\begin{figure}[H]
    \includegraphics[width = 0.95\textwidth]{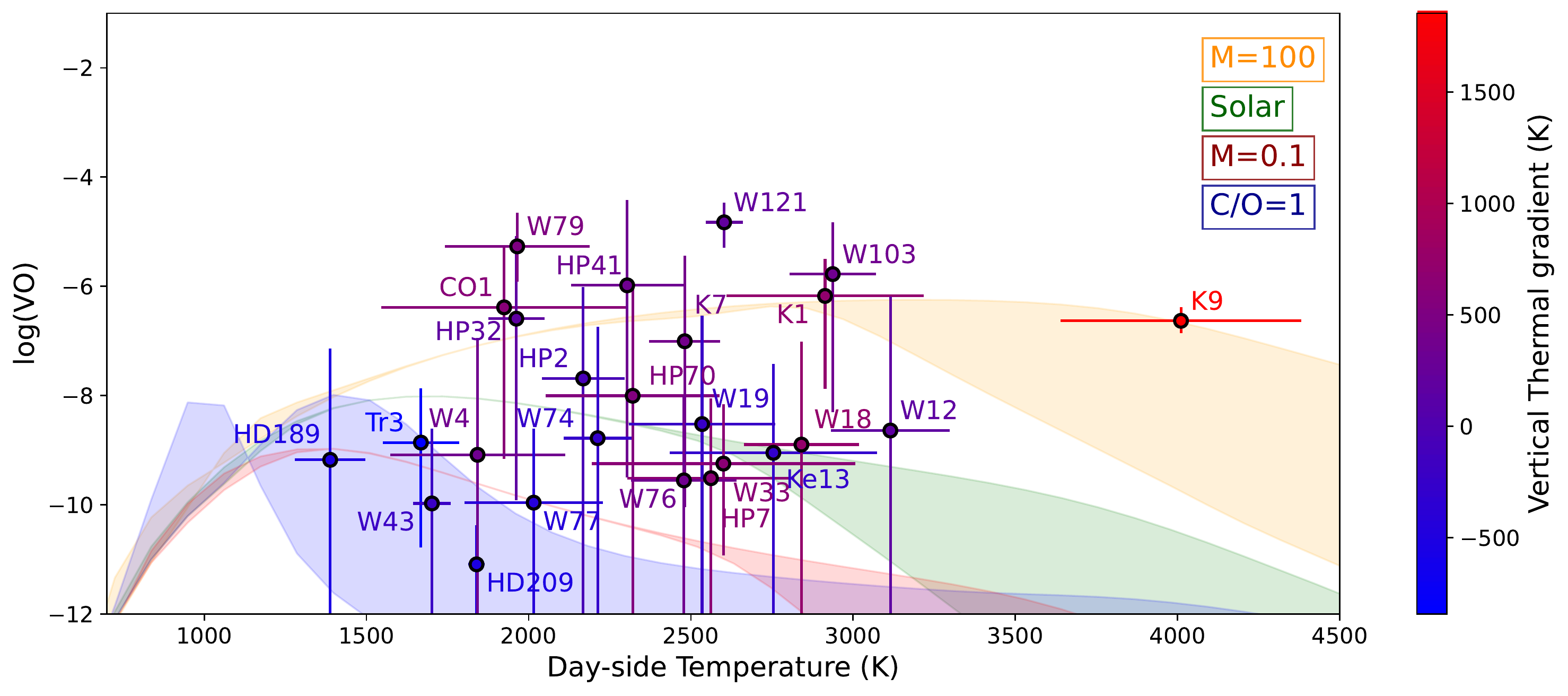}
    \includegraphics[width = 0.95\textwidth]{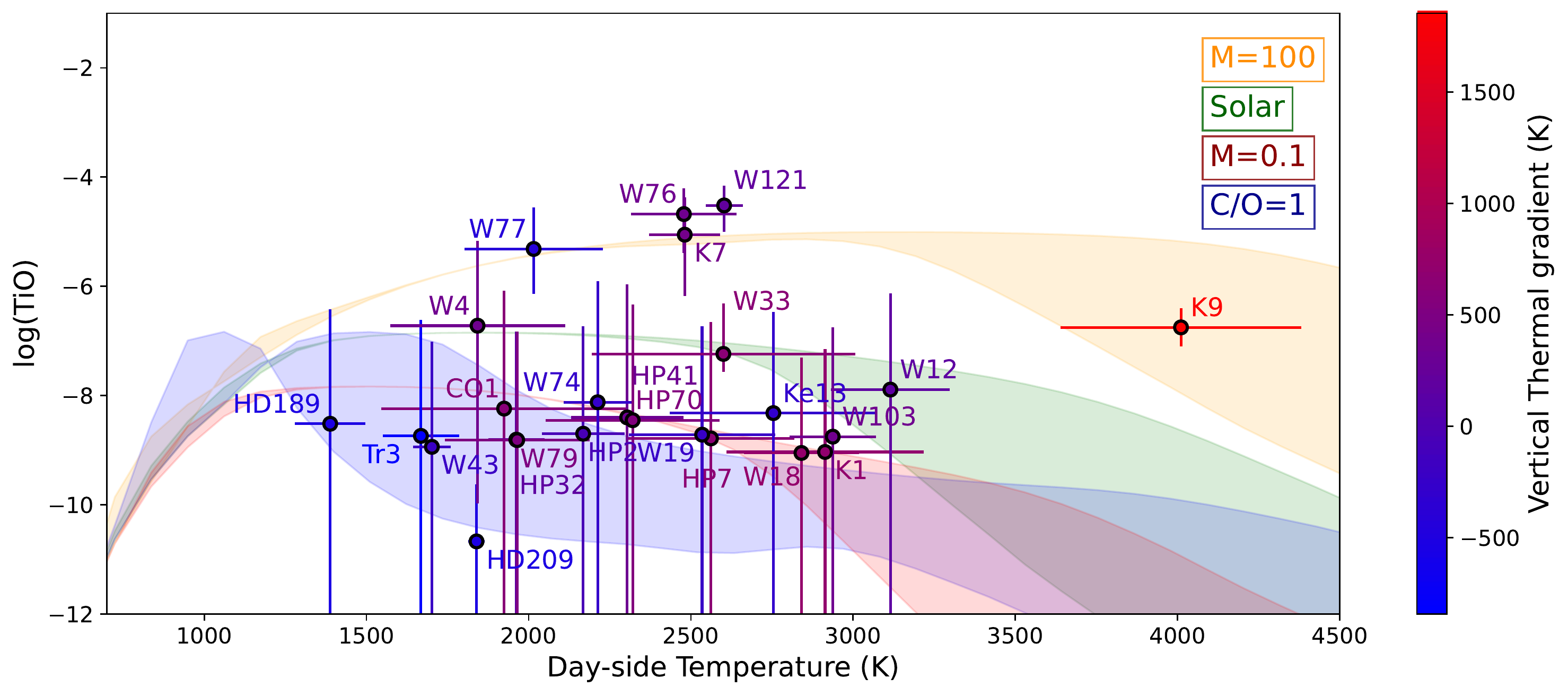}
    \caption{Correlations at the day-side between the VO (top) and TiO (bottom) abundances and the mean retrieved temperature weighted by the contribution function. The colours indicate the retrieved thermal gradient, defined as the temperature differences between the upper and lower quantiles of the contribution function. The shaded green region is the predicted abundances from equilibrium chemistry at solar metallity and C/O ratio between 1 bar and 0.01 bar. In orange, the metallicity is increased to 100 times solar. In red, the metallicity is decreased to 0.1 times solar. In blue, the metallicity is solar and the C/O ratio is increased to 1.}
    \label{fig:correlateVO_fullspz}
\end{figure}

\begin{figure}[H]
    \includegraphics[width = 0.95\textwidth]{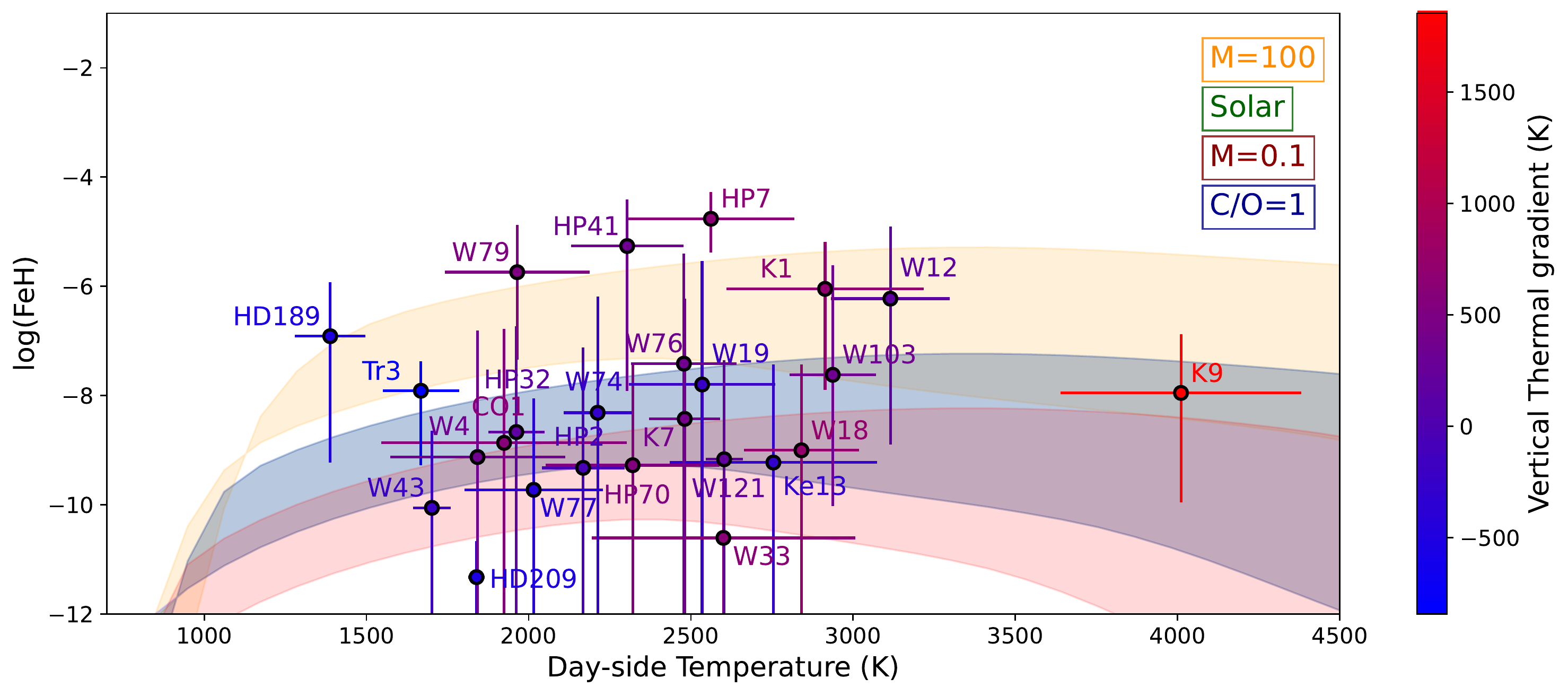}
    \includegraphics[width = 0.95\textwidth]{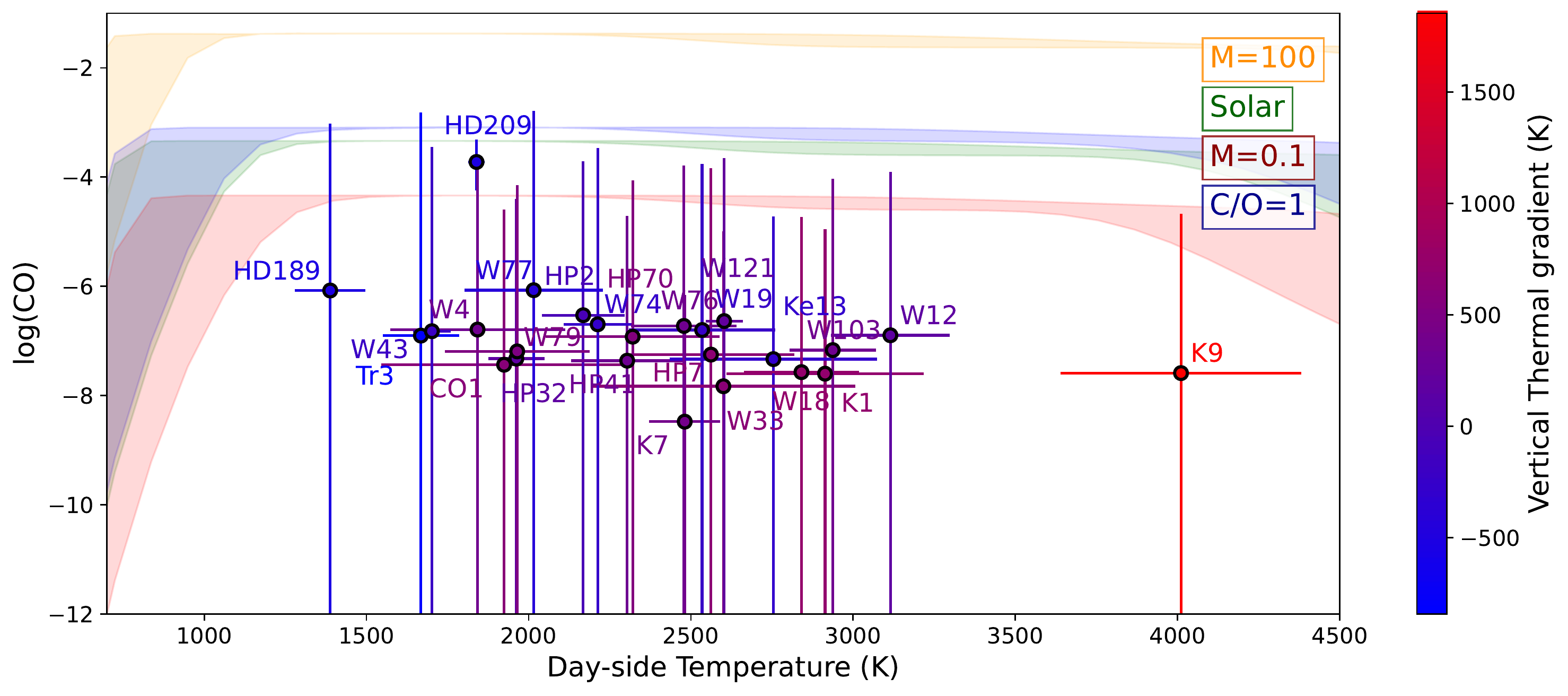}
    \caption{Correlations at the day-side between the FeH (top) and CO (bottom) abundances and the mean retrieved temperature weighted by the contribution function. The colours indicate the retrieved thermal gradient, defined as the temperature differences between the upper and lower quantiles of the contribution function. The shaded green region is the predicted abundances from equilibrium chemistry at solar metallity and C/O ratio between 1 bar and 0.01 bar. In orange, the metallicity is increased to 100 times solar. In red, the metallicity is decreased to 0.1 times solar. In blue, the metallicity is solar and the C/O ratio is increased to 1. Note that in the FeH plot (top), the Solar and C/O=1 cases overlap.}
    \label{fig:correlateCO_fullspz}
\end{figure}

\begin{figure}[h]
    \includegraphics[width = 0.95\textwidth]{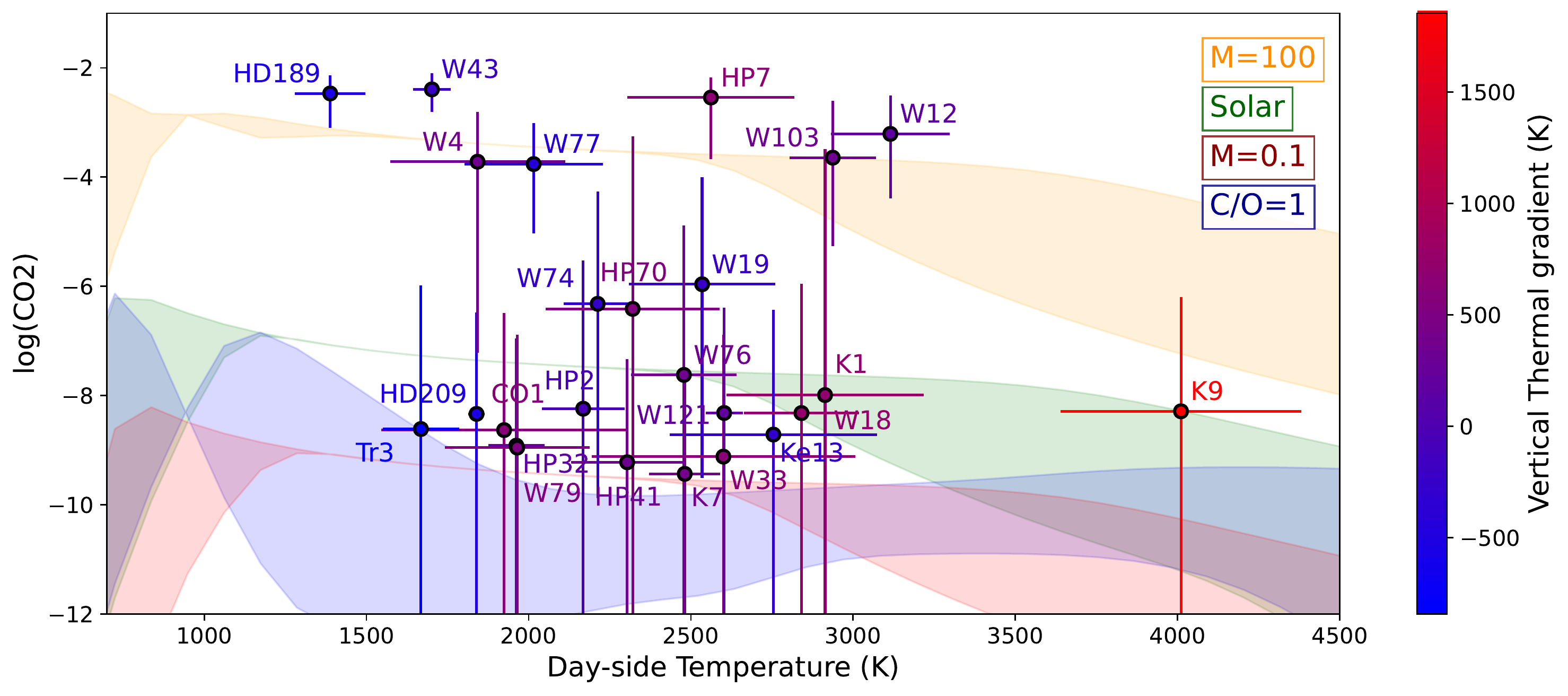}
    \includegraphics[width = 0.95\textwidth]{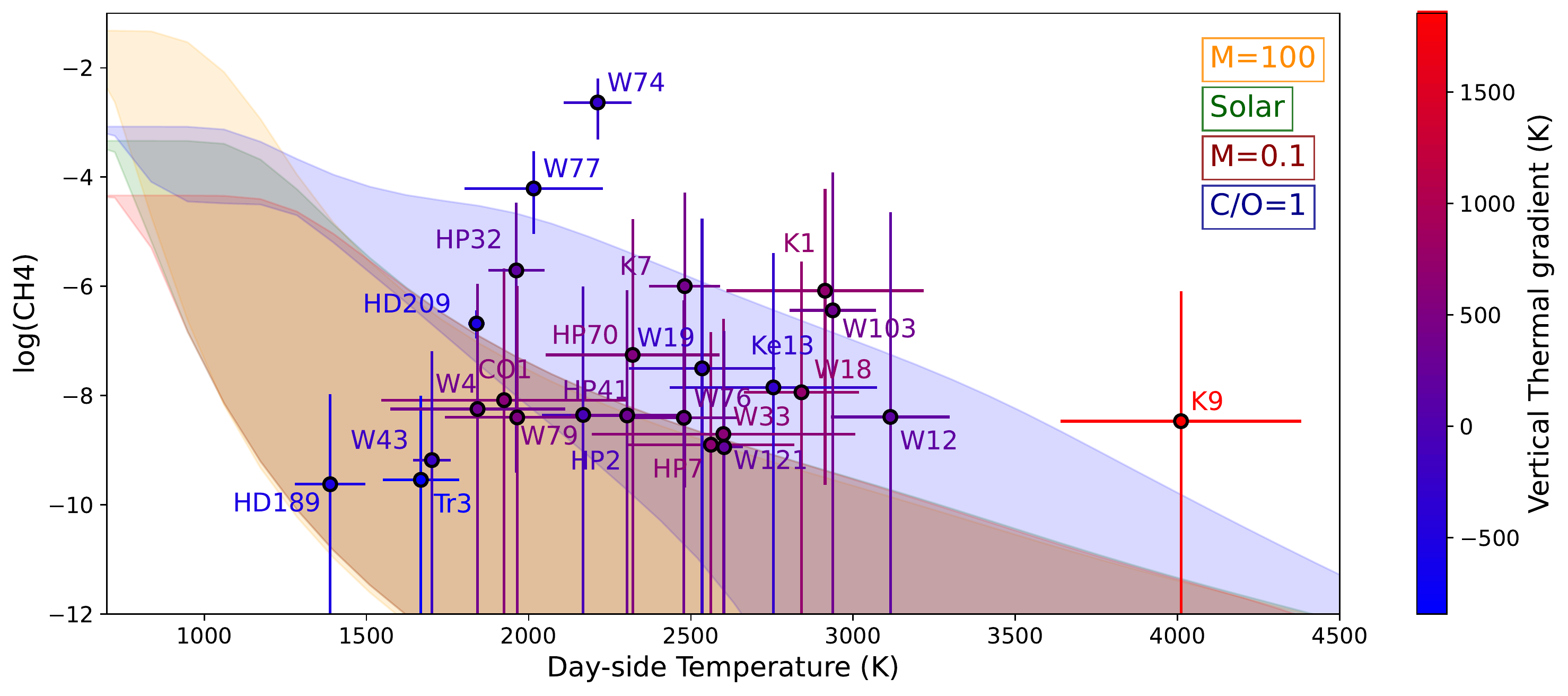}
    \caption{Correlations at the day-side between the CO$_2$ (top) and CH$_4$ (bottom) abundances and the mean retrieved temperature weighted by the contribution function. The colours indicate the retrieved thermal gradient, defined as the temperature differences between the upper and lower quantiles of the contribution function. The shaded green region is the predicted abundances from equilibrium chemistry at solar metallity and C/O ratio between 1 bar and 0.01 bar. In orange, the metallicity is increased to 100 times solar. In red, the metallicity is decreased to 0.1 times solar. In blue, the metallicity is solar and the C/O ratio is increased to 1.}
    \label{fig:correlateCH4_fullspz}
\end{figure}

\begin{figure*}
    \includegraphics[width = 0.95\textwidth]{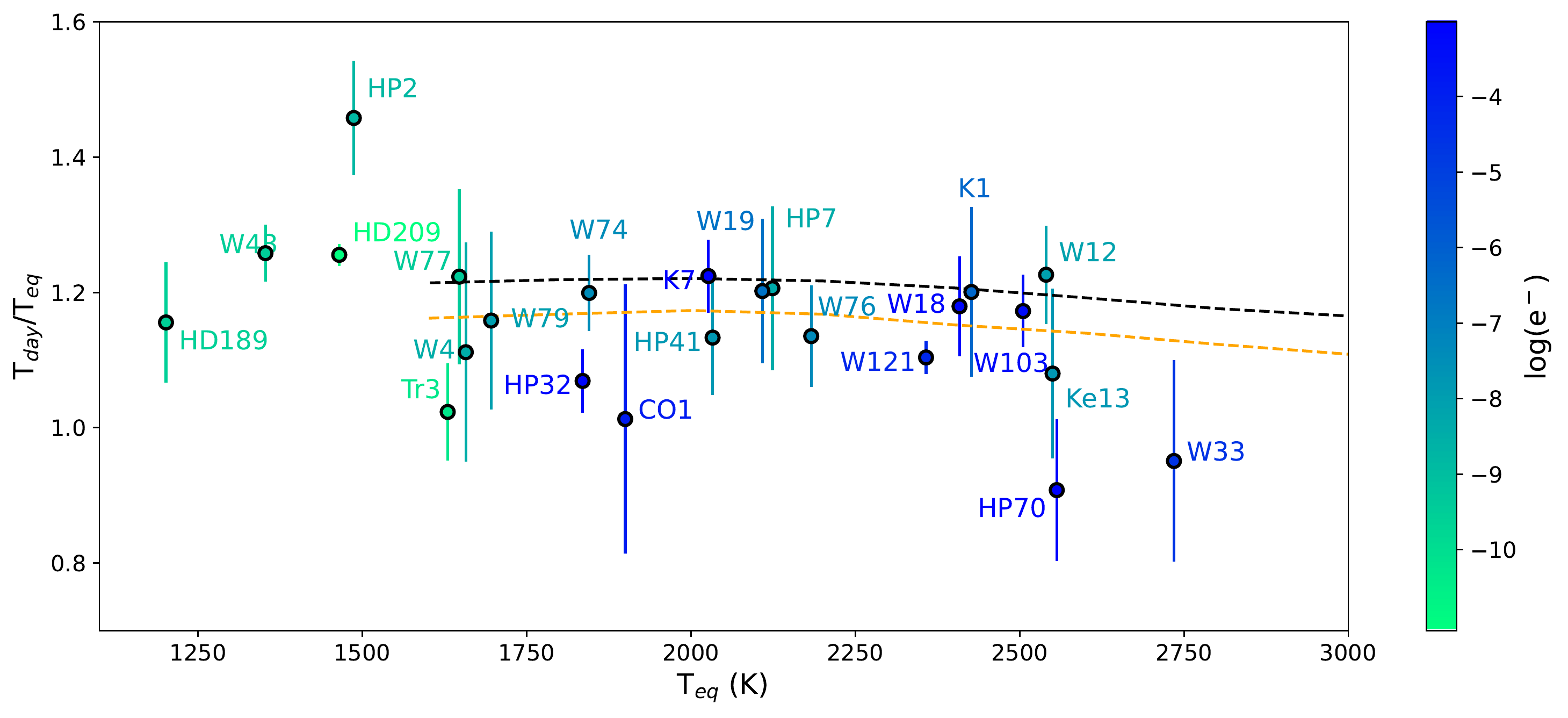}
    \includegraphics[width = 0.95\textwidth]{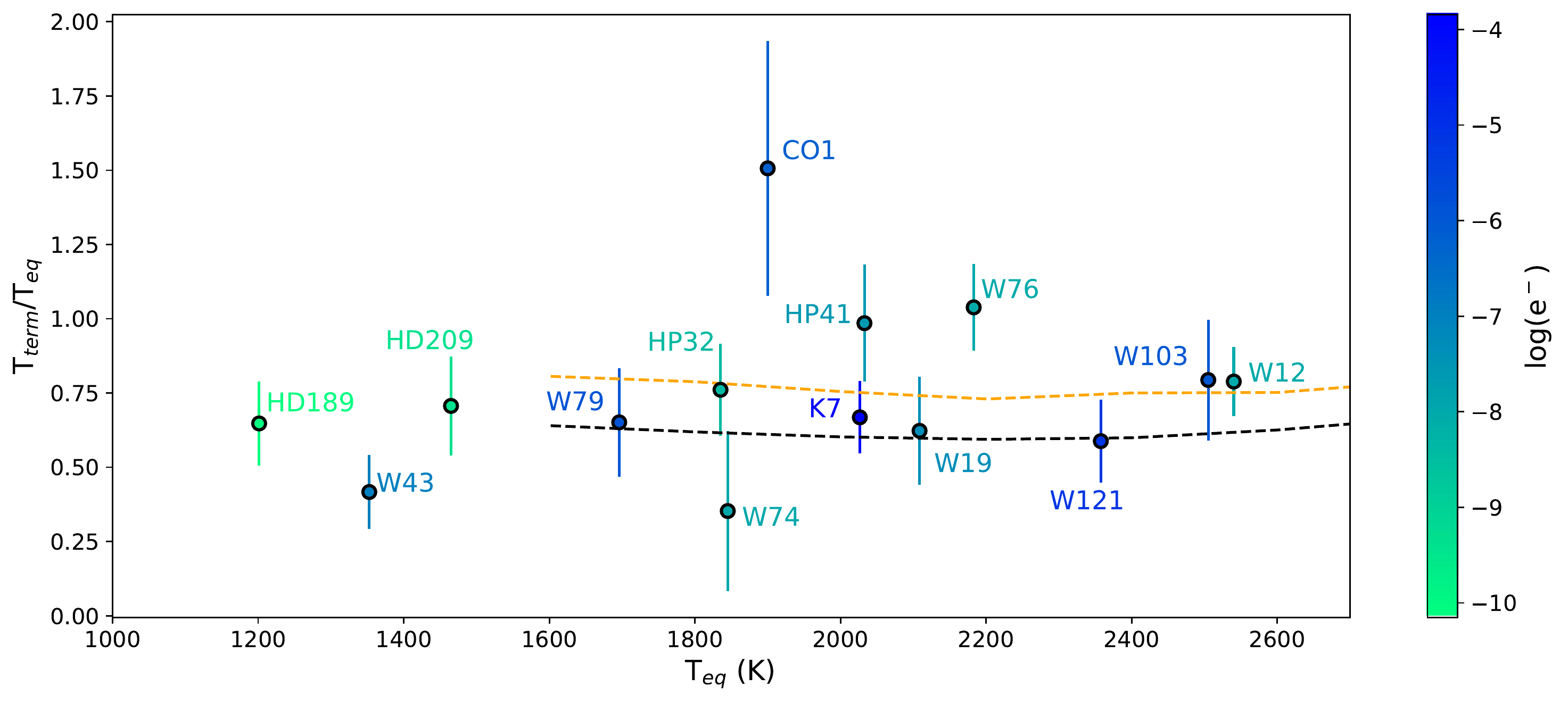}
    \caption{Weighted retrieved atmospheric temperature for the day-side (top) and terminator (bottom) as a function of the planet's equilibrium temperature. The equilibrium temperature is calculated for a planetary Albedo of 0. The colours corresponds to the abundance of e$^-$, which traces H$_2$ dissociation. As reference, we also show in dashed lines the GCM predictions from \cite{tan} with H$_2$ dissociation/recombination included for their slow (orange) and fast (black) drag timescales.}
    \label{fig:dynamics_day_term_e-_indiv}
\end{figure*}

\begin{figure*}
    \includegraphics[width = 0.95\textwidth]{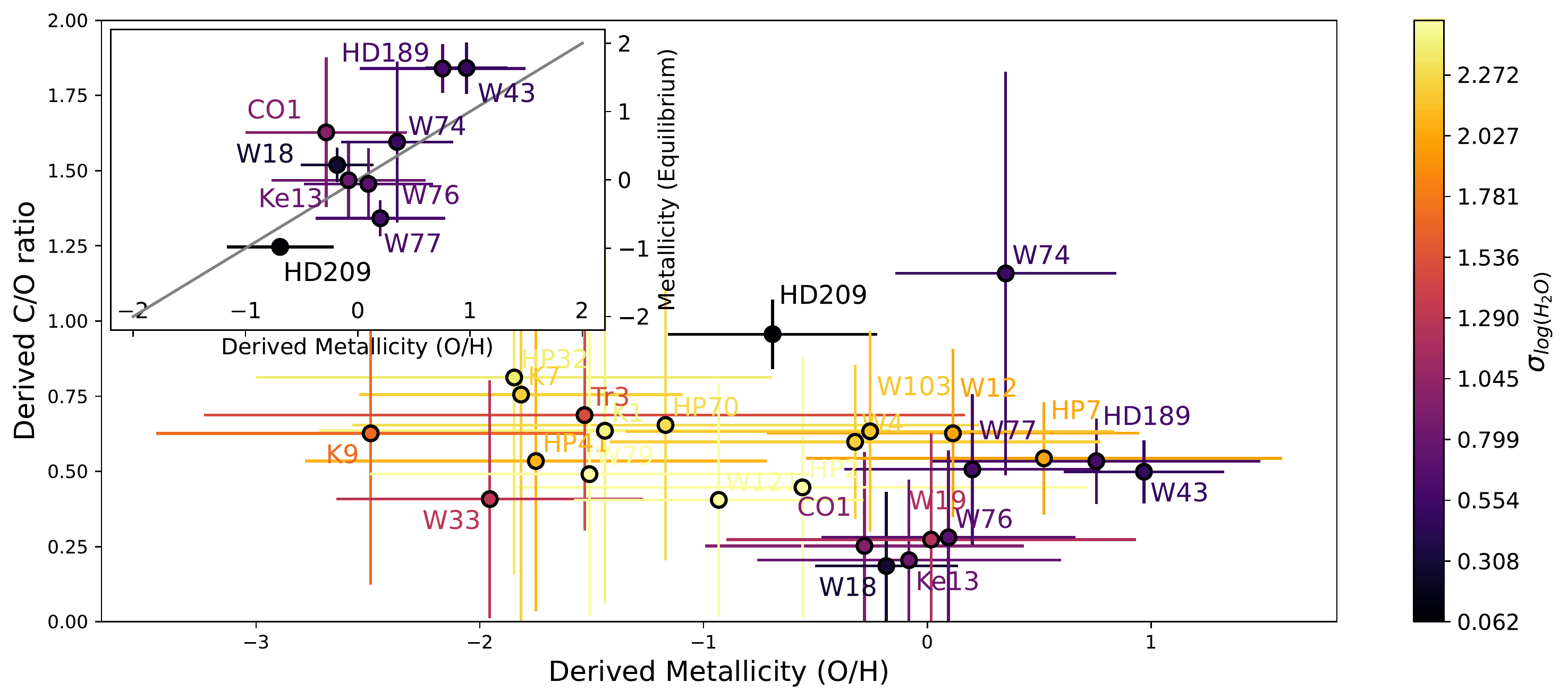}
    \caption{Metallicity (O/H) and C/O ratios recovered from our free chemistry retrievals. In colors, we highlight the error on the retrieved abundance of water $\sigma_{\mathrm{log(H}_2\mathrm{O)}}$, since only the cases where water is recovered are reliable (purple to black). C/O remains very difficult to constrain since our observations lack sensitivity to carbon bearing species. In the inset, we show the metallicity derived from free chemistry for the runs with $\sigma_{\mathrm{log(H}_2\mathrm{O)}} < 1$ versus the metallicity recovered in the equilibrium chemistry cases.}
    \label{fig:met_co_full_spz}
\end{figure*}

%

\clearpage

\section{Materials and Methods}

\renewcommand{\thefigure}{B\arabic{figure}}
\setcounter{figure}{0}

\renewcommand{\thetable}{B\arabic{table}}
\setcounter{table}{0}

\input{sec_description_obs}

\input{sec_materials_methods}

\section{Supplementary Text}

\renewcommand{\thefigure}{C\arabic{figure}}
\setcounter{figure}{0}

\renewcommand{\thetable}{C\arabic{table}}
\setcounter{table}{0}

\input{model_degen}

\subsection{On constraining metallicity and C/O ratio from free retrievals.}
\input{metallicity}

\section{Individual Planet Analyses}
\renewcommand{\thefigure}{D\arabic{figure}}
\setcounter{figure}{2}

\renewcommand{\thetable}{D\arabic{table}}
\setcounter{table}{0}

Individual exoplanet retrieval results are described in this sections. Figure Set D1 shows the eclipse results and Figure Set D2 shows the transit results. Table D1 summarizes the recovered parameters and provides the Bayesian Evidence of each model. For molecules, non-detections are characterized by a retrieved log(VMR) below -8 for the mean, and if the mean minus the sum of the 1$\sigma$ uncertainties (on both sides) is lower than -12. In this case, only the upper limit is provided.

\input{individual_figure_set}
\input{combined_retrieval_table}

\subsection{Individual analysis of CoRoT-1\,b}
\input{sec_CO1}

\subsection{Individual analysis of HAT-P-2\,b}

\input{sec_H2}

\subsection{Individual analysis of HAT-P-7\,b}

\input{sec_H7}

\subsection{Individual analysis of HAT-P-32\,b}
\input{sec_H32}

\subsection{Individual analysis of HAT-P-41\,b}
\input{sec_H41}

\subsection{Individual analysis of HAT-P-70\,b}

\input{sec_H70}

\subsection{Individual analysis of HD 189733\,b}

\input{sec_HD189}

\subsection{Individual analysis of HD 209458\,b}

\input{sec_HD209}

\subsection{Individual analysis of KELT-1\,b}

\input{sec_K1}

\subsection{Individual analysis of KELT-7\,b}

\input{sec_K7}

\subsection{Individual analysis of KELT-9\,b}

\input{sec_K9}

\subsection{Individual analysis of Kepler-13A\,b}

\input{sec_Ke13}

\subsection{Individual analysis of TrES-3\,b}

\input{sec_TR3}

\subsection{Individual analysis of WASP-4\,b}

\input{sec_W4}

\subsection{Individual analysis of WASP-12\,b}
\input{sec_W12}

\subsection{Individual analysis of WASP-18\,b}

\input{sec_W18}

\subsection{Individual analysis of WASP-19\,b}

\input{sec_W19}

\subsection{Individual analysis of WASP-33\,b}

\input{sec_W33}

\subsection{Individual analysis of WASP-43\,b}

\input{sec_W43}

\subsection{Individual analysis of WASP-74\,b}

\input{sec_W74}

\subsection{Individual analysis of WASP-76\,b}

\input{sec_W76}

\subsection{Individual analysis of WASP-77\,A\,b}
\input{sec_W77}

\subsection{Individual analysis of WASP-79\,b}
\input{sec_W79}

\subsection{Individual analysis of WASP-103\,b}

\input{sec_W103}

\subsection{Individual analysis of WASP-121\,b}

\input{sec_W121}

\clearpage



\end{document}

%% file: sec_description_obs.tex
\subsection{Description of the observations and their reduction}

For this study, we considered the planets that have been observed in eclipse using the Hubble Space Telescope (HST) WFC3 camera with the Grism G141. This constitutes a sample of 25 planets. For all planets except HAT-P-70\,b, data from the Spitzer Space Telescope is also available for at least the 3.6\,$\mu$m and the 4.5\,$\mu$m IRAC channels. When available, we also re-analysed the complementary HST WFC3 data from transit observations, which is the case for 17 planets of our sample. We describe below our reduction method for the HST observations and the sources for the Spitzer data. \\

\subsubsection{Hubble Space Telescope data}

For all planets, except for Kepler-13\,A\,b and WASP-33\,b, we downloaded the publicly available data from the Mikulski\footnote{https://archive.stsci.edu/} Archive (MAST). We note that data from HST also exist for the planet CoRoT-2\,b, but due to the very shallow eclipse for this planet \citep{Wilkins_Corot2b_spectrum_em}, we were not able to reliable reduced the observations and chose to not include this planet in the study. The publicly available data consist of a series of raw detector images.

We carried out the analysis of all HST WFC3 data using \emph{Iraclis}, our highly-specialised software for processing WFC3 spatially scanned spectroscopic images \citep{tsiaras_hd209} which has been used in a number of studies \citep{tsiaras_55cnce,damiano_h32,tsiaras_30planets,tsiaras_h2o,Anisman_2020_W117,Changeat_2020_K11,Edwards_2020_ares,skaf_aresII,pluriel_aresIII,Guilluy_2021,Mugnai_2021,Edwards_2021_LHS,Yip_2021_W96,Changeat_2021_k9,Libby_2021}. For this study and except for two planets (WASP-33\,b and Kepler-13A\,b), we acquired literature spectral data only if the reduction had been performed with the \emph{Iraclis} pipeline to ensure homogeneity between planets. In particular, Tsiaras et al. (2018), later referred to as T18 \citep{tsiaras_30planets}, already reduced a number of the datasets for the transits we consider here with \emph{Iraclis}. Else, we performed a reduction process that included the following steps: zero-read subtraction, reference-pixels correction, non-linearity correction, dark current subtraction, gain conversion, sky background subtraction, calibration, flat-field correction and bad-pixels/cosmic-rays correction. Then we extracted the white (1.088-1.68\,$\mu$m) and the spectral light curves from the reduced images, taking into account the geometric distortions caused by the tilted detector of the WFC3 infrared channel.
\begin{table*}\fontsize{4}{5}\selectfont
\centering
\begin{tabular}{llllllllllll}
\hline
Planet & R$_s$ & T$_s$ & log(g) & [M/H] & P & a/R$_s$ & e & i & t$_{mid}$ & M$_p$ & ref  \\ \hline \hline

CoRoT-1\,b & 1.11 & 5950 & 4.25 & -0.3 & 1.5089557 & 4.92 & 0.0 & 85.1  & 2454159.4532 &  1.03 & \cite{2008_barge_corot1b} \\ \hline

HAT-P-2\,b & 1.64 & 6290 & 4.16 & 0.14 & 5.6334729 & 8.99 & 0.5171 & 86.72  & 2454387.49375 & 9.09 & \cite{Pal_2010_hp2} \\ \hline

HAT-P-7\,b & 1.84 & 6350 & 4 & 0.32 & 2.2047354 & 4.1545 & 0.0 & 83.14  & 2454954.3585723 & 1.78 & \cite{esteves_hatp7} \\ \hline

HAT-P-32\,b & 1.367 & 6207 & 4.33 & -0.04 & 2.1500082 & 5.344 & 0.159 & 88.98 & 2455867.24 &  0.68 & \cite{2019_wang_hatp} \\ \hline

\multirow{2}{*}{HAT-P-41\,b} & \multirow{2}{*}{2.05} & \multirow{2}{*}{6390} & \multirow{2}{*}{4.14} & \multirow{2}{*}{0.21} & \multirow{2}{*}{2.694047} & \multirow{2}{*}{5.45} & \multirow{2}{*}{0.0} & \multirow{2}{*}{87.7} & \multirow{2}{*}{2454983.86167} & \multirow{2}{*}{1.19} & \cite{Stassun_planetparam} \\
 &  &  &  &  &  &  &  &  &  &  & \cite{2017_johnson_hp41} \\ \hline

HAT-P-70\,b & 1.86 & 8450 & 4.18 & -0.059 & 2.74432452 & 5.45 & 0 & 96.50  & 2458439.57519 & $<$6.78 & \cite{Zhou_2019_hp70} \\ \hline

\multirow{2}{*}{HD\,189733\,b} & \multirow{2}{*}{0.75} & \multirow{2}{*}{5052} & \multirow{2}{*}{4.49} & \multirow{2}{*}{-0.02} & \multirow{2}{*}{2.218577} & \multirow{2}{*}{8.84} & \multirow{2}{*}{0.0} & \multirow{2}{*}{85.69} & \multirow{2}{*}{2458334.990899} & \multirow{2}{*}{1.13} & \cite{Stassun_planetparam} \\
 &  &  &  &  &  &  &  &  &  &  & \cite{2019_addison_Minerva} \\ \hline

\multirow{2}{*}{HD\,209458\,b} & \multirow{2}{*}{1.20} & \multirow{2}{*}{6092} & \multirow{2}{*}{4.28} & \multirow{2}{*}{0.0} & \multirow{2}{*}{3.524750} & \multirow{2}{*}{8.87} & \multirow{2}{*}{0.0} & \multirow{2}{*}{86.78} & \multirow{2}{*}{2454560.80588} & \multirow{2}{*}{0.64} & \cite{Boyajian_2015} \\ 
 &  &  &  &  &  &  &  &  &  &  & \cite{Evans_2015_hd209} \\ \hline

KELT-1\,b & 1.462 & 6518 & 4.337 & 0.052 & 1.217514 & 3.60 & 0.0099 & 87.8 & 2455914.1628 & 27.23 & \cite{2012_siverd_kelt1} \\ \hline

KELT-7\,b & 1.732 & 6789 & 4.149 & 0.139 & 2.7347749 & 5.50 & 0.0 & 86.79 & 2457095.68572 & 2.88 & \cite{gaudi_k9} \\ \hline

KELT-9\,b & 2.362 & 10170 & 4.093 & -0.03 & 1.4811235 & 3.153 & 0.0 & 83.76 & 2456355.229809 & 1.28 & \cite{2015_bieryla_kelt7} \\ \hline

Kepler-13A\,b & 1.74 & 7650 & 4.2 & 0.2 & 1.763588 & 4.5007 & 0.00064 & 86.77 & 2454953.56596 & 9.28 & \cite{esteves} \\ \hline

\multirow{2}{*}{TrES-3\,b} & \multirow{2}{*}{0.817} & \multirow{2}{*}{5650} & \multirow{2}{*}{4.581} & \multirow{2}{*}{-0.19} & \multirow{2}{*}{1.30618608} & \multirow{2}{*}{6.01} & \multirow{2}{*}{0} & \multirow{2}{*}{81.99} & \multirow{2}{*}{2454538.58069} & \multirow{2}{*}{1.910} & \cite{2011_christiansen_tres} \\
 &  &  &  &  &  &  &  &  &  &  & \cite{2011_southworth_30planets} \\ \hline

WASP-4\,b & 0.893 & 5400 & 4.47 & -0.07 & 1.3382299 & 5.451 & 0.0 & 89.06 & 2455804.515752 & 1.186 & \cite{2019_Bouma_W4} \\ \hline

WASP-12\,b & 1.657 & 6360 & 4.157 & 0.33 & 1.0914203 & 3.039 & 0.0 & 83.37 & 2456176.66825800 & 1.47 & \cite{collins_w12} \\ \hline

WASP-18\,b & 1.26 & 6431 & 4.47 & 0.11 & 0.9414526 & 3.562 & 0.0091 & 84.88 & 2458375.169883 & 10.4 & \cite{shporer_w18} \\ \hline

WASP-19\,b & 1.004 & 5568 & 4.45 & 0.15 & 0.788838989 & 3.46 & 0.002 & 78.78 & 2455708.534626 & 1.114 & \cite{Wong_HATP7b_w19_em} \\ \hline

\multirow{2}{*}{WASP-33\,b} & \multirow{2}{*}{1.444} & \multirow{2}{*}{7430} & \multirow{2}{*}{4.30} & \multirow{2}{*}{0.10} & \multirow{2}{*}{1.2198675} & \multirow{2}{*}{3.68} & \multirow{2}{*}{0.0} & \multirow{2}{*}{87.90} & \multirow{2}{*}{2455507.5222} & \multirow{2}{*}{3.266} & \cite{collier_2010_w33} \\ 
 &  &  &  &  &  &  &  &  &  &  & \cite{von_essen_2014_w33} \\ \hline

\multirow{2}{*}{WASP-43\,b} & \multirow{2}{*}{0.6} & \multirow{2}{*}{4400} & \multirow{2}{*}{4.65} & \multirow{2}{*}{-0.05} & \multirow{2}{*}{0.813473978} & \multirow{2}{*}{4.872} & \multirow{2}{*}{0.0} & \multirow{2}{*}{82.1} & \multirow{2}{*}{2455528.868634} & \multirow{2}{*}{1.78} & \cite{hellier_w43} \\ \
 &  &  &  &  &  &  &  &  &  &  & \cite{kreidberg_w43} \\ \hline

WASP-74\,b & 1.42 & 5990 & 4.39 & 0.39 & 2.137750 & 4.86 & 0.0 & 79.81 & 2456506.892589  & 0.72 & \cite{Stassun_planetparam} \\ \hline

WASP-76\,b & 1.73 & 6250 & 4.128 & 0.23 & 1.809886 & 3.9997 & 0.0 & 89.9 & 2456107.85507 & 0.92 & \cite{2016_west_W76-82-90} \\ \hline

WASP-77\,A\,b & 0.955 & 5500 & 4.33 & 0.00 & 1.3600309 & 5.40 & 0 & 89.4 & 2455870.44977 & 1.76 & \cite{Maxted_2013_w77} \\ \hline

WASP-79\,b & 1.51 & 6600 & 4.226 & 0.03 & 3.662392 & 7.41 & 0.0 & 86.1 & 2456215.4556  & 0.85 & \cite{2017_brown_wasp} \\ \hline

WASP-103\,b & 1.436 & 6110 & 4.22 & 0.06 & 0.925542 & 2.978 & 0.0 & 86.3 & 2456459.59957  & 1.490 & \cite{Gillon_Wasp103b_em} \\ \hline

WASP-121\,b & 1.458 & 6459 & 4.242 & 0.13 & 1.2749255 & 3.754 & 0.0 & 87.6 & 2456635.70832    & 1.183 & \cite{Delrez_Wasp121b_em} \\ \hline
\hline
\end{tabular}%
\caption{List of the orbital parameters used in this paper and the associated literature references. R$_s$ is the star radius in Sun radius. T$_s$ is the star temperature in Kelvin. g is the stellar surface gravity in cm/s$^2$. [M/H] is the stellar metallicity in dex, P is the orbital period of the planet in days. a is the semi-major axis. e is the eccentricity. i is the inclination in degrees. t$_{mid}$ is the mid-transit time is days. M$_p$ is the planetary mass in masses of Jupiter. For HAT-P-70\,b only a upper limit on the planetary mass is available. } 
\label{tab:planet_params}
\end{table*}

We fitted the light curves using our light curve modelling package \emph{PyLightcurve} \citep{tsiaras_plc} with the parameters from Table \ref{tab:planet_params}. During our fitting of the white light curve, the planet-to-star flux ratio and the mid-eclipse time were the only free parameters, along with a model for the systematics \citep{Kreidberg_GJ1214b_clouds,tsiaras_hd209}. It is common for WFC3 exoplanet observations to be affected by two kinds of time-dependent systematics: the long-term and short-term `ramps'. The first affects each HST visit and is modelled by a linear function, while the second affects each HST orbit and has an exponential behaviour. The formula we used for the white light curve systematics ($R_w$) was the following:
\begin{equation}
    R_w(t) = n^{scan}_w(1-r_a(t - T_0))(1-r_{b1}e^{-r_{b2}(t-t_)})
\end{equation}
\noindent where $t$ is time, $n^{scan}_w$ is a normalization factor, $T_0$ is the mid-eclipse time, $t_o$ is the time when each HST orbit starts, $r_a$ is the slope of a linear systematic trend along each HST visit and ($r_{b1},r_{b2}$) are the coefficients of an exponential systematic trend along each HST orbit. The normalization factor we used ($n^{scan}_w$) was changed to $n^{for}_w$ for upward scanning directions (forward scanning) and to $n^{rev}_w$) for downward scanning directions (reverse scanning). The reason for using separate normalization factors is the slightly different effective exposure time due to the known upstream/downstream effect \citep{McCullough_wfc3}. 

We fitted the white light curves using the formulae above and the uncertainties per pixel, as propagated through the data reduction process. However, it is common in HST/WFC3 data to have additional scatter that cannot be explained by the ramp model. For this reason, we scaled up the uncertainties in the individual data points, for their median to match the standard deviation of the residuals, and repeated the fitting \citep{tsiaras_30planets}. The white light-curve fits obtained for the eclipse spectra are shown in Figure Set B1.

\begin{figure*}
\figurenum{B1}
\plotone{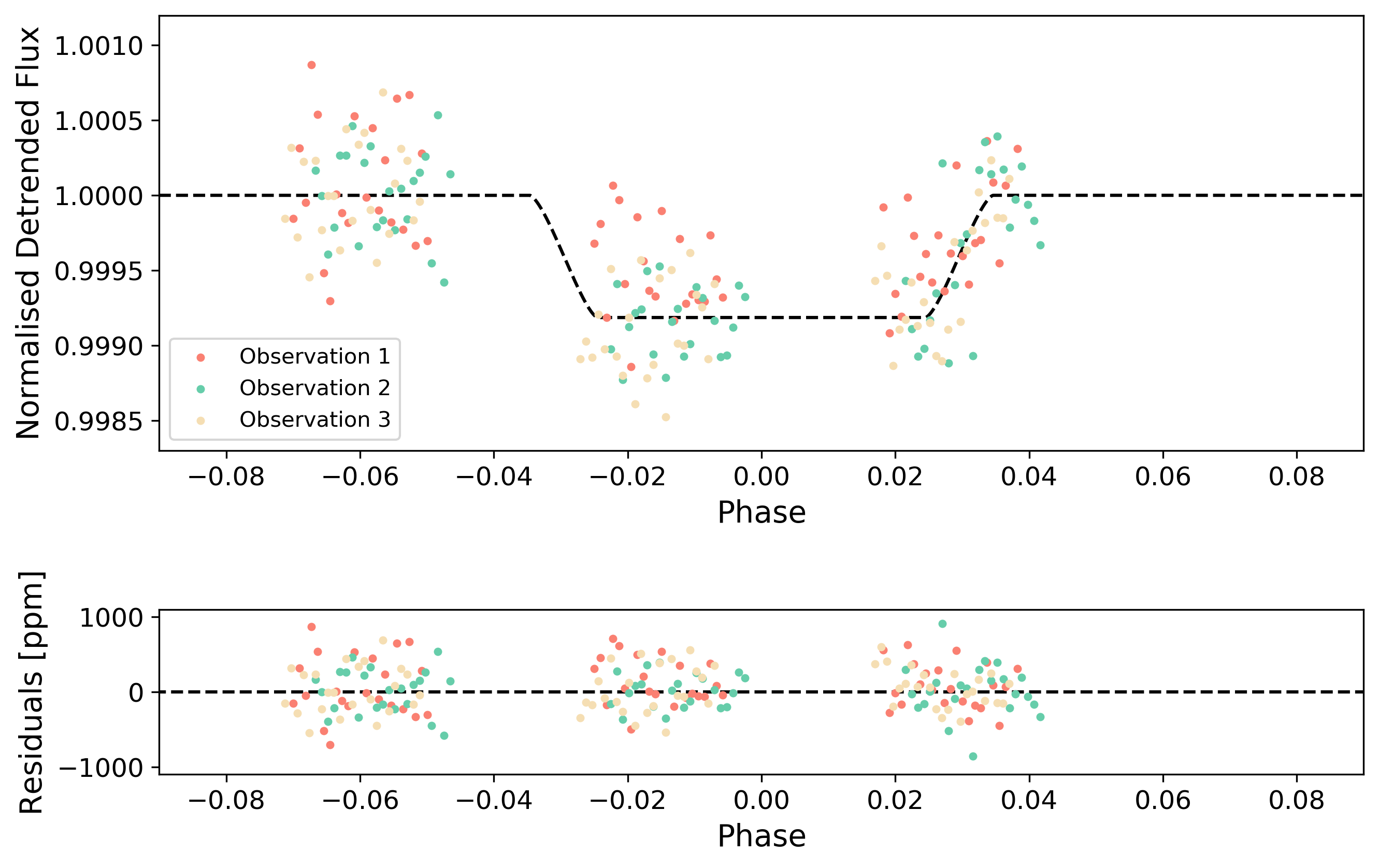}
\caption{White light-curve fit of CoroT-1 b with \emph{Iraclis} (Top) and residuals (Bottom). This figure is a sample from the Figure Set B1 that contains the same information for all the planets. Figure Set B1 is available in the electronic edition of the {\it Astrophysical Journal}.}\label{figset:LC}
\end{figure*}

\setcounter{figure}{1}

Next, we fitted the spectral light curves with a transit model (with the planet-to-star flux ratio being the only free parameter) along with a model for the systematics ($R_\lambda$) that included the white light curve (divide-white method \citep{Kreidberg_GJ1214b_clouds}) and a wavelength-dependent, visit-long slope \citep{tsiaras_hd209}. 

\begin{equation}
    R_\lambda(t) = n^{scan}_\lambda(1-\chi_\lambda(t-T_0))\frac{LC_w}{M_w}
\end{equation}{}

\noindent where $\chi_\lambda$ is the slope of a wavelength-dependent linear systematic trend along each HST visit, $LC_w$ is the white light curve and $M_w$ is the best-fit model for the white light curve. Again, the normalisation factor we used ($n^{scan}_\lambda$) was changed to ($n^{for}_\lambda$) for upward scanning directions (forward scanning) and to ($n^{for}_\lambda$) for downward scanning directions (reverse scanning). Also, in the same way as for the white light curves, we performed an initial fit using the pipeline uncertainties and then refitted while scaling these uncertainties up, for their median to match the standard deviation of the residuals.

We note that several of these datasets were acquired using the staring mode which is usually less efficient than the spatial scanning technique. Furthermore, the signal to noise ratio achieved on the eclipse varied between planets. We used the SNR on the white light curve to dictate the resolution of the spectral fitting. For those with an SNR \textgreater 7, a high resolution spectrum was extracted (25 bins) while a lower resolution binning was chosen for all others (18 bins). The recovered spectral light-curve for the planets reduced with \emph{Iraclis} are shown in Figure Set B2.

\begin{figure*}
\figurenum{B2}
\plotone{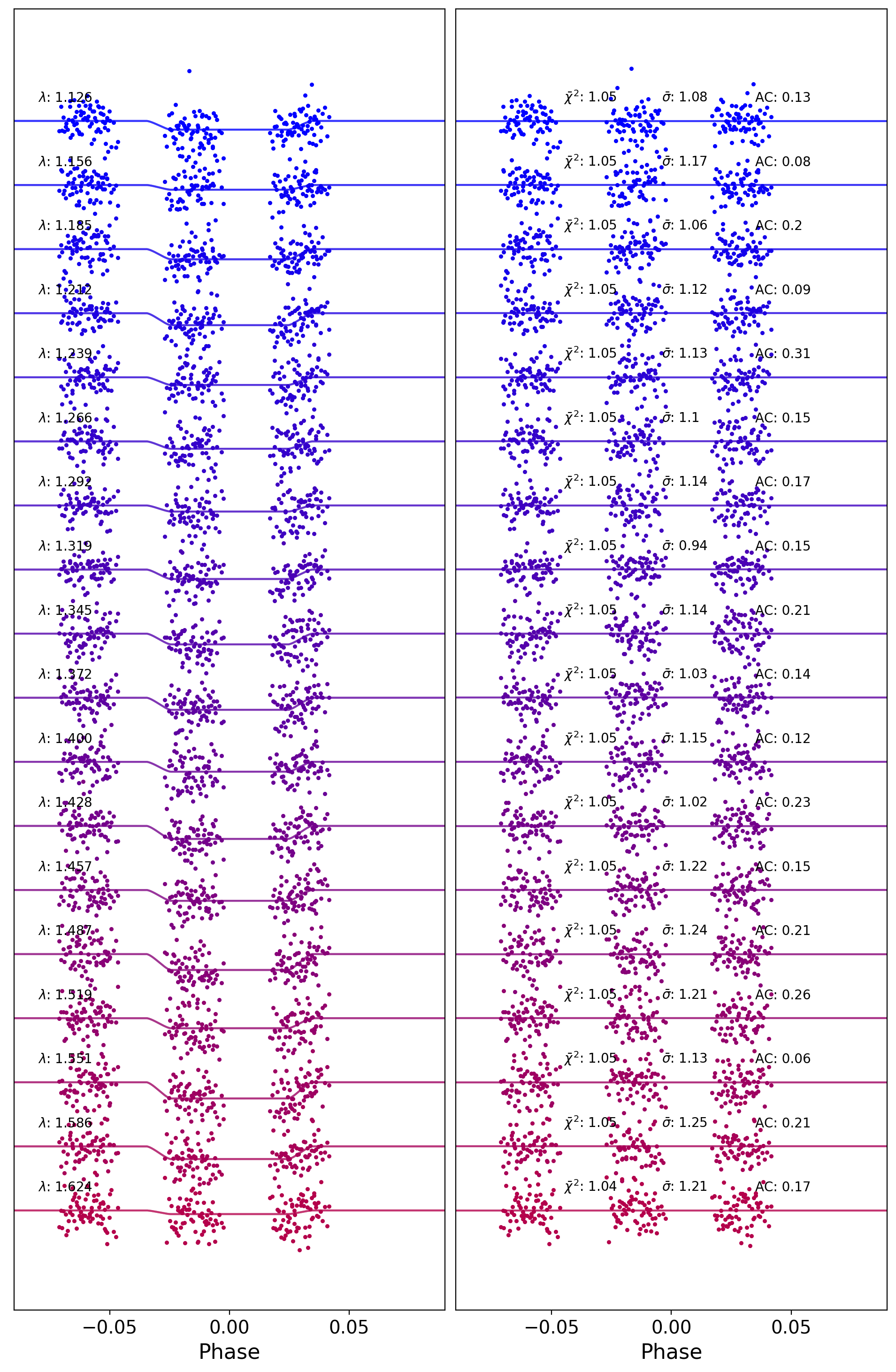}
\caption{Spectral light-curve of CoRoT-1\,b obtained with \emph{Iraclis}. This figure is a sample from the Figure Set B2 that contains the same information for all the planets. Figure Set B2 is available in the electronic edition of the {\it Astrophysical Journal}.}\label{figset:SLC}
\end{figure*}
\setcounter{figure}{2}

\begin{figure*}
    \centering
    \includegraphics[width=\textwidth]{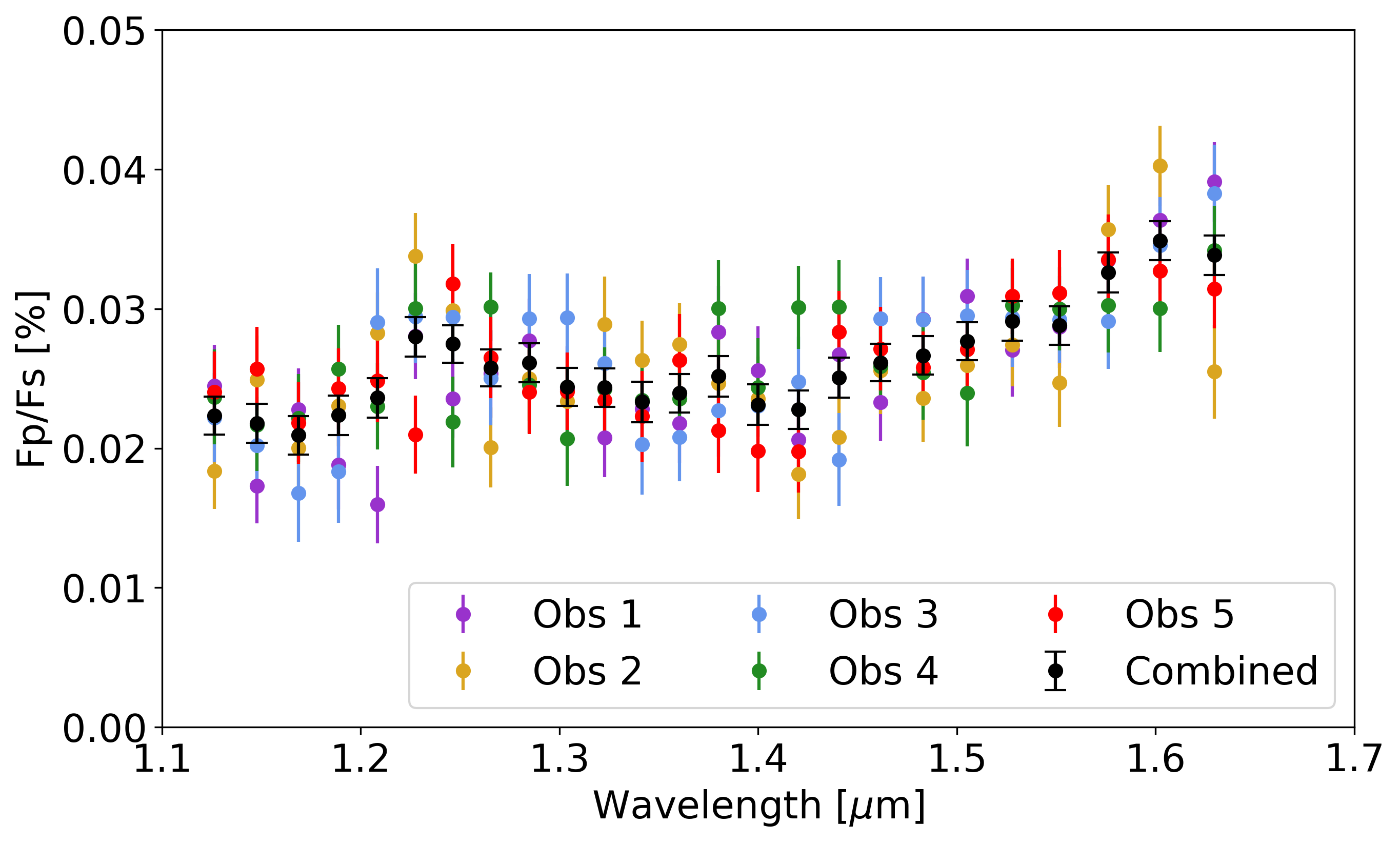}
    \includegraphics[width=\textwidth]{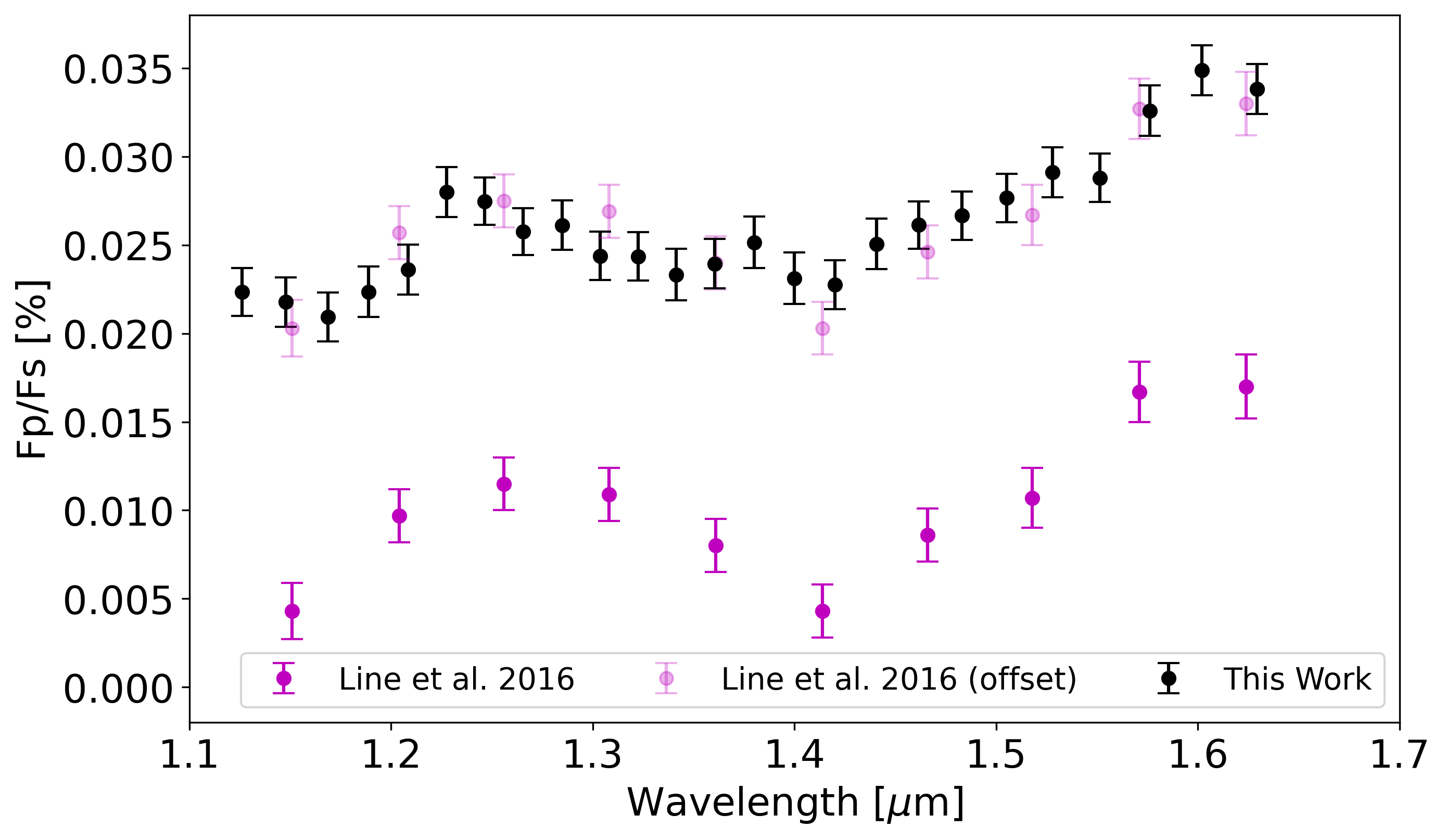}
    \caption{\textbf{Top:} recovered spectra from each visit and the final, averaged spectrum (black). \textbf{Bottom:} comparison of the spectrum recovered here to that of \cite{Line_HD209_spectrum_em}. We note that, while the spectra have similar shapes, they are offset by around 0.016\% (160 ppm).  }
    \label{fig:hd209_spec}
\end{figure*}

For planets with multiple visits, we correct for offsets by subtracting each spectrum by the corresponding white light curve depth, $(F_p/F_\star)^2_{w,v}$, and adding the weighted average eclipse depth of all white light-curves, $(F_p/F_\star)^2_w$. Finally, we compute the weighted average from all the eclipse observations which we use for all subsequent analysis. Due to possible variabilities in the star, the planet or the instrument systematics, the measured white eclipse depths could change between the different visits, so we choose not perform a joint-fit of these datasets as done in many previous studies \citep{cartier_wasp103, Wakeford_2018_w39, arcangeli_wasp18_em}. However, except for HD\,209458\,b, when multiple visits are available, the individual eclipse depths are all consistent within 1-sigma, meaning that a joint-fit would be equivalent. For HD\,209458\,b, we show the extracted spectra in Figure \ref{fig:hd209_spec} along with the final spectrum obtained by averaging all five visits. We note that the spectral features are consistent in the five observations.

While most of our spectra are consistent with the literature, HD\,209458\,b differs significantly from the spectrum derived in \cite{Line_HD209_spectrum_em}, as shown in Figure \ref{fig:hd209_spec}. The shape of the spectra is highly similar, but we observe a $\sim$160 ppm vertical offset. We identified that the difference comes from the assumption of a quadratic systematic for the long-term ramp in \cite{Line_HD209_spectrum_em}, while we here assume a linear behaviour. We note that \cite{Line_HD209_spectrum_em} only managed to fit four of the five eclipse observations while we fitted all five.  Additionally, when fitting these observations with a quadratic trend we found the white light curve depth, and associated errors, were far more unstable than the linear fits. The spectral shape recovered with both trends was very similar. HD\,209458\,b is the planet for which the most significant difference to the literature can be observed. Other spectra from the literature are compared to our \emph{Iraclis} pipeline reductions in the Figure Set B4. Overall, we find that except for HD\,209458\,b and for TrES-3\,b, we do not find significant differences in spectral shape and observe only minor offsets.

We also considered the transmission spectra, which are available for 17 planets. These were reduced using the same process as for the emission data.

For two systems, WASP-12 and WASP-103, a close stellar companion contaminated the data from HST WFC3. For exoplanet spectroscopy, this third light modifies the transit/eclipse depth. To account for this, we used the freely available WFC3 simulator \emph{Wayne} \footnote{\emph{Wayne}: https://github.com/ucl-exoplanets/wayne}.

\emph{Wayne} is capable of producing grism spectroscopic frames, both in staring and in spatial scanning modes \citep{Varley_2017}. We utilised \emph{Wayne} to model the contribution of each companion star to the spectral data obtained. We created simulated detector images of both the main and companion star, using these to extract the flux contribution in each spectral bin of each star. The correction to the spectra is then applied as a wavelength dependent dilution factor which is derived as a ratio of extracted flux between the stars. Such an approach has previously been used on WFC3 data, e.g. for WASP-76 b \citep{Edwards_2020_ares}. \\

\begin{figure*}
\figurenum{B4}
\epsscale{1.15}
\plotone{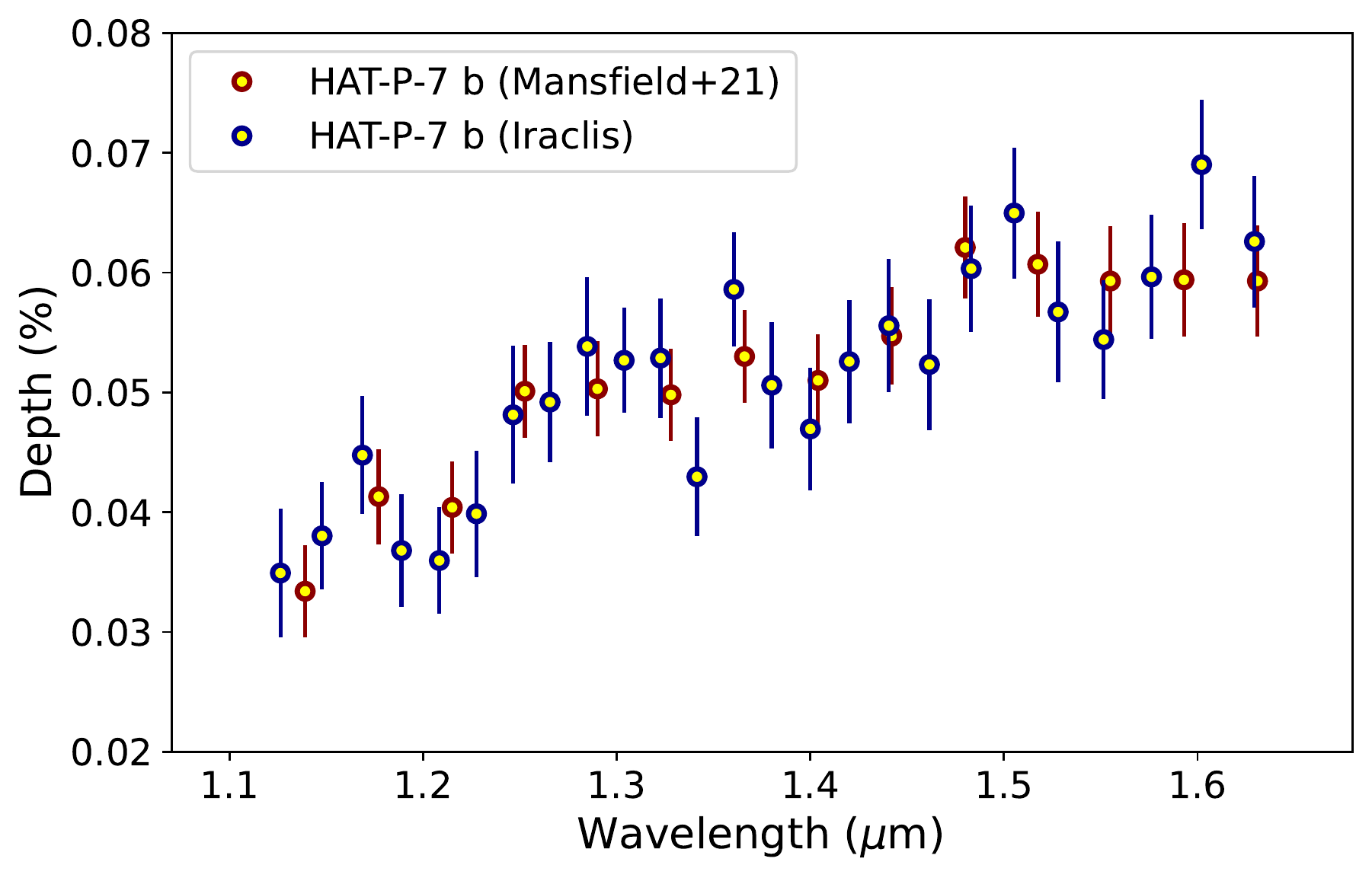}
\caption{Comparison of the \emph{Iraclis} final reduction (Blue) with the literature for HAT-P-7\,b. This figure is a sample from the Figure Set B4 that contains the same information for all the planets with literature data. Figure Set B4 is available in the electronic edition of the {\it Astrophysical Journal}.  
}
\end{figure*}
\setcounter{figure}{4}


\begin{table*}
\centering
\resizebox{0.65\textwidth}{!}{%
\begin{tabular}{cccc}\hline\hline
Planet & Type & Instruments & Sources \\\hline\hline
\multirow{3}{*}{CoRoT-1\,b} & Eclipse &  HST G141 & 12181 (Deming)  \\ 
    & Eclipse & Spz 3.6, 4.5 &  \cite{Deming_2010_co1b} \\
    & Transit & HST G141 & 12181 (Deming)  \\
\hline
\multirow{2}{*}{HAT-P-2\,b} & Eclipse &  HST G141 & 16194 (Desert)  \\ 
    & Eclipse & Spz 3.6, 4.5 & \cite{Lewis_2013_hp2_spz} \\
\hline
\multirow{3}{*}{HAT-P-7\,b} & Eclipse & HST G141 & 14782 (Bean)  \\
    & Eclipse & Spz 3.6, 4.5, 5.8, 8 & \cite{Christiansen_hatp7}  \\
    & Transit & HST G141 & 12181 (Deming)  \\
\hline
\multirow{3}{*}{HAT-P-32\,b} & Eclipse &  HST G141 & 14767 (Sing)  \\ 
    & Eclipse & Spz 3.6, 4.5 &  \cite{Zhao_2014_hp32} \\
    & Transit & HST G141 & 14260 (Deming)  \\
\hline
\multirow{3}{*}{HAT-P-41\,b} & Eclipse &  HST G141 & 14767 (Sing)  \\ 
    & Eclipse & Spz 3.6, 4.5 &  \cite{Garhart_2020} \\
    & Transit & HST G141 & 14767 (Sing)  \\
\hline
\multirow{1}{*}{HAT-P-70\,b} & Eclipse &  HST G141 & 16307 (Fu)  \\ 
\hline
\multirow{3}{*}{HD-189733\,b} & Eclipse &  HST G141 & 12881 (McCullough) \\ 
    & Eclipse & Spz 3.6, 4.5, 5.8, 8, 16, 24 &  \cite{Charbonneau_2008} \\
    & Transit & HST G141 & 12881 (McCullough)  \\
\hline    
\multirow{3}{*}{HD\,209458\,b} & Eclipse &  HST G141 & 13467 (Bean)  \\ 
    & Eclipse & Spz 3.6, 4.5, 5.8, 8, 11.35, 24 &  \cite{Diamond_Lowe_2014, Swain_2008_hd209} \\
    & Transit & HST G141 & 12181 (Deming)  \\        
\hline
\multirow{3}{*}{KELT-1\,b} & Eclipse &  HST G141 & 14664 (Beatty) \\ 
    & Eclipse & Spz 3.6, 4.5 &  \cite{Garhart_2020} \\
    & Transit & HST G141 & 14664 (Beatty) \\        
\hline
\multirow{3}{*}{KELT-7\,b} & Eclipse &  HST G141 & 14767 (Sing)  \\ 
    & Eclipse & Spz 3.6, 4.5 &  \cite{Garhart_2020} \\
    & Transit & HST G141 & 14767 (Sing)  \\        
\hline
\multirow{2}{*}{KELT-9\,b} & Eclipse &  HST G141 & 15820 (Pino)  \\ 
    & Eclipse & Spz 4.5 & \cite{mansfield_k9} \\
\hline
\multirow{2}{*}{Kepler-13\,A\,b} & Eclipse &  HST G141 & \cite{Beatty_Kepler13_spectrum_em}  \\ 
    & Eclipse & Spz 3.6, 4.5 & \cite{Shporer_Kepler13_em} \\
\hline
\multirow{2}{*}{TrES-3\,b} & Eclipse &  HST G141 & 12181 (Deming)  \\ 
    & Eclipse & Spz 3.6, 4.5, 5.8, 8 & \cite{Fressin_2010_tr3} \\
\hline
\multirow{2}{*}{WASP-4\,b} & Eclipse &  HST G141 & 12181 (Deming)  \\ 
    & Eclipse & Spz 3.6, 4.5 & \cite{Beerer_2010} \\
\hline
\multirow{3}{*}{WASP-12\,b} & Eclipse &  HST G141 & 12230 (Swain)  \\ 
    & Eclipse & Spz 3.6, 4.5, 5.8, 8 & \cite{Campo_2011} \\
    & Transit & HST G141 & 13467 (Bean)  \\  
\hline
\multirow{3}{*}{WASP-18\,b} & Eclipse &  HST G141 & 13467 (Bean)  \\ 
    & Eclipse & Spz 3.6, 4.5, 5.8, 8 & \cite{nymeyer_w18_spit} \\
    & Transit & HST G141 & 13467 (Bean)  \\  
\hline
\multirow{3}{*}{WASP-19\,b} & Eclipse &  HST G141 & 13431 (Huitson)  \\ 
    & Eclipse & Spz 3.6, 4.5, 5.8, 8 & \cite{anderson_wasp19_spitzer} \\
    & Transit & HST G141 & 12181 (Deming) + 13431 (Huitson)  \\  
\hline
\multirow{2}{*}{WASP-33\,b} & Eclipse &  HST G141 & \cite{Haynes_Wasp33b_spectrum_em} \\ 
    & Eclipse & Spz 3.6, 4.5 & \cite{Deming_2012_w33} \\
\hline
\multirow{3}{*}{WASP-43\,b} & Eclipse &  HST G141 & 13467 (Bean)  \\ 
    & Eclipse & Spz 3.6, 4.5 & \cite{Garhart_2020}  \\
    & Transit & HST G141 & 13467 (Bean)  \\  
\hline
\multirow{3}{*}{WASP-74\,b} & Eclipse &  HST G141 & 14767 (Sing)  \\ 
    & Eclipse & Spz 3.6, 4.5 & \cite{Garhart_2020}  \\
    & Transit & HST G141 & 14767 (Sing)  \\  
\hline
\multirow{3}{*}{WASP-76\,b} & Eclipse &  HST G141 & 14767 (Sing)  \\ 
    & Eclipse & Spz 3.6, 4.5 & \cite{Garhart_2020}  \\
    & Transit & HST G141 & 14260 (Deming)  \\  
\hline
\multirow{2}{*}{WASP-77\,A\,b} & Eclipse &  HST G141 & 16168 (Mansfield)  \\ 
    & Eclipse & Spz 3.6, 4.5 & \cite{Garhart_2020} \\
\hline
\multirow{3}{*}{WASP-79\,b} & Eclipse &  HST G141 & 14767 (Sing)   \\ 
    & Eclipse & Spz 3.6, 4.5 & \cite{Garhart_2020}  \\
    & Transit & HST G141 & 14767 (Sing)  \\  
\hline
\multirow{3}{*}{WASP-103\,b} & Eclipse &  HST G141 & 14050 (Kreidberg) + 13660 (Zhao)  \\ 
    & Eclipse & Spz 3.6, 4.5 & \cite{Garhart_2020}  \\
    & Transit & HST G141 & 14050 (Kreidberg) \\  
\hline
\multirow{4}{*}{WASP-121\,b} & Eclipse &  HST G102 & 15135 (Mikal-Evans)  \\ 
    & Eclipse &  HST G141 & 14767 (Sing) + 15134 (Mikal-Evans)  \\ 
    & Eclipse & Spz 3.6, 4.5 & \cite{Garhart_2020}  \\
    & Transit & HST G141 & 14468 + 15134 (Mikal-Evans)  \\  
\hline
\end{tabular}
}
\caption{Summary of the observations considered in this paper. HST: Hubble Space Telescope; Spz: Spitzer Space Telescope. We used observations from the Grisms G102 and G141 aboard HST and photometric channels from 3.6 to 24 $\mu$m aboard Spitzer. } \label{tab:observations}
\end{table*}

\subsubsection{Spitzer data}

For the Spitzer data, we used the data from the literature directly. For all the planets in our sample, we find that at least the 3.6\,$\mu$m and the 4.5\,$\mu$m channels from the Infrared Array Camera (IRAC) are available. Adding Spitzer data significantly increases the wavelength coverage of our observations, but can also lead to biases when the observations are not compatible. This well-known issue is discussed further in Complementary Text. In order to be as consistent as possible, we prioritised the inclusion of the Spitzer data from the population study of \cite{Garhart_2020}, later referred as G20. A few planets possess additional channels from legacy Spitzer observations with the Infrared Array Camera (IRAC: 5.8\,$\mu$m and 8\,$\mu$m channels), the InfraRed Spectrograph (IRS: 16\,$\mu$m) and the Multiband Imaging Photometer for Spitzer (MIPS: 24\,$\mu$m). In this case, we add the most complete reduction, which allows to greatly increase the information content for these planets. For HD-209458\,b, we took the Spitzer data from \cite{Diamond_Lowe_2014}. We noted, however, that the 16\,$\mu$m point used, in fact, refers to the Spitzer IRS  low spectral resolution observations \citep{Swain_2008_hd209}, which span the wavelength range from 7.46\,$\mu$m to 15.25\,$\mu$m. We therefore chose to include the derived broad band eclipse depth from the original study. In Spitzer, photometric channels are large and can cover entire or multiple broadband molecular features. The shape of the spectral response function is therefore important and has to be accounted for during binning of the forward model. Spitzer spectral responses functions for the IRAC, IRS and MIPS instruments can be found at the NASA/IPAC Infrared Science Archive \footnote{https://irsa.ipac.caltech.edu/data/SPITZER/docs/}. \\

The full list of observations that are considered in the retrievals is provided in Table \ref{tab:observations}. We detail the spectra recovered from this reduction step in Tables \ref{tab:table_of_spectra1} and \ref{tab:table_of_spectra7} for, respectively, eclipse and transit observations. The full tables are available in machine-readable format. In those tables, we also include for convenience the Spitzer datasets that were obtained from the literature. These spectra constitute the inputs of our retrieval analysis, further described in the next section.

\begin{table*}
    \centering
	\tiny
    \begin{tabular}{|l|l|l|l|l|l|l|}
    \hline
        $\lambda$ ($\mu$m) & $\Delta\lambda$ & CoRoT-1b & HD\,189733 b & TrES-3 b & WASP-4 b & WASP-12 b   \\ \hline
        1.1262 & 0.0308 & 0.0682 $\pm$ 0.0149 & 0.0101 $\pm$ 0.0056 & 0.0515 $\pm$ 0.0209 & 0.0378 $\pm$ 0.0183 & 0.1547 $\pm$ 0.0255 \\ \hline
        1.1563 & 0.0293 & 0.0367 $\pm$ 0.0149 & 0.0038 $\pm$ 0.0038 & 0.0263 $\pm$ 0.0188 & 0.0271 $\pm$ 0.0177 & 0.1492 $\pm$ 0.0239 \\ \hline
        1.1849 & 0.0279 & 0.0797 $\pm$ 0.0159 & 0.0123 $\pm$ 0.0044 & 0.0088 $\pm$ 0.0137 & 0.0206 $\pm$ 0.0152 & 0.1612 $\pm$ 0.0228 \\ \hline
        1.2123 & 0.0269 & 0.0937 $\pm$ 0.0167 & 0.0112 $\pm$ 0.0054 & 0.0288 $\pm$ 0.0206 & 0.0858 $\pm$ 0.0208 & 0.1253 $\pm$ 0.0226 \\ \hline
        1.2390 & 0.0265 & 0.0600 $\pm$ 0.0158 & 0.0023 $\pm$ 0.0036 & 0.0210 $\pm$ 0.0185 & 0.0491 $\pm$ 0.0171 & 0.1781 $\pm$ 0.0224 \\ \hline
        1.2657 & 0.0269 & 0.0578 $\pm$ 0.0164 & 0.0153 $\pm$ 0.0049 & 0.0169 $\pm$ 0.0157 & 0.0808 $\pm$ 0.0174 & 0.1632 $\pm$ 0.0232 \\ \hline
        1.2925 & 0.0267 & 0.0477 $\pm$ 0.0158 & 0.0141 $\pm$ 0.0051 & 0.0174 $\pm$ 0.0162 & 0.0394 $\pm$ 0.0169 & 0.1762 $\pm$ 0.0274 \\ \hline
        1.3190 & 0.0263 & 0.0740 $\pm$ 0.0141 & 0.0198 $\pm$ 0.0049 & 0.0351 $\pm$ 0.0182 & 0.0249 $\pm$ 0.0178 & 0.1739 $\pm$ 0.0216 \\ \hline
        1.3454 & 0.0265 & 0.0840 $\pm$ 0.0172 & 0.0122 $\pm$ 0.0052 & 0.1050 $\pm$ 0.0267 & 0.0318 $\pm$ 0.0165 & 0.1769 $\pm$ 0.0226 \\ \hline
        1.3723 & 0.0274 & 0.0943 $\pm$ 0.0155 & 0.0150 $\pm$ 0.0046 & 0.0863 $\pm$ 0.0205 & 0.0264 $\pm$ 0.0147 & 0.1447 $\pm$ 0.0256 \\ \hline
        1.4000 & 0.0280 & 0.0772 $\pm$ 0.0161 & 0.0060 $\pm$ 0.0044 & 0.1188 $\pm$ 0.0227 & 0.0496 $\pm$ 0.0173 & 0.1937 $\pm$ 0.0234 \\ \hline
        1.4283 & 0.0285 & 0.1016 $\pm$ 0.0157 & 0.0188 $\pm$ 0.0048 & 0.1343 $\pm$ 0.0276 & 0.0969 $\pm$ 0.0185 & 0.1958 $\pm$ 0.0232 \\ \hline
        1.4572 & 0.0294 & 0.0845 $\pm$ 0.0182 & 0.0157 $\pm$ 0.0052 & 0.1228 $\pm$ 0.0217 & 0.0848 $\pm$ 0.0210 & 0.2375 $\pm$ 0.0254 \\ \hline
        1.4873 & 0.0308 & 0.1243 $\pm$ 0.0186 & 0.0227 $\pm$ 0.0053 & 0.0498 $\pm$ 0.0185 & 0.0723 $\pm$ 0.0165 & 0.2661 $\pm$ 0.0237 \\ \hline
        1.5186 & 0.0318 & 0.0787 $\pm$ 0.0185 & 0.0277 $\pm$ 0.0054 & 0.0510 $\pm$ 0.0232 & 0.0642 $\pm$ 0.0196 & 0.2197 $\pm$ 0.0243 \\ \hline
        1.5514 & 0.0337 & 0.1265 $\pm$ 0.0168 & 0.0228 $\pm$ 0.0043 & 0.0511 $\pm$ 0.0187 & 0.0307 $\pm$ 0.0171 & 0.2019 $\pm$ 0.0264 \\ \hline
        1.5862 & 0.0360 & 0.0993 $\pm$ 0.0188 & 0.0184 $\pm$ 0.0056 & 0.0276 $\pm$ 0.0204 & 0.0511 $\pm$ 0.0178 & 0.1687 $\pm$ 0.0247 \\ \hline
        1.6237 & 0.0390 & 0.0286 $\pm$ 0.0162 & 0.0150 $\pm$ 0.0055 & 0.0502 $\pm$ 0.0215 & 0.0812 $\pm$ 0.0176 & 0.2099 $\pm$ 0.0260 \\ \hline
        3.6000 & Spz3 & 0.4150 $\pm$ 0.0420 & 0.2560 $\pm$ 0.0140 & 0.3560 $\pm$ 0.0350 & 0.3190 $\pm$ 0.0310 & 0.4210 $\pm$ 0.0110 \\ \hline
        4.5000 & Spz8 & 0.4820 $\pm$ 0.0420 & 0.2140 $\pm$ 0.0200 & 0.3720 $\pm$ 0.0540 & 0.3430 $\pm$ 0.0270 & 0.4280 $\pm$ 0.0120 \\ \hline
        5.8000 & Spz5 & - & 0.3100 $\pm$ 0.0340 & 0.4490 $\pm$ 0.0970 & - & 0.6960 $\pm$ 0.0600 \\ \hline
        8.0000 & Spz8 & - & 0.3910 $\pm$ 0.0220 & 0.4750 $\pm$ 0.0460 & - & 0.6960 $\pm$ 0.0960  \\ \hline
        16.0000 & Spz16 & - & 0.5190 $\pm$ 0.0200 & - & - & -  \\ \hline
        24.0000 & Spz24 & - & 0.5980 $\pm$ 0.0380 & - & - & - \\ \hline
    \end{tabular}
\caption{List of eclipse spectra used in this population study. $\lambda$ refers to the central wavelength of the bin and $\Delta\lambda$ is the bin width. For Spitzer, the bin widths are irrelevant since we consider the spectral response of the channel.}\label{tab:table_of_spectra1}\tablecomments{Table B3 is published in its entirety in the electronic edition of the {\it Astrophysical Journal}. A portion is shown here.}
\end{table*}

\begin{table*}
    \centering
    \tiny
    \begin{tabular}{|l|l|l|l|l|l|l|l|l|l|l|l|l|l|l|l|}
    \hline
        $\lambda$ ($\mu$m) & $\Delta\lambda$ & HAT-P-7 b & HAT-P-32 b & HAT-P-41 b & HD-189733 b & HD-209458 b & KELT-1 b  \\ \hline
        1.1262 & 0.0219 & 0.5210 $\pm$ 0.0353 & 2.3076 $\pm$ 0.0115 & 1.0104 $\pm$ 0.0097 & 2.4271 $\pm$ 0.0066 & 1.4591 $\pm$ 0.0029 & 0.5613 $\pm$ 0.0092  \\ \hline
        1.1478 & 0.0211 & 0.5171 $\pm$ 0.0323 & 2.2925 $\pm$ 0.0113 & 1.0328 $\pm$ 0.0090 & 2.4319 $\pm$ 0.0076 & 1.4565 $\pm$ 0.0041 & 0.5553 $\pm$ 0.0076  \\ \hline
        1.1686 & 0.0206 & 0.6135 $\pm$ 0.0311 & 2.2939 $\pm$ 0.0092 & 1.0222 $\pm$ 0.0099 & 2.4430 $\pm$ 0.0074 & 1.4581 $\pm$ 0.0037 & 0.5680 $\pm$ 0.0085 \\ \hline
        1.1888 & 0.0198 & 0.4841 $\pm$ 0.0297 & 2.3032 $\pm$ 0.0102 & 1.0268 $\pm$ 0.0088 & 2.4304 $\pm$ 0.0082 & 1.4542 $\pm$ 0.0040 & 0.5872 $\pm$ 0.0076 \\ \hline
        1.2084 & 0.0193 & 0.5820 $\pm$ 0.0342 & 2.3107 $\pm$ 0.0106 & 1.0194 $\pm$ 0.0103 & 2.4300 $\pm$ 0.0082 & 1.4551 $\pm$ 0.0033 & 0.5785 $\pm$ 0.0069 \\ \hline
        1.2275 & 0.0190 & 0.5471 $\pm$ 0.0310 & 2.3040 $\pm$ 0.0104 & 1.0285 $\pm$ 0.0087 & 2.4219 $\pm$ 0.0092 & 1.4605 $\pm$ 0.0039 & 0.5717 $\pm$ 0.0075 \\ \hline
        1.2465 & 0.0189 & 0.5870 $\pm$ 0.0341 & 2.2861 $\pm$ 0.0098 & 1.0202 $\pm$ 0.0086 & 2.4115 $\pm$ 0.0072 & 1.4518 $\pm$ 0.0031 & 0.5834 $\pm$ 0.0088 \\ \hline
        1.2655 & 0.0192 & 0.5291 $\pm$ 0.0381 & 2.2834 $\pm$ 0.0098 & 1.0187 $\pm$ 0.0098 & 2.4198 $\pm$ 0.0065 & 1.4512 $\pm$ 0.0034 & 0.5764 $\pm$ 0.0074 \\ \hline
        1.2847 & 0.0193 & 0.5533 $\pm$ 0.0377 & 2.3034 $\pm$ 0.0140 & 1.0372 $\pm$ 0.0107 & 2.4168 $\pm$ 0.0072 & 1.4587 $\pm$ 0.0039 & 0.5881 $\pm$ 0.0102 \\ \hline
        1.3038 & 0.0188 & 0.5214 $\pm$ 0.0337 & 2.2908 $\pm$ 0.0117 & 1.0174 $\pm$ 0.0080 & 2.4325 $\pm$ 0.0072 & 1.4552 $\pm$ 0.0035 & 0.5812 $\pm$ 0.0090 \\ \hline
        1.3226 & 0.0188 & 0.5341 $\pm$ 0.0318 & 2.3030 $\pm$ 0.0108 & 1.0164 $\pm$ 0.0090 & 2.4283 $\pm$ 0.0068 & 1.4528 $\pm$ 0.0041 & 0.5809 $\pm$ 0.0086 \\ \hline
        1.3415 & 0.0189 & 0.5084 $\pm$ 0.0358 & 2.3352 $\pm$ 0.0125 & 1.0394 $\pm$ 0.0082 & 2.4318 $\pm$ 0.0068 & 1.4639 $\pm$ 0.0037 & 0.5822 $\pm$ 0.0077 \\ \hline
        1.3605 & 0.0192 & 0.6058 $\pm$ 0.0330 & 2.3386 $\pm$ 0.0103 & 1.0503 $\pm$ 0.0096 & 2.4446 $\pm$ 0.0065 & 1.4708 $\pm$ 0.0033 & 0.5781 $\pm$ 0.0088 \\ \hline
        1.3800 & 0.0199 & 0.6097 $\pm$ 0.0345 & 2.3320 $\pm$ 0.0093 & 1.0271 $\pm$ 0.0102 & 2.4388 $\pm$ 0.0066 & 1.4683 $\pm$ 0.0039 & 0.5672 $\pm$ 0.0095 \\ \hline
        1.4000 & 0.0200 & 0.5537 $\pm$ 0.0345 & 2.3139 $\pm$ 0.0131 & 1.0412 $\pm$ 0.0083 & 2.4367 $\pm$ 0.0077 & 1.4752 $\pm$ 0.0033 & 0.5731 $\pm$ 0.0085 \\ \hline
        1.4202 & 0.0203 & 0.5620 $\pm$ 0.0340 & 2.3399 $\pm$ 0.0094 & 1.0463 $\pm$ 0.0081 & 2.4458 $\pm$ 0.0079 & 1.4631 $\pm$ 0.0038 & 0.5700 $\pm$ 0.0083 \\ \hline
        1.4406 & 0.0206 & 0.5891 $\pm$ 0.0324 & 2.3443 $\pm$ 0.0105 & 1.0451 $\pm$ 0.0091 & 2.4385 $\pm$ 0.0066 & 1.4638 $\pm$ 0.0048 & 0.5806 $\pm$ 0.0077 \\ \hline
        1.4615 & 0.0212 & 0.5259 $\pm$ 0.0340 & 2.3096 $\pm$ 0.0120 & 1.0457 $\pm$ 0.0099 & 2.4315 $\pm$ 0.0067 & 1.4584 $\pm$ 0.0040 & 0.5747 $\pm$ 0.0085 \\ \hline
        1.4831 & 0.0220 & 0.5254 $\pm$ 0.0343 & 2.3156 $\pm$ 0.0129 & 1.0457 $\pm$ 0.0093 & 2.4341 $\pm$ 0.0076 & 1.4629 $\pm$ 0.0036 & 0.5890 $\pm$ 0.0080 \\ \hline
        1.5053 & 0.0224 & 0.6097 $\pm$ 0.0345 & 2.3149 $\pm$ 0.0180 & 1.0481 $\pm$ 0.0113 & 2.4271 $\pm$ 0.0075 & 1.4614 $\pm$ 0.0032 & 0.5701 $\pm$ 0.0082 \\ \hline
        1.5280 & 0.0230 & 0.5915 $\pm$ 0.0418 & 2.3031 $\pm$ 0.0129 & 1.0421 $\pm$ 0.0104 & 2.4338 $\pm$ 0.0071 & 1.4606 $\pm$ 0.0038 & 0.5628 $\pm$ 0.0084 \\ \hline
        1.5516 & 0.0241 & 0.5425 $\pm$ 0.0361 & 2.3110 $\pm$ 0.0113 & 1.0345 $\pm$ 0.0090 & 2.4123 $\pm$ 0.0061 & 1.4565 $\pm$ 0.0031 & 0.5545 $\pm$ 0.0094 \\ \hline
        1.5762 & 0.0253 & 0.5407 $\pm$ 0.0380 & 2.3116 $\pm$ 0.0122 & 1.0282 $\pm$ 0.0099 & 2.4209 $\pm$ 0.0069 & 1.4584 $\pm$ 0.0035 & 0.5726 $\pm$ 0.0093 \\ \hline
        1.6021 & 0.0264 & 0.6047 $\pm$ 0.0384 & 2.2796 $\pm$ 0.0100 & 1.0172 $\pm$ 0.0105 & 2.4223 $\pm$ 0.0075 & 1.4504 $\pm$ 0.0044 & 0.5686 $\pm$ 0.0094 \\ \hline
        1.6294 & 0.0283 & 0.5383 $\pm$ 0.0397 & 2.2666 $\pm$ 0.0141 & 1.0189 $\pm$ 0.0112 & 2.4141 $\pm$ 0.0074 & 1.4577 $\pm$ 0.0037 & 0.5676 $\pm$ 0.0078 \\ \hline
    \end{tabular}
\caption{List of transit spectra used in this population study. $\lambda$ refers to the central wavelength of the bin and $\Delta\lambda$ is the bin width.}\label{tab:table_of_spectra7}\tablecomments{Table B4 is published in its entirety in the electronic edition of the {\it Astrophysical Journal}. A portion is shown here.}
\end{table*}

%% file: sec_materials_methods.tex
\subsection{Standardised retrievals with \emph{Alfnoor}}

To analyse the spectra obtained by \emph{Iraclis}, we perform atmospheric retrievals using the \emph{Alfnoor} tool \citep{Changeat_2020_alfnoor}. \emph{Alfnoor} extends the capabilities of the Bayesian retrieval suite \emph{TauREx3} \citep{2019_al-refaie_taurex3, al-refaie_2021_taurex3.1} and automate retrievals to large exoplanet populations. It was first built for simulations in the context of the ESA space mission Ariel \citep{Tinetti_ariel, Tinetti_2021_redbook} to enable retrieval studies of the mission's entire target list \citep{edwards_ariel}, but it can also perform standardised retrievals from any real data observation. We use \emph{Alfnoor} to extract the information content of our planetary atmospheres (chemical composition and temperature structure) separately for the transit and eclipse scenarios. For spectroscopic data, such as what is presented here, Bayesian retrievals are the currently most adopted analysis technique to extract un-biased information from atmospheric spectra. Another method that has been used to investigate population trends in a few studies \citep[e.g.][]{Crossfield_2017_amplitude, mansfield_2021_pop} consists of measuring the signal amplitude of the molecules of interest, for instance water. The technique compares the observed flux in the HST water band, which is defined as between 1.35\,$\mu$m and 1.48\,$\mu$m for instance in \cite{mansfield_2021_pop}, with a reference flux taken outside the band to evaluate deviations from the expected signal. Such a method, however, does not rely on a full exploration of the parameter space. More importantly, the method requires the definition of signal and reference bands, both of which can contain additional molecular features as shown in our work. Here, we prefer to extract trends in our population by atmospheric retrievals with a Nested Sampling optimizer and by testing a variety of atmospheric scenarios.\\

\subsubsection{Emission and Transmission forward models from \emph{TauREx3}}

For both transit and eclipse scenarios, the atmosphere is modeled assuming 1D layers (default 100 layers). In the transit case \citep{Waldmann_taurex1}, the total transit depth at wavelength $\lambda$ is given by:
\begin{equation}
    \Delta_{\lambda} = \frac{R_p^2 + a_{\lambda}}{R_s^2},
\end{equation}

\noindent where $R_p$ is the planet radius and $R_s$ is the parent star radius. $a_{\lambda}$ is the wavelength contribution from the atmosphere (transit depth), which takes the form:
\begin{equation}
    a_{\lambda} = 2 \int_0^{z_{max}}(R_p+z)(1-e^{-\tau_{\lambda}(z)}) dz.
\end{equation}
where $z_{max}$ is the altitude at the top of the atmosphere and $\tau_{\lambda}(z)$ is the wavelength dependant optical depth. It is evaluated by:

\begin{equation}
    \label{eq:tot_opt_dpth}
    \tau_{\lambda}(z) = \sum_i \tau_{\lambda,i}(z),
\end{equation}

with $\tau_{\lambda,i}$ is the optical depth of each absorber $i$.

In eclipse \citep{Waldmann_taurex2}, the emission from each layer is integrated to produce the final spectrum. The wavelength dependant intensity at the top of the atmosphere from a viewing angle $\theta$ is:
    
\begin{equation}
    I(\tau = 0,\mu) = B_{\lambda}(T_s) e^{-\frac{\tau_s}{\mu}} + \int_0^{1}\int_0^{\tau_s}B_{\lambda}(T_{\tau}) e^{-\frac{\tau}{\mu}} d\tau d\mu,
\end{equation}

\noindent where $\mu=cos(\theta)$, $B_{\lambda}(T)$ is the plank function at a given temperature T, T$_s$ denotes the temperature at maximum atmospheric pressure and $\tau_s$ is the total optical depth from the planetary surface to the top of the atmosphere. Then the flux is integrated for the cosine viewing angle $\mu$ using an $N$ point Gauss–Legendre quadrature scheme:
\begin{equation}
    I(\tau = 0) = \sum_i^{N} I(\tau=0,x_i)x_i w_i,
\end{equation}
\noindent where $w_i$ and $x_i$ are our weights and abscissas respectively.

The final emission spectrum is expressed as:
\begin{equation}
    \frac{F_p}{F_s} = \frac{I(\tau = 0) }{I_s} \times \left(\frac{R_p}{R_s}\right)^2
\end{equation}
\noindent where $I_s$ is the specific intensity from the star. In this work modelled using PHOENIX spectra \citep{phoenix}. \\

\subsubsection{Opacity sources}

Molecular opacity sources are included in both transmission and emission using their cross sections $\zeta_{i,\lambda}$. Their contribution to the optical depth is then given by:

\begin{equation}
\label{eq:opt_dpth}
    \tau_{\lambda,i}(z) = \int_z^{z_{max}}\zeta_{i,\lambda}(z') \chi_{i}(z') \rho(z') dz',
\end{equation}

\noindent where $\chi_{i}$ is the column density of the species $i$ and $\rho$, is the number density of the atmosphere. The contributions are integrated along the line of sight parametrised by d$z'$.

For the chemistry, we use the molecular line-lists from the Exomol project \citep{Tennyson_exomol, Chubb_2021_exomol}, HITEMP \citep{rothman} and HITRAN \citep{gordon}. In the free retrievals, we consider molecular cross-sections at resolution R=15,000 for H$_2$O \citep{barton_h2o,polyansky_h2o}, CH$_4$ \citep{hill_xsec,exomol_ch4}, CO \citep{li_co_2015}, CO$_2$ \citep{Yurchenko_2020}, TiO \citep{McKemmish_TiO_new}, VO \citep{McKemmish_2016_vo}, FeH \citep{Bernath_2020_FeH}, and H- \citep{John_1988_hmin,Edwards_2020_ares} depending on the model considered. In the chemical equilibrium runs, additional molecules are considered: HCN \citep{Barber_2013_HCN}, H$_2$CO \citep{Al_refaie_2015_h2co}, H$_2$S \citep{Azzam_2016_h2s}, CN \citep{Syme_2021_CN}, CP \citep{RAM_2014_CP, Bernath_2020_FeH, QIN_2021_CP}, C$_2$H$_2$ \citep{Chubb_2020_c2h2}, C$_2$H$_4$ \citep{2018_Mant_C2H4}, NH$_3$ \citep{al-derzi_2015_nh3, Coles_2019_nh3}, MgO \citep{Li_2019_mgo}, AlO \citep{bowesman_2021_alo, Patrascu_2015_alo}, SiO \citep{Barton_2013_sio,Yurchenko_2021_sio}, PH$_3$ \citep{sousa_silva_2015_ph3}, ScH \citep{Lodi_2015_ScH} and CrH \citep{Bernath_2020_FeH}. Atomic and ionic species were not included as they do not absorb in the wavelength range considered here. \\

In this work, we assumed that the planets are mainly composed of hydrogen and helium with a ratio He/H$_2$ = 0.17. In addition to molecular opacities, we included collision induced absorption (CIA) of the H$_2$-H$_2$ \citep{abel_h2-h2,fletcher_h2-h2} and H$_2$-He \citep{abel_h2-he} pairs as well as opacities induced by Rayleigh scattering \citep{cox_allen_rayleigh}. \\ 

For planets with temperatures higher than 2000K, the day-side is expected to be cloud free. For planets in lower temperature regimes, we tested the presence of clouds on the day-side using two parametric models. We first tested a fully opaque grey cloud layer with a single free parameter, the cloud deck top pressure. Such model is often used when analyzing HST data due to the small wavelength coverage in near-infrared. Since we also include the infrared bands from Spitzer in the eclipse analyses, we also tested a more realistic Mie model from \cite{Lee_2013_clouds}. The retrieved parameters for this model are cloud top pressure, cloud particle radius, and cloud mixing ratio. Comparing the Bayesian evidence of both of the cloudy retrievals with the cloud-free case, we did not find evidence in flavor of clouds in any of the eclipse spectra. Therefore, for the rest of the paper, we did not include clouds for eclipse retrievals. We note, however, that tested cloud models only simulate extinction processes and do not account for complex scattering. In transit, since the observations are more sensitive to clouds, we included the fully opaque grey cloud model in all retrievals.  \\


\subsubsection{Retrieval model setups}

In order to build our comparative study, we perform the same standardised retrievals for all the planets using \emph{Alfnoor}. All the planets in our study possess eclipse observations. The planet atmospheres are modelled using plane-parallel assumption and composed of 100 layers spaced in log-pressures from 10$^6$ Pa to 10$^{-5}$ Pa. The stellar parameters and planetary mass are always fixed to literature values as dedicated studies provide better constraints than what can be recovered from our data \citep{Changeat_2020_mass}.

In eclipse, due to the stronger degeneracies with the thermal profile, we fix the planetary radius to the literature values. We parametrise the temperature profile using a N-point point free profile containing three nodes. This is a heuristic profile that linearly interpolates between N freely moving points. We perform a total of eight retrievals using four different models: \\

$\bullet$ {\bf Reduced:} This conservative model uses free chemistry with constant with altitude mixing ratios. It includes the molecular opacities from H$_2$O, CH$_4$, CO and CO$_2$ for a total of 9 free parameters. \\

$\bullet$ {\bf Full:} This model uses free chemistry with constant with altitude mixing ratios. In addition to the molecules of the reduced run, it includes the optical absorbers TiO, VO, FeH and H$^-$ for a total of 13 free parameters. \\

$\bullet$ {\bf Equilibrium (Eq):} This model uses equilibrium chemistry from the \emph{GGChem} model \citep{Woitke_2018_GG}. All the species available in \emph{GGChem} were used to compute the chemistry and all the available opacity sources (see Opacity section) were included. To explore the thermal profile, we use the same N-Point parametrization as for the free cases, so the total number of free parameters is 7.  \\

$\bullet$ {\bf Blackbody:} For reference we perform a retrieval with no molecules and a simple isothermal profile (1 free parameter). This is used as a reference to compare the Bayesian evidence of the other models. \\

For each of those different models, we run an HST only case and an HST+Spitzer scenario.  When Spitzer is included, we use the calibrated Spitzer instrument responses \citep{Reach_2005} to account for the wavelength dependant behaviour of the detector.

When available, we complement our eclipse retrievals with an analysis of the HST transit spectrum. In this case, we also retrieve the planet radius R$_p$ and a Grey opaque cloud deck as these observations are more sensitive to those properties. Since in transmission the impact of the temperature is mostly on the scale height, we restrain our model to an isothermal temperature structure \citep{Changeat_2019_2l,Rocchetto_2016}. Here we only analyse the HST-WFC3 spectra and do not consider the additional Spitzer data. We run two separated retrievals: \\

$\bullet$ {\bf Full:} This model uses free chemistry with constant with altitude mixing ratios. We include all the absorbers: H$_2$O, CH$_4$, CO, CO$_2$, TiO, VO, FeH and H$^-$, for a total of 11 free parameters. \\

$\bullet$ {\bf Equilibrium (Eq):} This model uses equilibrium chemistry from the \emph{GGChem} model \citep{Woitke_2018_GG} and the same isothermal temperature profile. The number of free parameters is 5. \\

$\bullet$ {\bf Featureless fit:} For reference, we performed a transit fit with no molecule, fitting only for a flat line. The number of free parameters is 1. \\

\subsubsection{Exploration of the parameter space}

All the free parameters considered in this study are explored using uniform, non-informative priors. These are listed in more details in Table \ref{tab:priors}. For the chemistry, since all the planets considered are large hot Jupiters, we assume a primary envelope and therefore limit the retrieved maximum bound of the considered active molecules to 0.01. We imposed stricter priors on the optical absorbers, limiting their abundance to a maximum of 10$^{-4}$ as their abundance is expected to be much less \citep{Woitke_2018_GG}. For the lower bounds, we select 10$^{-12}$ as such abundance is low enough to not leave any spectral features at the resolution and signal-to-noise considered. The sampling is done using the nested sampling algorithm \emph{MultiNest} \citep{Feroz_multinest} with 750 live points and a log-likelihood tolerance of 0.5. This ensures an optimal free exploration of the parameter space.

\begin{figure}
\centering
\scriptsize
\begin{tabular}{|l|lll|}

\hline
Type & Parameter & Mode & priors \\ \hline \hline

\multirow{3}{*}{Transmission only} & planet radius (R$_p$) & linear & 0.1 - 2.0 (R$_J$) \\ 
& temperature (T) & linear & 100 - 4000 (K) \\
& cloud pressure (P$_{clouds}$) & log & 10$^6$ - 1 (Pa) \\
\hline
\hline

\multirow{3}{*}{Emission only} & temperature points i ($T_{i}$) & linear & 100 - 4000 (K) \\
& pressure points 1 ($P_{1}$) & log & 10$^6$ - 100 (Pa) \\ 
& pressure points 2 ($P_{2}$) & log & 10$^4$ - 0.01 (Pa) \\ 

\hline
\hline

\multirow{4}{*}{Both} & abundances (H$_2$O, CH$_4$, CO, CO$_2$, H$^-$) & log & 10$^{-12}$ - 0.01 (VMR) \\ 
& abundances (TiO, VO, FeH) & log & 10$^{-12}$ - 10$^{-4}$ (VMR) \\ 
& Metallicity (Eq) & log & 0.1 - 100 (to solar) \\
& C/O ratio (Eq) & linear & 0.1 - 2 \\

\hline
\hline
\end{tabular}%

\caption{List of the free parameters and their uniform priors in the retrievals.}
\label{tab:priors}

\end{figure}

Since we use the nested sampling algorithm \emph{MultiNest}, the computation of the Bayesian Evidence for each model, here denoted E, is automatic. Then, the difference in ln(E) between two models M1 and M2 can be used for model selection and to compare the ability of the two models to explain the observed spectra \citep{jeffreys1961theory, kass_raft, Feroz_multinest}. The original Jeffreys' scale that we employ in this study is summarised in Table \ref{tab:jeffrey}.

\begin{table*} 
\centering
\resizebox{0.7\textwidth}{!}{%
\begin{tabular}{|c|c|c|c|}\hline\hline
Grade  & B & $\Delta$ln(E) & Interpretation \\ \hline\hline 
Grade 0 & B $<$1 & $\Delta$ln(E) $<$0 & Evidence against M1 \\    
Grade 1 & 1 $<$ B $<$3 & 0 $<$ $\Delta$ln(E) $<$ 1 & Evidence for M1 not worth mentioning \\ 
Grade 2 & 3 $<$ B $<$20 & 1 $<$ $\Delta$ln(E) $<$ 3 & Positive Evidence for M1 \\ 
Grade 3 & 20 $<$ B $<$150 & 3 $<$ $\Delta$ln(E) $<$ 5 & Strong Evidence for M1 \\ 
Grade 4 & B $>$150 & $\Delta$ln(E) $>$ 5 & Decisive Evidence for M1 \\ 
\hline\hline
\end{tabular}%
}
\caption{Table of interpretation for the Bayes Factor adapted from \cite{kass_raft}. The Bayes Factor B is defined as the evidence ratios of the models M1 and M2. In particular ln(B) = $\Delta$ln(E). Note that we use ln(E), the natural logarithm of the Evidence, which is the standard \emph{MultiNest} outcome and is often noted log(E) in the literature.}
\label{tab:jeffrey}
\end{table*}

%% file: model_degen.tex
\subsection{Model choices and degeneracies}


Atmospheric retrieval techniques were historically developed for Earth and Solar System applications, where in-situ and high quality measurements provide a wealth of information. In this context, retrievals are used to infer probability distributions on a set of parameters for which priors knowledge is already available and tightly constraining. On the opposite, exoplanet retrievals must deal with low information content data sets and physical systems that are poorly understood. In this context, model assumptions are crucial \citep{Min_2020}. These assumptions are user defined and should be selected carefully, as a retrieval will always provide a model dependent solution \citep{Rocchetto_2016, Changeat_2019_2l, Changeat_2020_alfnoor}. In general, free assumptions attempt to heuristically describe properties of exoplanets, such as thermal or chemical profiles, using as few assumptions as possible. On the opposite, self-consistent models describe the given property using physically driven assumptions, for example chemical equilibrium or radiative equilibrium. In a low prior knowledge case, free retrievals should be favored until the understanding of the system is high enough to justify the assumptions of self-consistent models. Another issue of self-consistent modeling in retrievals is that the addition of more physical descriptions comes at the cost of computational resources, which often limits the level of details that can be included. In this work, we chose to explore both options, using free and equilibrium chemistry retrievals. Often, we find that the free approach leads to a better Bayesian evidence, thus demonstrating that the added flexibility in the free models provides a better explanation of the datasets considered here. \\

In this work, we focus on eclipse spectra, which as compared to transit observations, provide more constraints on the thermal profile \citep{Burrows_spectra_windows, Rocchetto_2016}. Due to the narrow wavelength range of HST, only one broadband spectral feature is usually observed, which can be difficult to interpret without further constraints. In the HST only case, the retrievals can easily be misguided, confusing for instance whether a single feature should be seen in emission or absorption. This is the case for HD\,209458\,b (see corresponding section), where the HST only retrievals converge to two different solutions depending on the opacities that are considered. When optical absorbers are included, the retrieval interprets the spectrum with an emission feature, while it is simply fit with absorption of H$_2$O in the reduced run. Adding the Spitzer data, here, provides confidence that the second scenario is more likely, but such degeneracies are often difficult to disentangle. In this work, we performed retrievals using a reduced list of opacities and a full list, as well as HST only and HST+Spitzer scenarios. In general, we find that a larger wavelength coverage is often required to extract information regarding the chemistry and thermal profile of the observed atmospheres in eclipse and we therefore focus on the HST+Spitzer runs for the main part of this study. \\  




\subsection{Biases arising from the data reduction}

In previous works, data reduction pipelines have led to different answers when used on the same raw data. When considering HST, different pipelines usually recover the same spectral shape, but differences can appear as flat offsets \citep[e.g.]{Changeat_2020_K11}. See also Figure Set B5, which compares our spectra with the ones obtained in the literature.

\begin{figure*}
    \includegraphics[width = 0.69\textwidth]{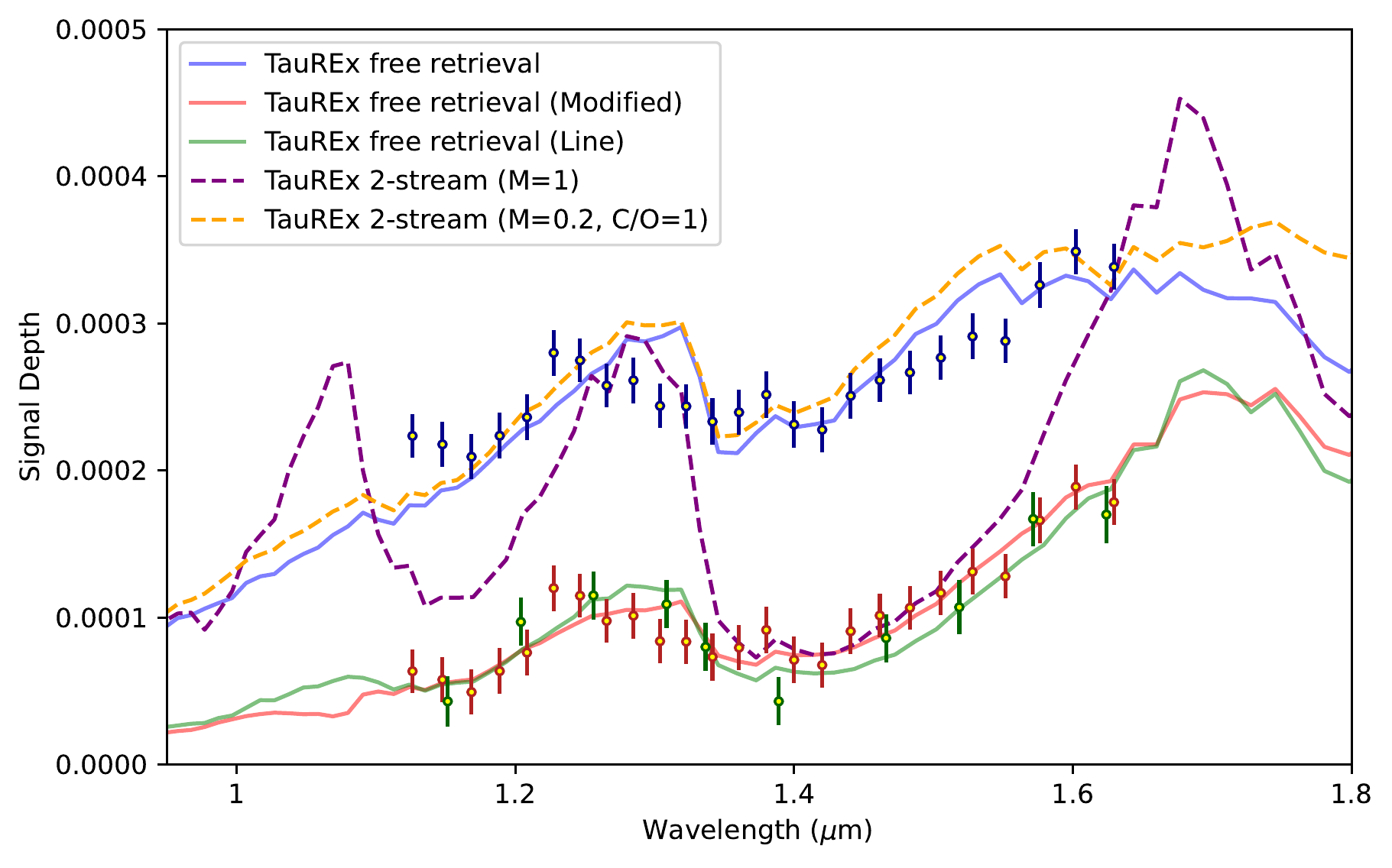}
    \includegraphics[width = 0.3\textwidth]{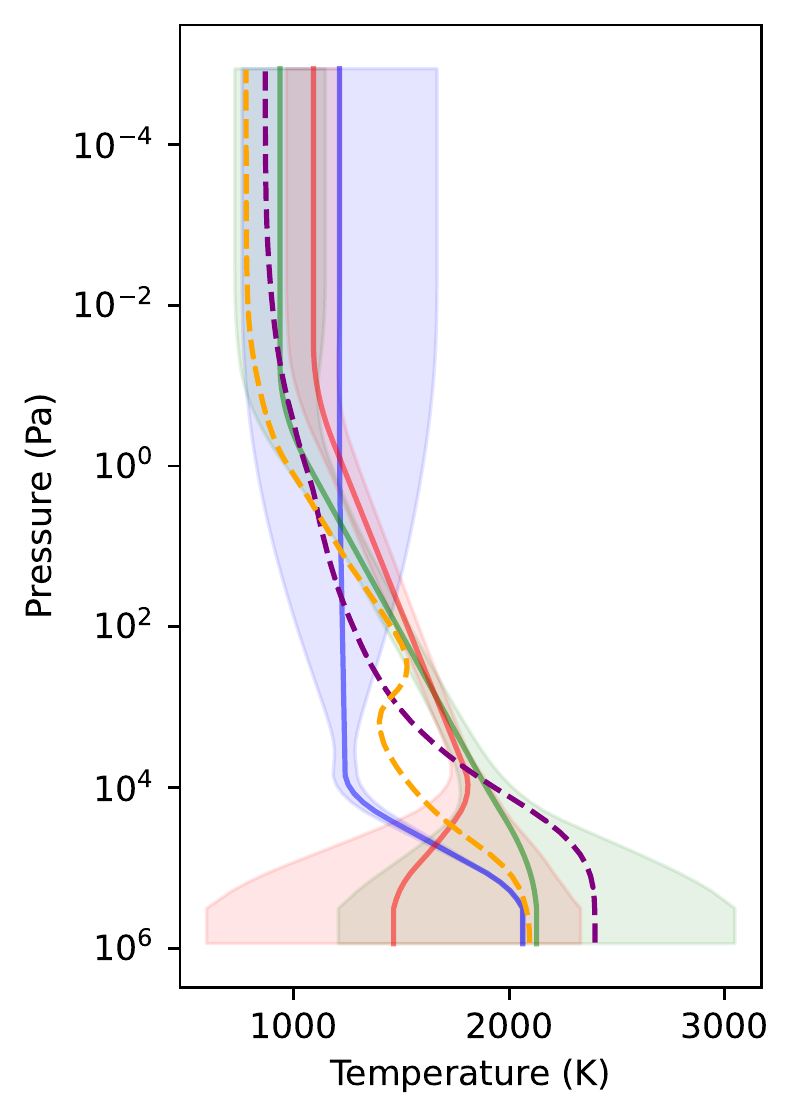}
    \caption{Simulated spectra (left) and thermal structure (right) for HD\,209458\,b. The blue datapoints are our original \emph{Iraclis} reduced spectrum. The green are the \cite{Line_HD209_spectrum_em} data. The red are the \emph{Iraclis} observations offset by 160 ppm to match the \cite{Line_HD209_spectrum_em} absolute depth. The same color scheme is applied for our retrievals on those different datasets. We also added forward model spectra computed from two-stream self-consistent approximations in dashed lines: solar in purple; metallicity 0.2 and C/O ratio 1 in orange. }
    \label{fig:spec_tp_2s}
\end{figure*}

The observed flat offsets are usually attributed to differences in the background subtraction steps or different choices to model the HST ramps. A particularly striking example of such offset can be seen in our reduction of the HD\,209458\,b spectrum, which differs by about 160 ppm with the one obtained by \cite{Line_HD209_spectrum_em}. Such large offsets are difficult to explain by differences in the background substraction alone. However, performing the reduction of the spectra using a quadratic trend for the long-term ramp, we obtained a similar flux level to \cite{Line_HD209_spectrum_em}. We also note that \cite{Line_HD209_spectrum_em} fitted all the visits with a single transit model, which could have led to different results as their reduction did not rely on a normalization of the white light curve depths. While our pipeline does not offer this option at the moment, future investigations could assess the impact of performing joint or separate fits. In order to investigate further this particular case, we performed comparative retrievals on the \cite{Line_HD209_spectrum_em} spectrum as well as our own spectrum offset by -160 ppm. In Figure \ref{fig:spec_tp_2s} we show the results of this exercise.

In the same figure, we also show self-consistent estimates of the HD\,209458\,b emission using a two-stream model available in \emph{TauREx}. While not suitable for direct data analyses studies, self-consistent forward models provide important benchmark to compare retrieval outcomes. This forward model computes the thermal profile and chemistry in radiative and chemical equilibrum from the bulk and orbital parameters of the system, which are here taken from the literature. We show two models, one with solar metallicity and C/O ratio and one with a metallicity 0.2 and a C/O ratio of 1.0, which roughly matches the derived values of our full HST+Spitzer run. Those forward models predicts a surface emission that might be compatible with the reduced observation we obtained from \emph{Iraclis}. This exercise is by no mean proving that our recovered spectrum is correct, but this highlight systematic biases that can arise from different assumptions in HST reduction pipeline.

These offsets can even be more difficult to handle when combining observations from different instruments in transit \citep{Yip_2020_LC, Changeat_2020_K11, Yip_2021_W96, Saba_2021}. In eclipse or phase-curve \citep{changeat_2021_phasecurve2}, similar incompatibilities can occur depending on the different reduction techniques or input parameters. Similarly, if the observations are not carried out simultaneously, one encounters the risk of contamination from stellar or even planetary variabilities. Indeed, dynamic simulations predict that exoplanet atmospheres are subject to large variabilities, with time-dependent storms and modons \citep{Cho_2003, Cho_2021}. In this work, we do not investigate those phenomena further, and for simplicity, we assume that the obtained HST and Spitzer observations are compatible. We, however, performed a few sensitivity tests.

\begin{figure*}
    \includegraphics[width = 0.5\textwidth]{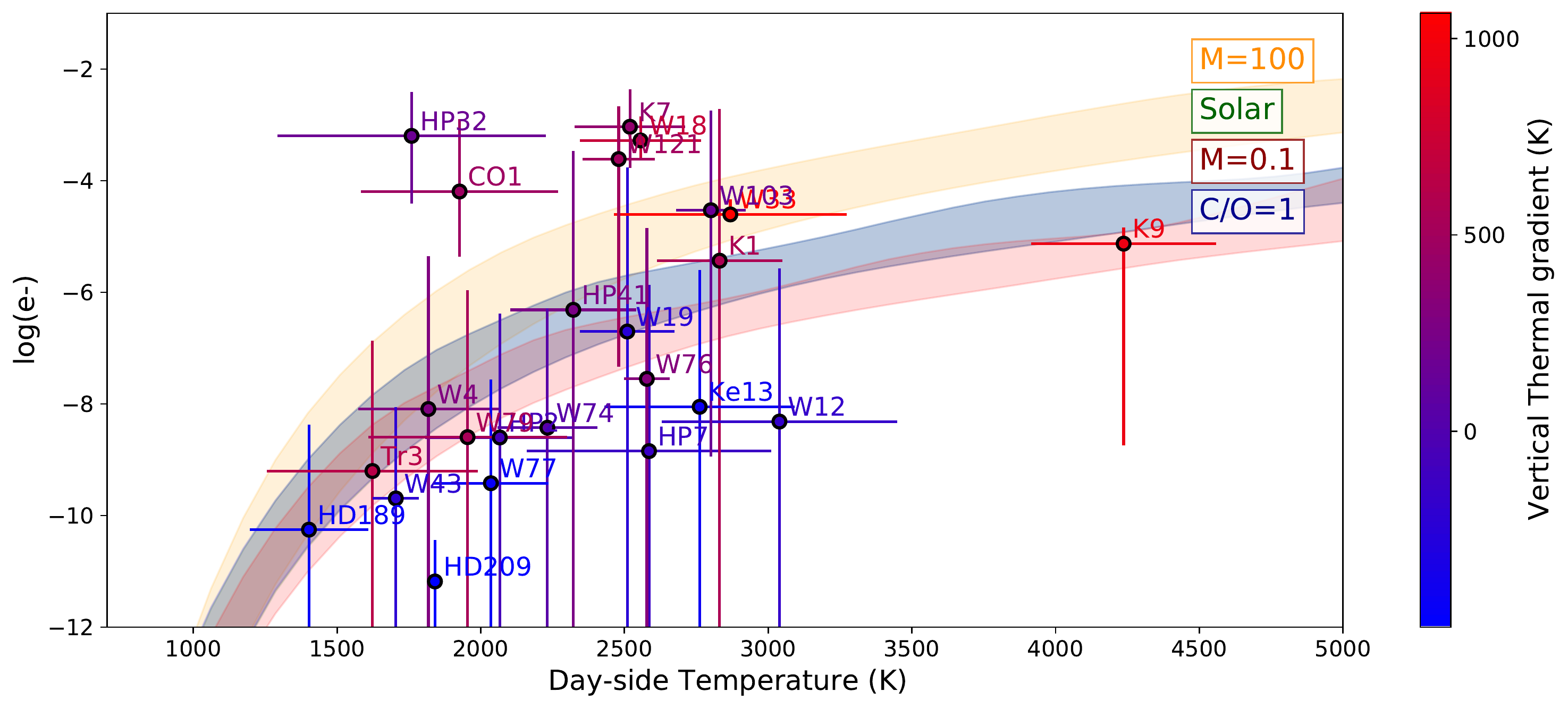}
    \includegraphics[width = 0.5\textwidth]{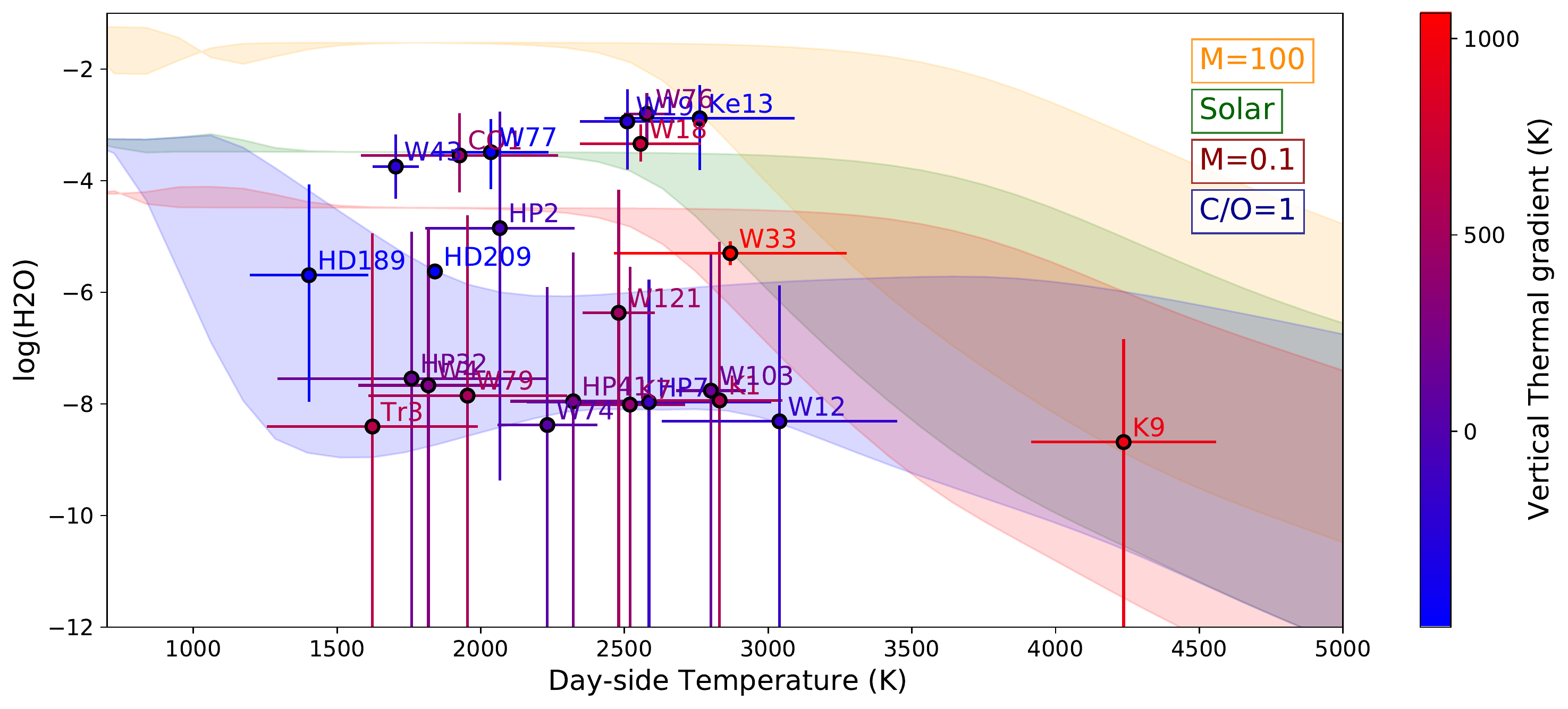}
    \includegraphics[width = 0.5\textwidth]{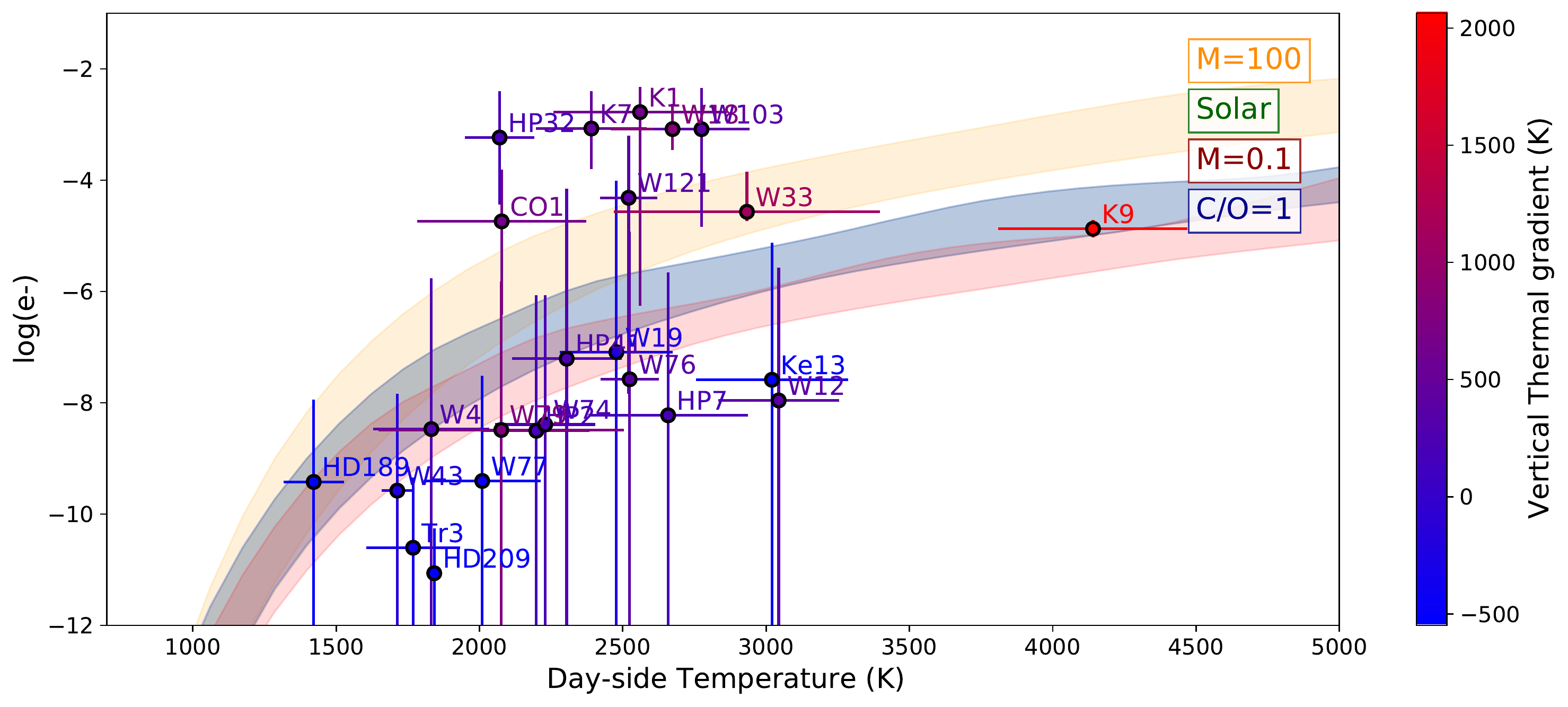}
    \includegraphics[width = 0.5\textwidth]{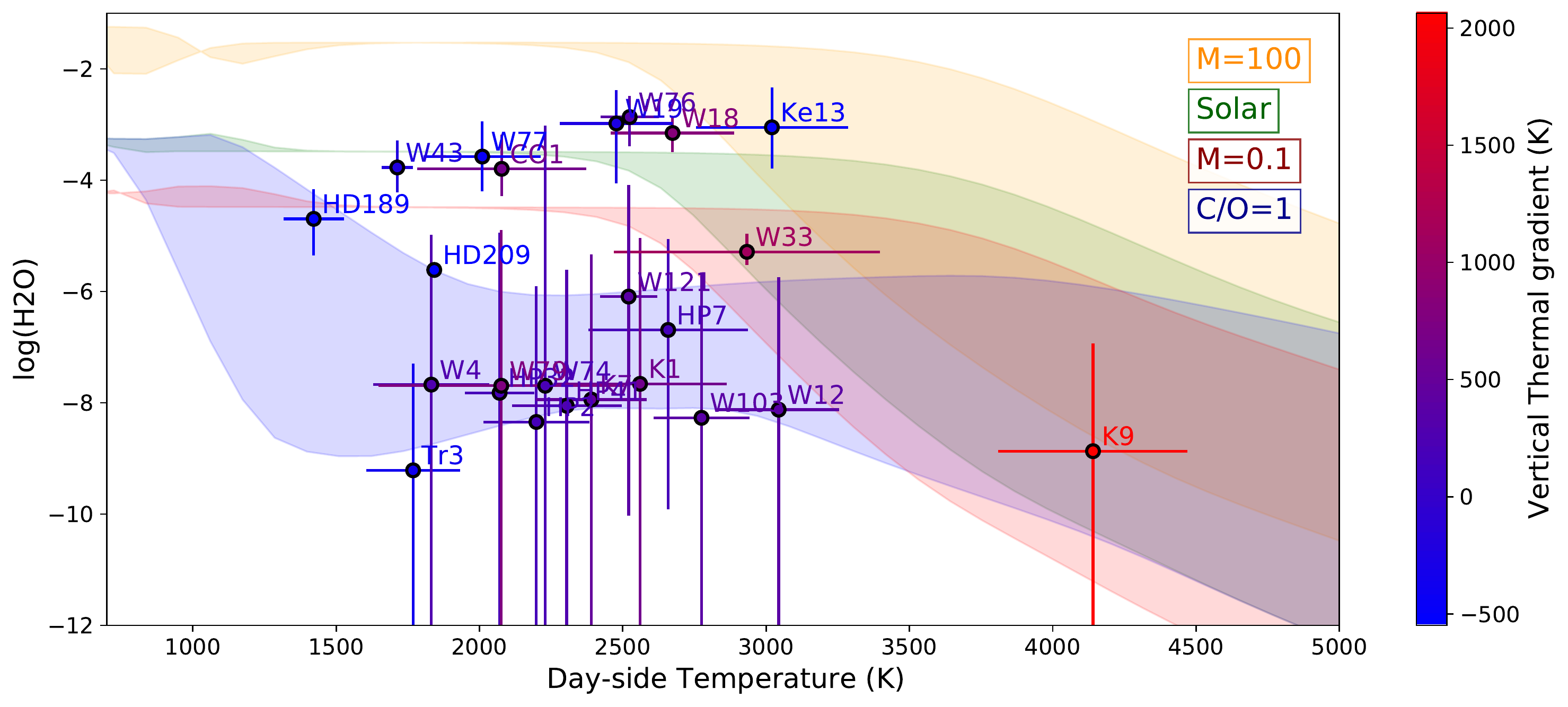}
    \caption{Recovered e$^-$ (left) and H$_2$O (right) maps of our full retrieval performed on intentionally biased Spitzer data. Top: The noise of the Spitzer points has been doubled. Bottom: The eclipsed depth of the Spitzer points is offset by +100ppm. While individual planet results can variate, the overall trends obtained from the unbiased Spitzer retrievals are conserved.}
    \label{fig:spitzer_combinations}
\end{figure*}

In order to verify the stability of our results to potential offset biases in the Spitzer data, we performed complementary retrievals on the whole population. We applied the full scenario on the HST+Spitzer data, modifying the observed Spitzer data. In our first test, we doubled the error of the Spitzer data for all planets, which is equivalent to adding another source of unknown Gaussian noise. In the two supplementary tests, we tested potential biases in the Spitzer data, by offsetting the eclipse depths of the Spitzer data by $\pm$ 100 ppm. In those three tests, we occasionally found different solutions for individual planets but overall the trends reported using the un-modified Spitzer sets remained unchanged. This is shown in Figure \ref{fig:spitzer_combinations}, which displays the water-temperature map for our population in the case where the Spitzer noise has been doubled, and the case where Spitzer has been offset by +100 ppm.

%% file: metallicity.tex
 In our free runs, we also compute the metallicity as O/H and the C/O ratios following the standard practice in the field \citep{Lee_2013_clouds, Macdonald_2019_hp29}. For each sample obtained during our retrieval exploration, we compute the metallicity, defined as the ratio O/H normalised to the solar value of 4.9$\times$10$^{-4}$ \citep{Asplund_2009}, and the C/O ratio. Constraints on those parameters are important for planetary formation and evolution models. 

The results of these calculations on our population are shown in Figure \ref{fig:met_co_full_spz}. We find that the estimates of metallicity and C/O ratios from free retrievals heavily rely on the detected molecules (see behaviour of the C/O ratio in the individual planet analyses) and come with many simplifying assumptions. Often, when a single molecule is detected in the retrieval, the derived parameter will be heavily biased towards the elemental ratio of the detected molecule. For example, detecting only CO$_2$ will lead to a derived C/O ratio of 0.5, which is spurious.

By comparing the derived metallicity to the one obtained by the equilibrium runs, we however believe this parameter is stable in the case where H$_2$O is clearly detected. This is shown in the top left panel of Figure \ref{fig:met_co_full_spz}, where we show that the retrievals detecting H$_2$O to high accuracies lead to similar estimates of the metallicity in free and equilibrium runs. We attribute this to the good capabilities of HST to detect this particular molecule, and our better understanding of the class of planets between 1000K and 2000K, which are believed to more closely following equilibrium chemistry.

\clearpage

%% file: individual_figure_set.tex
\begin{figure}[H]
\figurenum{D1}
\epsscale{1.15}
\plotone{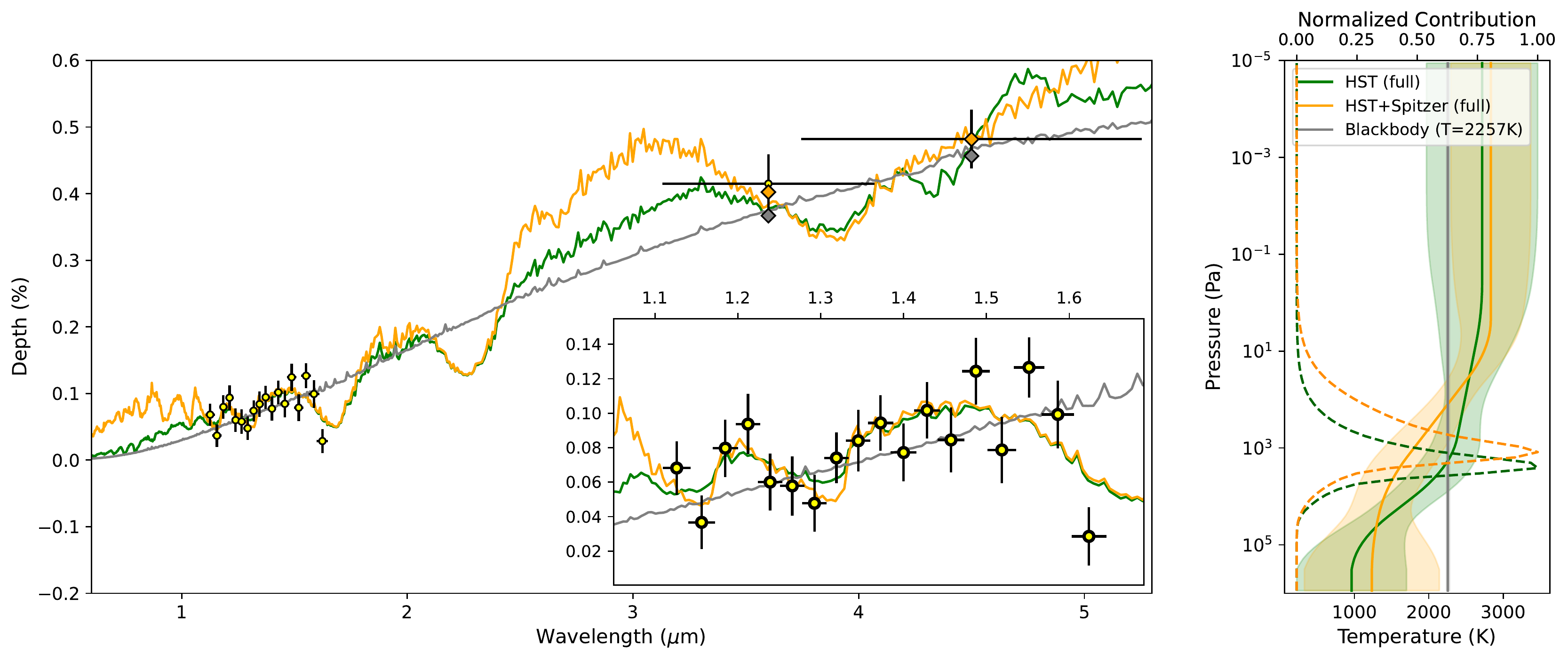}
\plotone{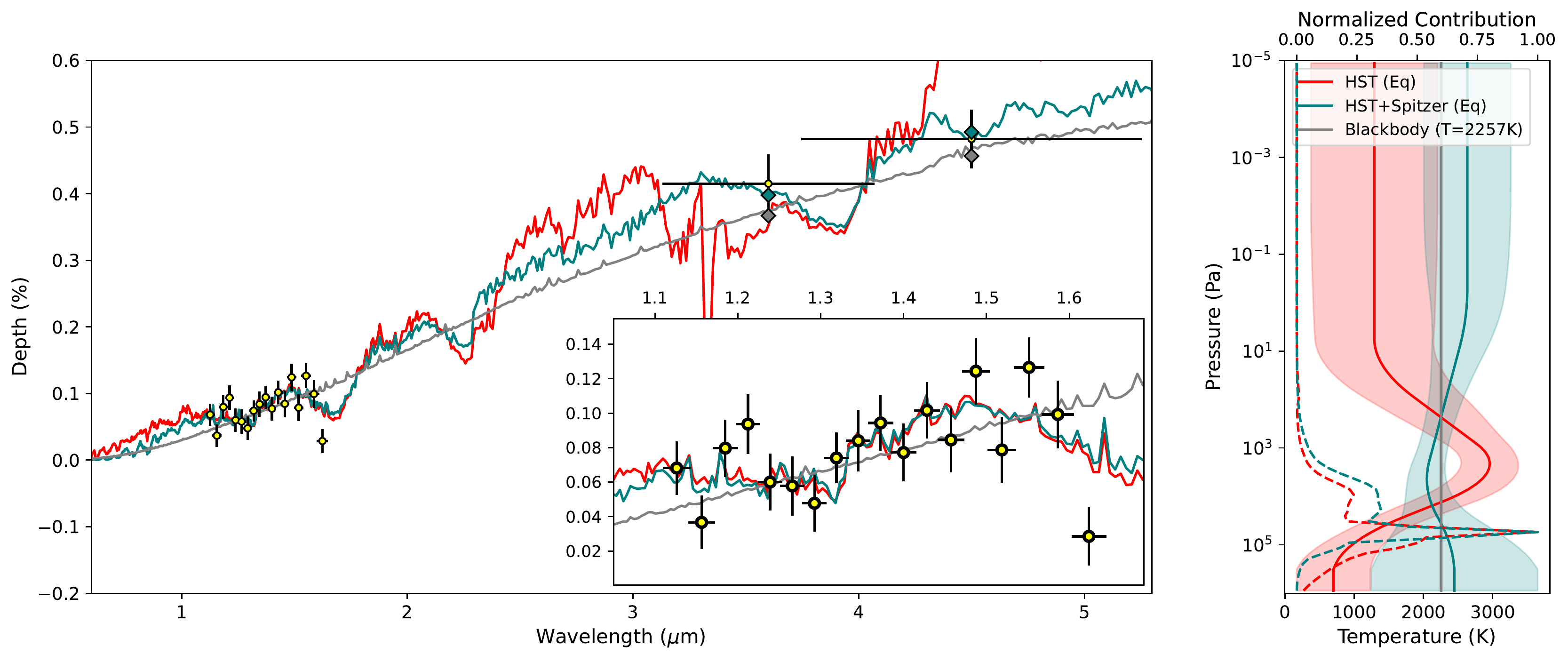}
\plotone{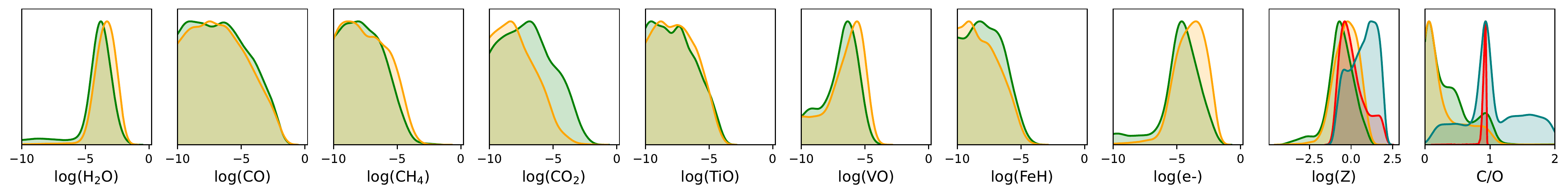}
\caption{Retrieval results for CoRoT-1\,b. Top: Best-fit eclipse spectra (left) and thermal structure (right) for the free retrievals. Middle: Best-fit eclipse spectra (left) and thermal structure (right) for the equilibrium retrievals. Bottom: Posterior distributions of these runs. This figure is a sample from the Figure Set D1 that contains the same information for all the planets. Figure Set D1 is available in the electronic edition of the {\it Astrophysical Journal}.
}
\end{figure}\label{fig:fig_set1}

\begin{figure}[H]
\figurenum{D2}
\epsscale{1.15}
\plotone{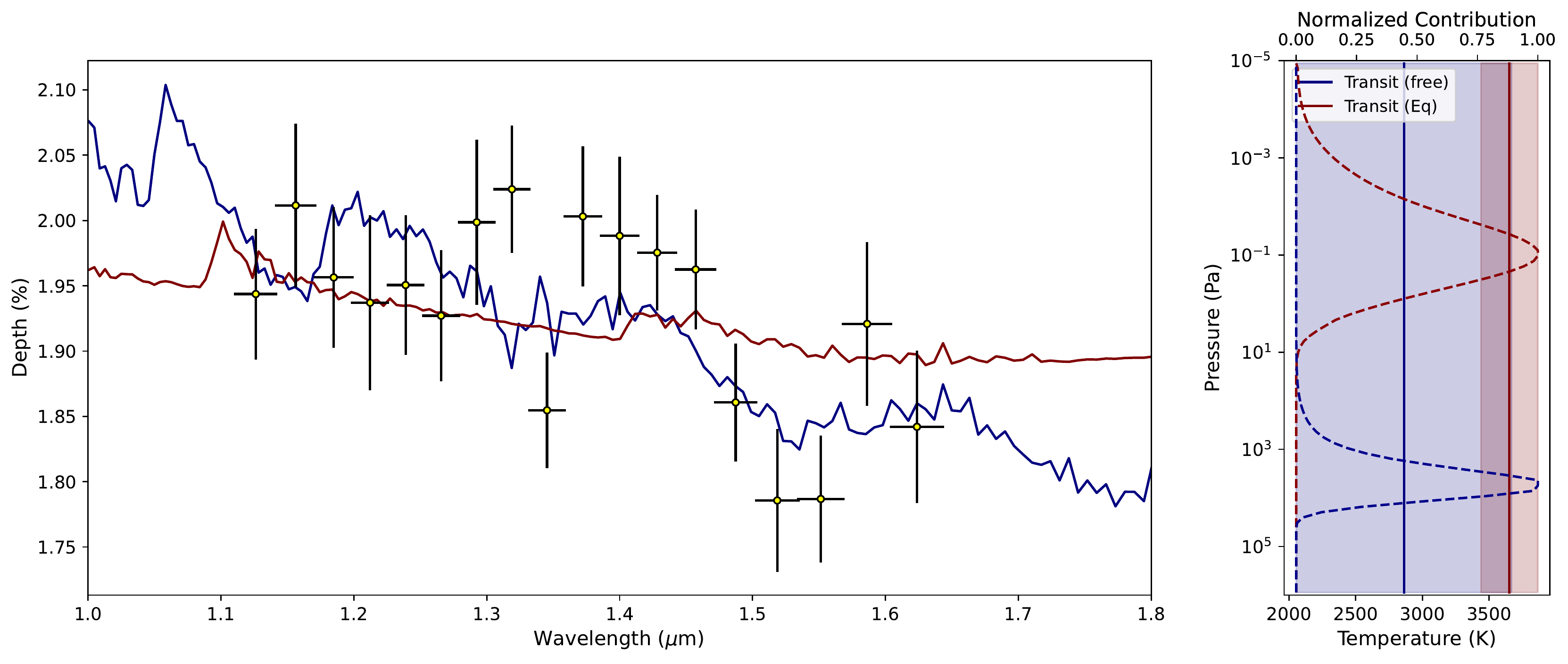}
\plotone{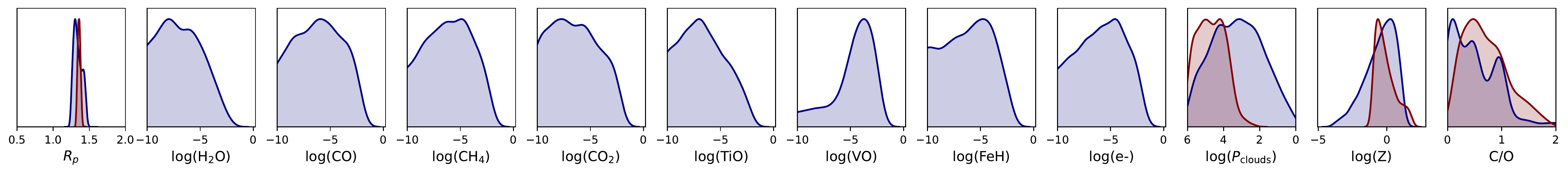}
\caption{Top: Best-fit spectra (left) and thermal structure (right) for the transit retrievals. Bottom: Posterior distributions of these runs. This figure is a sample from the Figure Set D2 that contains the same information for all the planets. Figure Set D2 is available in the electronic edition of the {\it Astrophysical Journal}. 
}
\end{figure}\label{fig:fig_set2}

%% file: combined_retrieval_table.tex
\startlongtable
\begin{deluxetable}{ccccccccc}
\tablecaption{Table of the retrieval results for the 24 planets. This table is published in a machine-readable format.}
\tablewidth{0pt}
\tabletypesize{\scriptsize}
\tablehead{
\colhead{Parameter} & \colhead{HST} & \colhead{HST+Spitzer} &
\colhead{HST} & \colhead{HST+Spitzer} & 
\colhead{HST} & \colhead{HST+Spitzer} &
\colhead{Transit} & \colhead{Transit} \\
\colhead{} & \multicolumn{2}{c}{(red)} & \multicolumn{2}{c}{(full)} &
\multicolumn{2}{c}{(eq)} & \colhead{(free)} & \colhead{(eq)}
}
\startdata
\cutinhead{CoRoT-1\,b\tablenotemark{a}} \\
log(H$_2$O) & $< -7.3$ & $< -7.0$ & $-3.9^{+0.7}_{-0.8}$ & $-3.4^{+0.7}_{-0.7}$ & -  & -  & $< -4.9$ & - \\  
log(CH$_4$) & $-5.1^{+2.0}_{-4.1}$ & $< -4.5$ & $< -6.2$ & $< -5.7$ & -  & -  & $< -3.9$ & - \\  
log(CO) & $< -3.9$ & $< -3.5$ & $< -4.5$ & $< -4.6$ & -  & -  & $< -3.8$ & - \\  
log(CO$_2$) & $< -5.1$ & $< -3.6$ & $< -5.1$ & $< -6.5$ & -  & -  & $< -4.0$ & - \\  
log(TiO) & -  & -  & $< -6.2$ & $< -6.1$ & -  & -  & $< -4.5$ & - \\  
log(VO) & -  & -  & $-6.8^{+1.0}_{-2.5}$ & $-6.4^{+1.1}_{-2.7}$ & -  & -  & $-4.3^{+1.3}_{-3.0}$ & - \\  
log(FeH) & -  & -  & $< -6.5$ & $< -6.8$ & -  & -  & $< -3.9$ & - \\  
log(e-) & -  & -  & $-4.6^{+1.2}_{-1.5}$ & $-3.9^{+1.1}_{-1.2}$ & -  & -  & $-6.2^{+2.5}_{-3.3}$ & - \\  
log(Z) & $-1.7^{+1.4}_{-1.4}$ & $-1.2^{+1.5}_{-1.5}$ & $-0.6^{+0.8}_{-0.8}$ & $-0.3^{+0.7}_{-0.7}$ & $-0.1^{+1.0}_{-0.5}$ & $0.8^{+0.8}_{-1.1}$ & $-0.5^{+1.1}_{-1.1}$ & $-0.3^{+0.9}_{-0.5}$\\  
C/O & $1.0^{+0.5}_{-0.5}$ & $0.9^{+0.5}_{-0.5}$ & $0.3^{+0.3}_{-0.3}$ & $0.3^{+0.3}_{-0.3}$ & $0.9^{+0.0}_{-0.0}$ & $0.9^{+0.7}_{-0.3}$ & $0.5^{+0.5}_{-0.5}$ & $0.6^{+0.5}_{-0.3}$\\  
ln(E) & 118.1 & 125.6 & 119.1 & 132.4 & 116.8 & 127.7 & 100.0 & 96.9\\ 
\cutinhead{HAT-P-2\,b\tablenotemark{b}} \\
log(H$_2$O) & $< -4.3$ & $-5.8^{+1.9}_{-4.1}$ & $-5.5^{+2.0}_{-4.1}$ & $-3.7^{+1.1}_{-3.8}$ & -  & - & - & -\\  
log(CH$_4$) & $< -5.4$ & $< -7.1$ & $< -5.0$ & $< -6.1$ & -  & - & -& - \\  
log(CO) & $< -3.6$ & $< -4.8$ & $< -3.9$ & $< -3.7$ & -  & - & - & -\\  
log(CO$_2$) & $< -3.2$ & $< -6.6$ & $< -3.1$ & $< -5.5$ & -  & - & - & -\\  
log(TiO) & -  & -  & $< -7.0$ & $< -6.8$ & -  & - & - & -\\  
log(VO) & -  & -  & $< -6.8$ & $< -6.0$ & -  & - & - & -\\  
log(FeH) & -  & -  & $< -6.8$ & $< -7.1$ & -  & - & - & -\\  
log(e-) & -  & -  & $< -6.8$ & $< -6.7$ & -  & - & - & -\\  
log(Z) & $-0.8^{+1.3}_{-1.3}$ & $-1.6^{+1.3}_{-1.3}$ & $-0.8^{+1.4}_{-1.4}$ & $-0.6^{+1.3}_{-1.3}$ & $0.5^{+1.1}_{-1.2}$ & $0.3^{+1.0}_{-0.9}$ & - & -\\  
C/O & $0.7^{+0.4}_{-0.4}$ & $0.6^{+0.5}_{-0.5}$ & $0.6^{+0.4}_{-0.4}$ & $0.4^{+0.4}_{-0.4}$ & $0.6^{+0.3}_{-0.3}$ & $0.5^{+0.3}_{-0.2}$ & - & -\\  
ln(E) & 216.6 & 244.8 & 216.4 & 244.5 & 216.9 & 245.7 & - & -\\ 
\cutinhead{HAT-P-7\,b\tablenotemark{c}} \\
log(H$_2$O) & $< -6.1$ & $< -6.9$ & $< -3.3$ & $-6.0^{+1.0}_{-3.4}$ & -  & -  & $-5.3^{+2.3}_{-3.5}$ & - \\  
log(CH$_4$) & $< -6.5$ & $-5.9^{+0.7}_{-2.2}$ & $< -4.9$ & $< -6.9$ & -  & -  & $< -3.6$ & - \\  
log(CO) & $< -3.9$ & $< -4.8$ & $< -4.1$ & $< -3.8$ & -  & -  & $< -4.6$ & - \\  
log(CO$_2$) & $< -5.6$ & $< -6.2$ & $< -3.2$ & $-2.5^{+0.3}_{-1.1}$ & -  & -  & $< -3.6$ & - \\  
log(TiO) & -  & -  & $< -6.4$ & $< -6.7$ & -  & -  & $< -5.2$ & - \\  
log(VO) & -  & -  & $< -6.1$ & $< -8.1$ & -  & -  & $< -4.7$ & - \\  
log(FeH) & -  & -  & $< -5.1$ & $-4.8^{+0.5}_{-0.6}$ & -  & -  & $< -4.6$ & - \\  
log(e-) & -  & -  & $< -4.2$ & $< -5.9$ & -  & -  & $< -3.9$ & - \\  
log(Z) & $-1.8^{+1.2}_{-1.2}$ & $-2.4^{+1.1}_{-1.1}$ & $-0.8^{+1.4}_{-1.4}$ & $0.5^{+1.1}_{-1.1}$ & $0.8^{+0.6}_{-0.9}$ & $-0.4^{+1.1}_{-0.5}$ & $-0.5^{+1.2}_{-1.2}$ & $0.4^{+1.1}_{-0.9}$\\  
C/O & $0.8^{+0.4}_{-0.4}$ & $1.1^{+0.6}_{-0.6}$ & $0.6^{+0.5}_{-0.5}$ & $0.5^{+0.2}_{-0.2}$ & $1.6^{+0.2}_{-0.3}$ & $1.4^{+0.3}_{-0.3}$ & $0.5^{+0.5}_{-0.5}$ & $0.9^{+0.7}_{-0.5}$\\  
ln(E) & 208.4 & 230.6 & 208.1 & 231.6 & 207.9 & 228.8 & 156.7 & 157.1\\ 
\cutinhead{HAT-P-32\,b\tablenotemark{d}} \\
log(H$_2$O) & $< -7.4$ & $< -6.8$ & $< -5.7$ & $< -4.9$ & -  & -  & $-2.9^{+0.6}_{-0.7}$ & - \\  
log(CH$_4$) & $< -6.8$ & $< -7.8$ & $< -5.4$ & $-5.7^{+1.2}_{-3.7}$ & -  & -  & $< -6.3$ & - \\  
log(CO) & $< -3.9$ & $< -3.7$ & $< -4.2$ & $< -4.4$ & -  & -  & $< -4.5$ & - \\  
log(CO$_2$) & $-3.5^{+0.7}_{-0.8}$ & $-2.8^{+0.5}_{-0.5}$ & $< -4.5$ & $< -7.0$ & -  & -  & $< -4.9$ & - \\  
log(TiO) & -  & -  & $< -6.9$ & $< -6.8$ & -  & -  & $< -6.3$ & - \\  
log(VO) & -  & -  & $< -5.3$ & $-6.6^{+1.5}_{-3.3}$ & -  & -  & $< -7.1$ & - \\  
log(FeH) & -  & -  & $< -5.0$ & $< -6.8$ & -  & -  & $-7.9^{+1.4}_{-2.2}$ & - \\  
log(e-) & -  & -  & $-3.8^{+1.2}_{-4.6}$ & $-3.2^{+0.8}_{-1.0}$ & -  & -  & $< -5.8$ & - \\  
log(Z) & $-0.1^{+0.9}_{-0.9}$ & $0.6^{+0.6}_{-0.6}$ & $-1.6^{+1.3}_{-1.3}$ & $-1.8^{+1.1}_{-1.1}$ & $1.0^{+0.6}_{-1.1}$ & $0.4^{+1.0}_{-0.9}$ & $0.1^{+0.6}_{-0.6}$ & $-0.1^{+1.2}_{-0.7}$\\  
C/O & $0.6^{+0.2}_{-0.2}$ & $0.5^{+0.1}_{-0.1}$ & $0.7^{+0.4}_{-0.4}$ & $0.8^{+0.7}_{-0.7}$ & $1.6^{+0.3}_{-0.3}$ & $1.5^{+0.3}_{-0.3}$ & $0.2^{+0.2}_{-0.2}$ & $0.5^{+0.3}_{-0.3}$\\  
ln(E) & 181.8 & 192.3 & 179.9 & 193.4 & 176.8 & 189.9 & 179.4 & 181.7\\ 
\cutinhead{HAT-P-41\,b\tablenotemark{e}} \\
log(H$_2$O) & $< -6.9$ & $< -6.9$ & $< -4.8$ & $< -5.6$ & -  & -  & $-2.7^{+0.5}_{-0.6}$ & - \\  
log(CH$_4$) & $< -6.9$ & $< -6.7$ & $< -4.1$ & $< -6.1$ & -  & -  & $< -5.2$ & - \\  
log(CO) & $< -4.7$ & $< -5.5$ & $< -4.1$ & $< -4.8$ & -  & -  & $< -4.3$ & - \\  
log(CO$_2$) & $< -4.9$ & $< -7.6$ & $< -4.0$ & $< -7.3$ & -  & -  & $< -4.6$ & - \\  
log(TiO) & -  & -  & $< -7.1$ & $< -6.0$ & -  & -  & $< -6.8$ & - \\  
log(VO) & -  & -  & $-6.0^{+1.0}_{-2.8}$ & $-6.0^{+1.5}_{-3.5}$ & -  & -  & $< -7.2$ & - \\  
log(FeH) & -  & -  & $-5.7^{+1.2}_{-3.1}$ & $-5.3^{+0.8}_{-2.6}$ & -  & -  & $< -7.1$ & - \\  
log(e-) & -  & -  & $-5.9^{+2.6}_{-3.3}$ & $< -5.3$ & -  & -  & $< -4.6$ & - \\  
log(Z) & $-1.9^{+1.2}_{-1.2}$ & $-2.5^{+1.2}_{-1.2}$ & $-1.2^{+1.2}_{-1.2}$ & $-1.7^{+1.0}_{-1.0}$ & $0.5^{+0.9}_{-0.9}$ & $0.7^{+0.8}_{-0.9}$ & $0.3^{+0.6}_{-0.6}$ & $1.0^{+0.5}_{-1.1}$\\  
C/O & $0.8^{+0.4}_{-0.4}$ & $0.8^{+0.5}_{-0.5}$ & $0.7^{+0.5}_{-0.5}$ & $0.5^{+0.5}_{-0.5}$ & $1.3^{+0.4}_{-0.8}$ & $1.6^{+0.3}_{-0.3}$ & $0.2^{+0.2}_{-0.2}$ & $0.5^{+0.3}_{-0.2}$\\  
ln(E) & 187.2 & 200.9 & 188.4 & 200.3 & 186.4 & 200.3 & 190.6 & 193.4\\
\cutinhead{HAT-P-70\,b\tablenotemark{f}} \\
log(H$_2$O) & $< -5.8$ & - & $< -5.0$ & - & - & - & - & -\\  
log(CH$_4$) & $-4.4^{+0.6}_{-3.1}$ & - & $< -4.8$ & - & - & - & - & -\\  
log(CO) & $< -2.6$ & - & $< -4.1$ & - & - & - & - & -\\  
log(CO$_2$) & $< -3.6$ & - & $< -3.3$ & - & - & - & - & -\\  
log(TiO) & -  & - & $< -6.4$ & - & - & - & - & -\\  
log(VO) & -  & - & $< -6.0$ & - & - & - & - & -\\  
log(FeH) & -  & - & $< -7.5$ & - & - & - & - & -\\  
log(e-) & -  & - & $-3.1^{+0.7}_{-0.8}$ & - & - & - & - & -\\  
log(Z) & $-0.8^{+1.4}_{-1.4}$ & - & $-1.2^{+1.4}_{-1.4}$ & - & $-0.5^{+0.6}_{-0.4}$ & - & - & -\\  
C/O & $1.0^{+0.5}_{-0.5}$ & - & $0.7^{+0.4}_{-0.4}$ & - & $1.0^{+0.05}_{-0.05}$ & - & - & -\\  
ln(E) & 198.3 & - & 201.5 & - & 200.0 & - & - & -\\ 
\cutinhead{HD\,189733\,b\tablenotemark{g}} \\
log(H$_2$O) & $< -6.3$ & $-5.4^{+0.5}_{-0.5}$ & $< -4.2$ & $-4.6^{+0.5}_{-0.6}$ & -  & -  & $-2.6^{+0.4}_{-0.6}$ & - \\  
log(CH$_4$) & $-6.4^{+1.9}_{-3.6}$ & $< -8.3$ & $< -4.0$ & $< -8.0$ & -  & -  & $< -5.3$ & - \\  
log(CO) & $< -3.6$ & $< -3.7$ & $< -4.3$ & $< -3.1$ & -  & -  & $< -4.3$ & - \\  
log(CO$_2$) & $< -4.4$ & $-5.8^{+2.4}_{-0.6}$ & $< -4.0$ & $-2.5^{+0.3}_{-0.6}$ & -  & -  & $< -4.1$ & - \\  
log(TiO) & -  & -  & $< -5.5$ & $< -6.4$ & -  & -  & $< -5.2$ & - \\  
log(VO) & -  & -  & $< -5.8$ & $< -7.2$ & -  & -  & $-7.4^{+1.4}_{-2.6}$ & - \\  
log(FeH) & -  & -  & $< -6.0$ & $-6.9^{+1.0}_{-2.3}$ & -  & -  & $< -7.2$ & - \\  
log(e-) & -  & -  & $-3.5^{+1.0}_{-4.8}$ & $< -8.1$ & -  & -  & $< -8.2$ & - \\  
log(Z) & $-1.5^{+1.4}_{-1.4}$ & $-1.2^{+1.2}_{-1.2}$ & $-1.2^{+1.3}_{-1.3}$ & $0.8^{+0.7}_{-0.7}$ & $-0.4^{+1.2}_{-0.4}$ & $1.7^{+0.2}_{-0.3}$ & $0.5^{+0.5}_{-0.5}$ & $0.7^{+0.7}_{-1.0}$\\  
C/O & $0.8^{+0.5}_{-0.5}$ & $0.5^{+0.3}_{-0.3}$ & $0.6^{+0.5}_{-0.5}$ & $0.5^{+0.1}_{-0.1}$ & $1.4^{+0.4}_{-0.3}$ & $0.8^{+0.1}_{-0.2}$ & $0.2^{+0.2}_{-0.2}$ & $0.5^{+0.6}_{-0.3}$\\  
ln(E) & 147.5 & 177.1 & 146.3 & 178.1 & 146.6 & 176.3 & 193.0 & 195.1\\  
\cutinhead{HD\,209458\,b\tablenotemark{h}} \\
log(H$_2$O) & $-5.56^{+0.09}_{-0.08}$ & $-5.6^{+0.06}_{-0.06}$ & $< -8.43$ & $-5.61^{+0.06}_{-0.06}$ & -  & -  & $-2.69^{+0.41}_{-0.75}$ & - \\  
log(CH$_4$) & $< -7.58$ & $-6.66^{+0.22}_{-0.25}$ & $< -7.78$ & $-6.68^{+0.22}_{-0.25}$ & -  & -  & $< -4.7$ & - \\  
log(CO) & $-5.19^{+1.54}_{-4.42}$ & $-3.77^{+0.4}_{-0.53}$ & $< -6.27$ & $-3.72^{+0.38}_{-0.5}$ & -  & -  & $< -4.34$ & - \\  
log(CO$_2$) & $< -5.37$ & $< -6.44$ & $< -5.88$ & $< -6.51$ & -  & -  & $< -4.86$ & - \\  
log(TiO) & -  & -  & $< -9.05$ & $< -9.65$ & -  & -  & $< -5.57$ & - \\  
log(VO) & -  & -  & $< -9.41$ & $< -10.4$ & -  & -  & $< -7.66$ & - \\  
log(FeH) & -  & -  & $-9.56^{+0.3}_{-0.45}$ & $< -10.69$ & -  & -  & $< -7.69$ & - \\  
log(e-) & -  & -  & $-6.56^{+0.06}_{-0.06}$ & $< -10.33$ & -  & -  & $< -7.3$ & - \\  
log(Z) & $-1.52^{+0.85}_{-0.85}$ & $-0.74^{+0.48}_{-0.48}$ & $-2.74^{+0.74}_{-0.74}$ & $-0.69^{+0.46}_{-0.46}$ & $1.97^{+0.03}_{-0.05}$ & $-0.99^{+0.02}_{-0.01}$ & $0.3^{+0.53}_{-0.53}$ & $0.04^{+0.77}_{-0.69}$\\  
C/O & $0.66^{+0.35}_{-0.35}$ & $0.95^{+0.13}_{-0.13}$ & $0.73^{+0.4}_{-0.4}$ & $0.96^{+0.11}_{-0.11}$ & $0.77^{+0.03}_{-0.03}$ & $0.92^{+0.01}_{-0.01}$ & $0.18^{+0.21}_{-0.21}$ & $0.43^{+0.3}_{-0.23}$\\  
ln(E) & 198.76 & 236.92 & 224.61 & 232.58 & 187.72 & 193.11 & 208.8 & 210.8\\  
\cutinhead{KELT-1\,b\tablenotemark{i}} \\
log(H$_2$O) & $< -5.9$ & $< -6.0$ & $< -4.9$ & $< -4.5$ & -  & -  & $< -3.9$ & - \\  
log(CH$_4$) & $< -4.0$ & $< -7.1$ & $< -5.3$ & $-6.1^{+1.8}_{-3.5}$ & -  & -  & $< -4.2$ & - \\  
log(CO) & $< -2.8$ & $< -5.6$ & $< -4.0$ & $< -5.0$ & -  & -  & $< -4.0$ & - \\  
log(CO$_2$) & $-2.3^{+0.2}_{-0.2}$ & $-2.3^{+0.2}_{-0.2}$ & $< -4.0$ & $< -3.5$ & -  & -  & $< -4.0$ & - \\  
log(TiO) & -  & -  & $< -7.2$ & $< -7.1$ & -  & -  & $< -4.2$ & - \\  
log(VO) & -  & -  & $-6.1^{+0.6}_{-2.3}$ & $-6.2^{+0.6}_{-1.7}$ & -  & -  & $< -3.6$ & - \\  
log(FeH) & -  & -  & $-5.2^{+0.8}_{-1.1}$ & $-6.0^{+0.8}_{-1.8}$ & -  & -  & $< -3.9$ & - \\  
log(e-) & -  & -  & $< -3.0$ & $< -2.8$ & -  & -  & $< -4.1$ & - \\  
log(Z) & $1.1^{+0.2}_{-0.2}$ & $1.1^{+0.4}_{-0.4}$ & $-1.3^{+1.2}_{-1.2}$ & $-1.4^{+1.3}_{-1.3}$ & $-0.4^{+0.3}_{-0.2}$ & $1.0^{+0.5}_{-0.6}$ & $-0.6^{+1.2}_{-1.2}$ & $0.5^{+1.0}_{-0.9}$\\  
C/O & $0.5^{+0.1}_{-0.1}$ & $0.5^{+0.1}_{-0.1}$ & $0.6^{+0.4}_{-0.4}$ & $0.6^{+0.6}_{-0.6}$ & $1.8^{+0.1}_{-0.2}$ & $0.4^{+0.2}_{-0.2}$ & $0.5^{+0.5}_{-0.5}$ & $1.0^{+0.7}_{-0.5}$\\  
ln(E) & 189.3 & 199.8 & 193.2 & 206.9 & 186.9 & 198.3 & 191.1 & 190.5\\ 
\cutinhead{KELT-7\,b\tablenotemark{j}} \\
log(H$_2$O) & $< -6.9$ & $< -6.1$ & $< -5.3$ & $< -5.5$ & -  & -  & $-4.4^{+1.2}_{-3.8}$ & - \\  
log(CH$_4$) & $< -6.0$ & $< -7.3$ & $< -5.1$ & $-6.0^{+1.7}_{-3.7}$ & -  & -  & $< -5.9$ & - \\  
log(CO) & $< -3.5$ & $< -2.9$ & $< -3.9$ & $< -6.2$ & -  & -  & $< -4.4$ & - \\  
log(CO$_2$) & $< -4.1$ & $-2.6^{+0.4}_{-0.6}$ & $< -4.0$ & $< -7.7$ & -  & -  & $< -5.0$ & - \\  
log(TiO) & -  & -  & $< -6.0$ & $-5.1^{+0.7}_{-1.1}$ & -  & -  & $< -5.5$ & - \\  
log(VO) & -  & -  & $< -7.1$ & $-7.0^{+1.5}_{-3.0}$ & -  & -  & $< -7.1$ & - \\  
log(FeH) & -  & -  & $< -7.6$ & $< -6.2$ & -  & -  & $< -5.5$ & - \\  
log(e-) & -  & -  & $-2.9^{+0.6}_{-0.7}$ & $-3.0^{+0.6}_{-0.8}$ & -  & -  & $-3.8^{+1.2}_{-1.9}$ & - \\  
log(Z) & $-1.2^{+1.2}_{-1.2}$ & $0.7^{+0.7}_{-0.7}$ & $-1.4^{+1.3}_{-1.3}$ & $-1.8^{+0.7}_{-0.7}$ & $0.7^{+0.8}_{-0.8}$ & $0.7^{+0.7}_{-0.6}$ & $-0.9^{+1.1}_{-1.1}$ & $0.7^{+0.7}_{-0.9}$\\  
C/O & $0.8^{+0.4}_{-0.4}$ & $0.6^{+0.2}_{-0.2}$ & $0.7^{+0.4}_{-0.4}$ & $0.8^{+0.8}_{-0.8}$ & $1.0^{+0.0}_{-0.0}$ & $1.6^{+0.3}_{-0.3}$ & $0.5^{+0.4}_{-0.4}$ & $0.7^{+0.8}_{-0.4}$\\  
ln(E) & 208.5 & 218.1 & 209.7 & 222.7 & 204.6 & 215.3 & 200.6 & 185.8\\  
\cutinhead{KELT-9\,b\tablenotemark{k}} \\
log(H$_2$O) & $-5.5^{+0.1}_{-0.2}$ & $-3.9^{+0.5}_{-0.2}$ & $< -5.2$ & $< -7.0$ & -  & - & -  & - \\  
log(CH$_4$) & $< -4.1$ & $-2.9^{+0.2}_{-0.1}$ & $< -3.7$ & $< -6.1$ & -  & - & -  & - \\  
log(CO) & $-2.3^{+0.2}_{-0.4}$ & $< -3.2$ & $< -4.0$ & $< -4.7$ & -  & - & -  & - \\  
log(CO$_2$) & $-3.7^{+0.6}_{-0.2}$ & $-2.1^{+0.0}_{-0.2}$ & $< -6.2$ & $< -6.2$ & -  & - & -  & - \\  
log(TiO) & -  & -  & $-3.7^{+0.5}_{-0.8}$ & $-6.8^{+0.3}_{-0.3}$ & -  & - & -  & - \\  
log(VO) & -  & -  & $-3.8^{+0.5}_{-0.7}$ & $-6.6^{+0.2}_{-0.2}$ & -  & - & -  & - \\  
log(FeH) & -  & -  & $-5.1^{+1.3}_{-2.8}$ & $-8.0^{+1.1}_{-2.0}$ & -  & - & -  & - \\  
log(e-) & -  & -  & $< -4.2$ & $-4.9^{+0.2}_{-0.1}$ & -  & - & -  & - \\  
log(Z) & $0.9^{+0.2}_{-0.2}$ & $1.2^{+0.4}_{-0.4}$ & $-0.3^{+0.6}_{-0.6}$ & $-2.5^{+1.0}_{-1.0}$ & $-0.3^{+0.2}_{-0.2}$ & $1.6^{+0.2}_{-0.4}$ & -  & - \\  
C/O & $0.9^{+0.1}_{-0.1}$ & $0.7^{+0.4}_{-0.4}$ & $0.5^{+0.6}_{-0.6}$ & $0.6^{+0.5}_{-0.5}$ & $1.0^{+0.0}_{-0.0}$ & $1.1^{+0.6}_{-0.6}$ & -  & - \\  
ln(E) & 183.0 & 147.6 & 207.1 & 214.2 & 169.3 & -2844.3 & -  & - \\  
\cutinhead{Kepler-13A\,b\tablenotemark{l}} \\
log(H$_2$O) & $-4.1^{+1.5}_{-0.5}$ & $-3.9^{+0.8}_{-0.5}$ & $-2.7^{+0.5}_{-0.9}$ & $-3.2^{+0.8}_{-0.7}$ & -  & - & -  & - \\  
log(CH$_4$) & $< -5.7$ & $< -5.6$ & $< -4.8$ & $< -5.5$ & -  & - & -  & - \\  
log(CO) & $< -4.4$ & $< -5.2$ & $< -4.2$ & $< -4.7$ & -  & - & -  & - \\  
log(CO$_2$) & $< -4.6$ & $< -6.7$ & $< -4.3$ & $< -6.4$ & -  & - & -  & - \\  
log(TiO) & -  & -  & $-7.6^{+1.8}_{-2.3}$ & $< -6.5$ & -  & - & -  & - \\  
log(VO) & -  & -  & $< -7.2$ & $< -7.4$ & -  & - & -  & - \\  
log(FeH) & -  & -  & $< -7.2$ & $< -7.4$ & -  & - & -  & - \\  
log(e-) & -  & -  & $< -5.1$ & $< -5.4$ & -  & - & -  & - \\  
log(Z) & $-0.5^{+0.8}_{-0.8}$ & $-0.6^{+0.7}_{-0.7}$ & $0.3^{+0.6}_{-0.6}$ & $-0.1^{+0.7}_{-0.7}$ & $-0.2^{+1.1}_{-0.5}$ & $-0.1^{+0.8}_{-0.5}$ & -  & - \\  
C/O & $0.3^{+0.3}_{-0.3}$ & $0.2^{+0.3}_{-0.3}$ & $0.2^{+0.3}_{-0.3}$ & $0.2^{+0.3}_{-0.3}$ & $0.5^{+0.2}_{-0.2}$ & $0.4^{+0.2}_{-0.2}$ & -  & - \\  
ln(E) & 114.2 & 126.9 & 113.3 & 126.4 & 115.6 & 128.2 & -  & - \\ 
\cutinhead{TrES-3\,b\tablenotemark{m}} \\
log(H$_2$O) & $-3.6^{+1.1}_{-0.9}$ & $< -7.5$ & $-2.6^{+0.4}_{-0.6}$ & $< -7.7$ & -  & - & -  & - \\  
log(CH$_4$) & $< -5.2$ & $< -7.9$ & $< -4.5$ & $< -8.0$ & -  & - & -  & - \\  
log(CO) & $< -4.8$ & $< -5.1$ & $< -4.0$ & $< -2.8$ & -  & - & -  & - \\  
log(CO$_2$) & $< -4.8$ & $< -6.8$ & $< -4.3$ & $< -6.0$ & -  & - & -  & - \\  
log(TiO) & -  & -  & $< -5.9$ & $< -6.6$ & -  & - & -  & - \\  
log(VO) & -  & -  & $< -7.3$ & $-8.9^{+1.0}_{-1.9}$ & -  & - & -  & - \\  
log(FeH) & -  & -  & $-6.9^{+1.1}_{-2.6}$ & $-7.9^{+0.5}_{-1.3}$ & -  & - & -  & - \\  
log(e-) & -  & -  & $< -4.2$ & $< -9.1$ & -  & - & -  & - \\  
log(Z) & $-0.4^{+0.9}_{-0.9}$ & $-2.3^{+1.4}_{-1.4}$ & $0.4^{+0.5}_{-0.5}$ & $-1.5^{+1.7}_{-1.7}$ & $-0.3^{+0.9}_{-0.5}$ & $1.2^{+0.5}_{-0.6}$ & -  & - \\  
C/O & $0.2^{+0.3}_{-0.3}$ & $0.8^{+0.4}_{-0.4}$ & $0.2^{+0.3}_{-0.3}$ & $0.7^{+0.4}_{-0.4}$ & $0.6^{+0.2}_{-0.3}$ & $1.6^{+0.3}_{-0.3}$ & -  & - \\  
ln(E) & 116.5 & 130.8 & 117.7 & 131.2 & 118.7 & 129.7 & -  & - \\  
\cutinhead{WASP-4\,b\tablenotemark{n}} \\
log(H$_2$O) & $< -5.8$ & $< -6.4$ & $< -4.5$ & $< -5.2$ & -  & - & -  & - \\  
log(CH$_4$) & $< -3.5$ & $< -7.6$ & $< -4.3$ & $< -5.9$ & -  & - & -  & - \\  
log(CO) & $< -4.4$ & $< -3.9$ & $< -4.0$ & $< -3.9$ & -  & - & -  & - \\  
log(CO$_2$) & $< -5.1$ & $< -6.3$ & $< -3.3$ & $-3.7^{+0.9}_{-3.5}$ & -  & - & -  & - \\  
log(TiO) & -  & -  & $-6.2^{+1.3}_{-3.0}$ & $-6.7^{+1.5}_{-3.2}$ & -  & - & -  & - \\  
log(VO) & -  & -  & $< -6.5$ & $< -7.0$ & -  & - & -  & - \\  
log(FeH) & -  & -  & $< -6.3$ & $< -6.8$ & -  & - & -  & - \\  
log(e-) & -  & -  & $-5.7^{+2.6}_{-3.7}$ & $< -5.9$ & -  & - & -  & - \\  
log(Z) & $-1.7^{+1.4}_{-1.4}$ & $-2.0^{+1.3}_{-1.3}$ & $-1.0^{+1.3}_{-1.3}$ & $-0.3^{+1.1}_{-1.1}$ & $0.2^{+1.1}_{-0.8}$ & $1.2^{+0.5}_{-0.6}$ & -  & - \\  
C/O & $0.8^{+0.5}_{-0.5}$ & $0.8^{+0.3}_{-0.3}$ & $0.6^{+0.5}_{-0.5}$ & $0.6^{+0.3}_{-0.3}$ & $1.5^{+0.3}_{-0.4}$ & $1.5^{+0.3}_{-0.4}$ & -  & - \\  
ln(E) & 119.6 & 131.7 & 119.8 & 130.4 & 121.1 & 131.5 & -  & - \\ 
\cutinhead{WASP-12\,b\tablenotemark{o}} \\
log(H$_2$O) & $< -7.0$ & $< -7.7$ & $< -4.0$ & $< -5.7$ & -  & -  & $-2.8^{+0.5}_{-0.6}$ & - \\  
log(CH$_4$) & $< -5.2$ & $< -6.2$ & $-4.4^{+1.9}_{-4.7}$ & $< -4.7$ & -  & -  & $< -5.6$ & - \\  
log(CO) & $< -2.9$ & $< -5.1$ & $< -4.0$ & $< -3.9$ & -  & -  & $< -4.7$ & - \\  
log(CO$_2$) & $< -5.7$ & $-7.1^{+1.2}_{-1.5}$ & $< -3.9$ & $-3.2^{+0.7}_{-1.2}$ & -  & -  & $< -4.8$ & - \\  
log(TiO) & -  & -  & $< -5.8$ & $< -6.2$ & -  & -  & $< -7.1$ & - \\  
log(VO) & -  & -  & $-7.7^{+1.8}_{-2.3}$ & $< -6.2$ & -  & -  & $< -7.1$ & - \\  
log(FeH) & -  & -  & $-7.3^{+1.9}_{-2.6}$ & $-6.2^{+1.3}_{-2.6}$ & -  & -  & $< -5.8$ & - \\  
log(e-) & -  & -  & $-6.1^{+2.4}_{-3.1}$ & $< -5.6$ & -  & -  & $< -5.3$ & - \\  
log(Z) & $-1.3^{+1.5}_{-1.5}$ & $-2.6^{+1.0}_{-1.0}$ & $-1.1^{+1.3}_{-1.3}$ & $0.1^{+0.8}_{-0.8}$ & $-0.0^{+0.9}_{-0.6}$ & $-0.1^{+1.3}_{-0.7}$ & $0.2^{+0.6}_{-0.6}$ & $0.2^{+0.9}_{-0.8}$\\  
C/O & $0.9^{+0.4}_{-0.4}$ & $0.9^{+0.4}_{-0.4}$ & $0.7^{+0.6}_{-0.6}$ & $0.6^{+0.3}_{-0.3}$ & $0.7^{+0.2}_{-0.3}$ & $1.0^{+0.0}_{-0.0}$ & $0.2^{+0.2}_{-0.2}$ & $0.5^{+0.3}_{-0.3}$\\  
ln(E) & 119.7 & 139.3 & 119.2 & 140.1 & 118.5 & 132.6 & 196.0 & 197.3\\  
\cutinhead{WASP-18\,b\tablenotemark{p}} \\
log(H$_2$O) & $< -7.4$ & $< -7.4$ & $-3.3^{+0.4}_{-0.4}$ & $-3.3^{+0.2}_{-0.4}$ & -  & -  & $< -4.2$ & - \\  
log(CH$_4$) & $< -4.8$ & $< -7.3$ & $< -5.4$ & $< -5.5$ & -  & -  & $< -3.8$ & - \\  
log(CO) & $-2.1^{+0.1}_{-0.1}$ & $-2.1^{+0.1}_{-0.1}$ & $-7.9^{+1.8}_{-1.8}$ & $< -4.8$ & -  & -  & $< -4.2$ & - \\  
log(CO$_2$) & $-3.9^{+0.4}_{-5.2}$ & $-3.4^{+0.2}_{-0.2}$ & $< -4.4$ & $< -6.0$ & -  & -  & $< -4.5$ & - \\  
log(TiO) & -  & -  & $< -7.3$ & $< -7.4$ & -  & -  & $< -4.1$ & - \\  
log(VO) & -  & -  & $< -7.2$ & $< -7.0$ & -  & -  & $< -5.2$ & - \\  
log(FeH) & -  & -  & $-8.9^{+1.4}_{-1.7}$ & $< -7.5$ & -  & -  & $< -4.1$ & - \\  
log(e-) & -  & -  & $-3.2^{+0.4}_{-0.4}$ & $-3.2^{+0.3}_{-0.3}$ & -  & -  & $< -3.8$ & - \\  
log(Z) & $1.0^{+0.1}_{-0.1}$ & $1.0^{+0.1}_{-0.1}$ & $-0.2^{+0.4}_{-0.4}$ & $-0.2^{+0.3}_{-0.3}$ & $-0.1^{+0.3}_{-0.4}$ & $0.2^{+0.2}_{-0.2}$ & $-0.9^{+1.3}_{-1.3}$ & $0.4^{+1.1}_{-0.9}$\\  
C/O & $1.0^{+0.0}_{-0.0}$ & $0.9^{+0.0}_{-0.0}$ & $0.2^{+0.2}_{-0.2}$ & $0.2^{+0.2}_{-0.2}$ & $0.9^{+0.0}_{-0.0}$ & $0.9^{+0.0}_{-0.0}$ & $0.6^{+0.5}_{-0.5}$ & $1.1^{+0.6}_{-0.6}$\\  
ln(E) & 197.6 & 215.9 & 207.3 & 237.4 & 207.0 & 232.8 & 209.8 & 209.8\\   
\cutinhead{WASP-19\,b\tablenotemark{q}} \\
log(H$_2$O) & $< -4.8$ & $-5.4^{+0.8}_{-2.0}$ & $-3.2^{+0.7}_{-3.0}$ & $-3.0^{+0.6}_{-1.0}$ & -  & -  & $-2.7^{+0.5}_{-0.7}$ & - \\  
log(CH$_4$) & $< -6.1$ & $< -6.6$ & $< -3.9$ & $< -4.8$ & -  & -  & $< -5.6$ & - \\  
log(CO) & $< -4.2$ & $< -4.3$ & $< -4.2$ & $< -3.8$ & -  & -  & $< -3.8$ & - \\  
log(CO$_2$) & $< -5.2$ & $< -6.5$ & $< -4.2$ & $-6.0^{+1.9}_{-3.5}$ & -  & -  & $< -4.5$ & - \\  
log(TiO) & -  & -  & $< -5.9$ & $< -6.7$ & -  & -  & $< -6.5$ & - \\  
log(VO) & -  & -  & $< -6.4$ & $< -6.5$ & -  & -  & $< -7.2$ & - \\  
log(FeH) & -  & -  & $-7.0^{+1.8}_{-2.7}$ & $< -5.6$ & -  & -  & $< -5.6$ & - \\  
log(e-) & -  & -  & $-6.1^{+2.4}_{-3.2}$ & $< -3.8$ & -  & -  & $< -5.0$ & - \\  
log(Z) & $-1.6^{+1.2}_{-1.2}$ & $-1.8^{+1.0}_{-1.0}$ & $-0.2^{+1.1}_{-1.1}$ & $0.0^{+0.9}_{-0.9}$ & $0.7^{+0.8}_{-0.9}$ & $0.6^{+0.8}_{-0.8}$ & $0.3^{+0.6}_{-0.6}$ & $0.5^{+0.8}_{-0.9}$\\  
C/O & $0.7^{+0.4}_{-0.4}$ & $0.5^{+0.4}_{-0.4}$ & $0.4^{+0.4}_{-0.4}$ & $0.3^{+0.3}_{-0.3}$ & $0.7^{+0.7}_{-0.3}$ & $0.5^{+0.2}_{-0.2}$ & $0.2^{+0.2}_{-0.2}$ & $0.4^{+0.3}_{-0.2}$\\  
ln(E) & 170.9 & 194.9 & 171.4 & 195.0 & 172.0 & 196.3 & 169.4 & 170.4\\ 
\cutinhead{WASP-33\,b\tablenotemark{r}} \\
log(H$_2$O) & $-5.8^{+0.1}_{-0.1}$ & $-5.8^{+0.1}_{-0.1}$ & $-5.4^{+0.2}_{-0.2}$ & $-5.4^{+0.2}_{-0.3}$ & -  & - & -  & - \\  
log(CH$_4$) & $< -7.4$ & $< -7.3$ & $< -6.1$ & $< -6.6$ & -  & - & -  & - \\  
log(CO) & $-2.6^{+0.2}_{-0.2}$ & $-2.5^{+0.3}_{-0.3}$ & $< -5.0$ & $< -5.0$ & -  & - & -  & - \\  
log(CO$_2$) & $< -4.7$ & $-4.4^{+0.2}_{-0.3}$ & $< -5.7$ & $< -6.9$ & -  & - & -  & - \\  
log(TiO) & -  & -  & $-7.4^{+0.5}_{-0.3}$ & $-7.2^{+0.9}_{-0.3}$ & -  & - & -  & - \\  
log(VO) & -  & -  & $< -8.5$ & $-9.2^{+1.1}_{-1.7}$ & -  & - & -  & - \\  
log(FeH) & -  & -  & $< -9.8$ & $< -9.6$ & -  & - & -  & - \\  
log(e-) & -  & -  & $-4.7^{+0.3}_{-0.2}$ & $-4.6^{+0.6}_{-0.2}$ & -  & - & -  & - \\  
log(Z) & $0.5^{+0.2}_{-0.2}$ & $0.6^{+0.3}_{-0.3}$ & $-1.9^{+0.7}_{-0.7}$ & $-2.0^{+0.7}_{-0.7}$ & $-0.9^{+0.1}_{-0.1}$ & $-0.7^{+0.4}_{-0.2}$ & -  & - \\  
C/O & $1.0^{+0.0}_{-0.0}$ & $1.0^{+0.0}_{-0.0}$ & $0.5^{+0.4}_{-0.4}$ & $0.4^{+0.4}_{-0.4}$ & $0.9^{+0.0}_{-0.0}$ & $0.9^{+0.0}_{-0.0}$ & -  & - \\  
ln(E) & 15.7 & 26.0 & 54.8 & 67.4 & 67.9 & 77.0 & -  & - \\ 
\cutinhead{WASP-43\,b\tablenotemark{s}} \\
log(H$_2$O) & $-4.3^{+0.8}_{-0.6}$ & $-4.0^{+0.4}_{-0.4}$ & $< -7.5$ & $-3.7^{+0.5}_{-0.4}$ & -  & -  & $-4.5^{+2.0}_{-4.5}$ & - \\  
log(CH$_4$) & $< -3.9$ & $< -7.5$ & $< -6.6$ & $< -7.2$ & -  & -  & $< -4.8$ & - \\  
log(CO) & $< -3.3$ & $< -3.0$ & $< -4.0$ & $< -3.4$ & -  & -  & $< -4.1$ & - \\  
log(CO$_2$) & $< -3.0$ & $-2.6^{+0.4}_{-0.7}$ & $< -5.5$ & $-2.4^{+0.3}_{-0.4}$ & -  & -  & $< -4.5$ & - \\  
log(TiO) & -  & -  & $< -8.2$ & $< -7.0$ & -  & -  & $< -4.3$ & - \\  
log(VO) & -  & -  & $-9.0^{+1.0}_{-1.7}$ & $< -8.7$ & -  & -  & $< -5.8$ & - \\  
log(FeH) & -  & -  & $-7.8^{+0.7}_{-0.4}$ & $< -8.7$ & -  & -  & $< -5.1$ & - \\  
log(e-) & -  & -  & $< -7.7$ & $< -7.8$ & -  & -  & $< -4.1$ & - \\  
log(Z) & $-0.3^{+0.9}_{-0.9}$ & $0.8^{+0.5}_{-0.5}$ & $-1.9^{+1.3}_{-1.3}$ & $1.0^{+0.4}_{-0.4}$ & $0.5^{+1.2}_{-1.3}$ & $1.8^{+0.2}_{-0.4}$ & $-0.4^{+1.2}_{-1.2}$ & $0.4^{+0.9}_{-0.9}$\\  
C/O & $0.6^{+0.3}_{-0.3}$ & $0.5^{+0.2}_{-0.2}$ & $0.8^{+0.4}_{-0.4}$ & $0.5^{+0.1}_{-0.1}$ & $0.5^{+0.3}_{-0.2}$ & $0.7^{+0.1}_{-0.2}$ & $0.5^{+0.5}_{-0.5}$ & $0.6^{+0.8}_{-0.4}$\\  
ln(E) & 195.8 & 205.5 & 196.1 & 204.5 & 193.9 & 201.9 & 202.4 & 199.0\\  
\cutinhead{WASP-74\,b\tablenotemark{t}} \\
log(H$_2$O) & $-7.3^{+1.4}_{-3.0}$ & $-7.4^{+1.6}_{-2.9}$ & $< -5.1$ & $-2.8^{+0.5}_{-0.6}$ & -  & -  & $-3.9^{+1.3}_{-4.7}$ & - \\  
log(CH$_4$) & $< -6.3$ & $-6.4^{+0.6}_{-0.6}$ & $< -4.2$ & $-2.6^{+0.4}_{-0.7}$ & -  & -  & $< -4.6$ & - \\  
log(CO) & $< -4.9$ & $< -6.3$ & $< -3.9$ & $< -3.5$ & -  & -  & $< -4.1$ & - \\  
log(CO$_2$) & $< -5.5$ & $< -8.3$ & $< -3.9$ & $-6.3^{+2.0}_{-3.5}$ & -  & -  & $< -4.3$ & - \\  
log(TiO) & -  & -  & $< -5.9$ & $< -5.9$ & -  & -  & $< -4.8$ & - \\  
log(VO) & -  & -  & $< -6.4$ & $< -6.8$ & -  & -  & $< -6.1$ & - \\  
log(FeH) & -  & -  & $< -4.9$ & $< -6.2$ & -  & -  & $-6.0^{+1.9}_{-2.9}$ & - \\  
log(e-) & -  & -  & $-3.4^{+0.9}_{-1.8}$ & $< -4.6$ & -  & -  & $< -5.1$ & - \\  
log(Z) & $-2.1^{+1.1}_{-1.1}$ & $-2.7^{+0.8}_{-0.8}$ & $-1.2^{+1.3}_{-1.3}$ & $0.4^{+0.5}_{-0.5}$ & $1.0^{+0.6}_{-1.0}$ & $0.6^{+0.9}_{-1.2}$ & $-0.4^{+1.2}_{-1.2}$ & $0.2^{+1.1}_{-0.9}$\\  
C/O & $0.7^{+0.5}_{-0.5}$ & $0.8^{+0.7}_{-0.7}$ & $0.7^{+0.5}_{-0.5}$ & $1.2^{+0.7}_{-0.7}$ & $0.7^{+0.9}_{-0.4}$ & $1.4^{+0.4}_{-0.3}$ & $0.4^{+0.5}_{-0.5}$ & $0.6^{+0.7}_{-0.3}$\\  
ln(E) & 202.8 & 216.3 & 202.5 & 215.4 & 202.0 & 213.6 & 196.9 & 196.7\\ 
\cutinhead{WASP-76\,b\tablenotemark{u}} \\
log(H$_2$O) & $< -6.9$ & $< -5.4$ & $-2.9^{+0.4}_{-0.6}$ & $-3.1^{+0.4}_{-0.6}$ & -  & -  & $-2.9^{+0.4}_{-0.6}$ & - \\  
log(CH$_4$) & $-3.4^{+0.3}_{-0.3}$ & $-2.9^{+0.4}_{-0.2}$ & $< -4.5$ & $< -6.3$ & -  & -  & $< -5.5$ & - \\  
log(CO) & $-5.6^{+2.3}_{-2.5}$ & $< -4.6$ & $< -4.0$ & $< -3.8$ & -  & -  & $< -6.6$ & - \\  
log(CO$_2$) & $< -6.4$ & $< -6.9$ & $< -4.4$ & $< -4.9$ & -  & -  & $< -5.1$ & - \\  
log(TiO) & -  & -  & $-4.8^{+0.5}_{-1.0}$ & $-4.7^{+0.5}_{-0.7}$ & -  & -  & $-8.0^{+1.7}_{-2.0}$ & - \\  
log(VO) & -  & -  & $< -7.7$ & $< -8.1$ & -  & -  & $< -7.1$ & - \\  
log(FeH) & -  & -  & $< -6.6$ & $< -5.4$ & -  & -  & $< -7.3$ & - \\  
log(e-) & -  & -  & $< -3.7$ & $< -4.9$ & -  & -  & $< -5.2$ & - \\  
log(Z) & $-1.7^{+1.4}_{-1.4}$ & $-2.1^{+1.2}_{-1.2}$ & $0.2^{+0.5}_{-0.5}$ & $0.1^{+0.6}_{-0.6}$ & $0.8^{+0.8}_{-1.1}$ & $-0.1^{+0.6}_{-0.5}$ & $0.1^{+0.5}_{-0.5}$ & $1.7^{+0.2}_{-0.3}$\\  
C/O & $1.4^{+0.6}_{-0.6}$ & $1.7^{+0.5}_{-0.5}$ & $0.3^{+0.3}_{-0.3}$ & $0.3^{+0.3}_{-0.3}$ & $0.7^{+0.4}_{-0.4}$ & $1.2^{+0.4}_{-0.1}$ & $0.1^{+0.2}_{-0.2}$ & $0.5^{+0.2}_{-0.3}$\\  
ln(E) & 183.9 & 201.2 & 189.5 & 199.7 & 188.5 & 203.0 & 197.7 & 204.6\\ 
\cutinhead{WASP-77\,A\,b\tablenotemark{v}} \\
log(H$_2$O) & $-4.1^{+0.5}_{-0.4}$ & $-4.1^{+1.7}_{-0.6}$ & $-3.9^{+0.9}_{-0.5}$ & $-3.5^{+0.6}_{-0.6}$ & -  & - & -  & - \\  
log(CH$_4$) & $< -6.4$ & $-4.7^{+1.6}_{-1.0}$ & $< -5.9$ & $-4.2^{+0.7}_{-0.8}$ & -  & - & -  & - \\  
log(CO) & $< -4.3$ & $< -3.8$ & $< -3.9$ & $< -2.8$ & -  & - & -  & - \\  
log(CO$_2$) & $< -5.0$ & $-4.3^{+1.3}_{-1.9}$ & $< -3.5$ & $-3.8^{+0.7}_{-1.2}$ & -  & - & -  & - \\  
log(TiO) & -  & -  & $-6.0^{+0.9}_{-0.4}$ & $-5.3^{+0.7}_{-0.8}$ & -  & - & -  & - \\  
log(VO) & -  & -  & $< -9.0$ & $< -8.7$ & -  & - & -  & - \\  
log(FeH) & -  & -  & $< -8.3$ & $< -8.1$ & -  & - & -  & - \\  
log(e-) & -  & -  & $< -7.9$ & $< -7.6$ & -  & - & -  & - \\  
log(Z) & $-0.8^{+0.5}_{-0.5}$ & $-0.2^{+1.0}_{-1.0}$ & $-0.3^{+0.8}_{-0.8}$ & $0.2^{+0.6}_{-0.6}$ & $1.3^{+0.4}_{-0.5}$ & $-0.6^{+0.3}_{-0.2}$ & -  & - \\  
C/O & $0.3^{+0.3}_{-0.3}$ & $0.4^{+0.2}_{-0.2}$ & $0.4^{+0.3}_{-0.3}$ & $0.5^{+0.2}_{-0.2}$ & $0.5^{+0.2}_{-0.3}$ & $0.8^{+0.1}_{-0.1}$ & -  & - \\  
ln(E) & 197.0 & 206.0 & 202.5 & 212.5 & 200.6 & 209.0 & -  & - \\ 
\cutinhead{WASP-79\,b\tablenotemark{w}} \\
log(H$_2$O) & $-4.4^{+0.9}_{-0.7}$ & $-3.4^{+0.7}_{-0.8}$ & $< -4.8$ & $< -4.2$ & -  & -  & $-2.6^{+0.4}_{-0.5}$ & - \\  
log(CH$_4$) & $-5.6^{+2.2}_{-3.9}$ & $< -3.0$ & $< -5.0$ & $< -6.0$ & -  & -  & $< -5.9$ & - \\  
log(CO) & $< -3.1$ & $< -3.8$ & $< -3.9$ & $< -4.2$ & -  & -  & $< -4.6$ & - \\  
log(CO$_2$) & $< -3.4$ & $< -5.7$ & $< -4.2$ & $< -7.0$ & -  & -  & $< -4.9$ & - \\  
log(TiO) & -  & -  & $< -6.7$ & $< -6.8$ & -  & -  & $< -5.4$ & - \\  
log(VO) & -  & -  & $-5.1^{+0.6}_{-0.8}$ & $-5.3^{+0.6}_{-0.6}$ & -  & -  & $< -7.6$ & - \\  
log(FeH) & -  & -  & $-5.3^{+0.7}_{-1.2}$ & $-5.7^{+0.8}_{-1.6}$ & -  & -  & $< -5.6$ & - \\  
log(e-) & -  & -  & $< -5.6$ & $< -5.4$ & -  & -  & $-5.9^{+2.4}_{-2.8}$ & - \\  
log(Z) & $-0.4^{+0.9}_{-0.9}$ & $-0.1^{+0.7}_{-0.7}$ & $-1.1^{+1.0}_{-1.0}$ & $-1.5^{+1.0}_{-1.0}$ & $1.1^{+0.6}_{-0.9}$ & $0.2^{+0.8}_{-0.7}$ & $0.5^{+0.4}_{-0.4}$ & $0.5^{+0.9}_{-0.9}$\\  
C/O & $0.8^{+0.6}_{-0.6}$ & $0.8^{+0.7}_{-0.7}$ & $0.6^{+0.4}_{-0.4}$ & $0.5^{+0.5}_{-0.5}$ & $0.4^{+0.2}_{-0.2}$ & $0.3^{+0.2}_{-0.2}$ & $0.1^{+0.2}_{-0.2}$ & $0.3^{+0.3}_{-0.2}$\\  
ln(E) & 136.9 & 151.9 & 141.9 & 156.4 & 136.0 & 151.0 & 191.7 & 184.6\\ 
\cutinhead{WASP-103\,b\tablenotemark{x}} \\
log(H$_2$O) & $< -7.4$ & $< -6.9$ & $< -5.4$ & $< -5.4$ & -  & -  & $< -3.7$ & - \\  
log(CH$_4$) & $< -7.1$ & $< -5.8$ & $< -4.1$ & $-6.4^{+2.5}_{-3.1}$ & -  & -  & $< -5.4$ & - \\  
log(CO) & $< -5.4$ & $< -3.8$ & $< -3.8$ & $< -4.1$ & -  & -  & $< -4.7$ & - \\  
log(CO$_2$) & $< -4.5$ & $-2.3^{+0.2}_{-0.5}$ & $< -4.2$ & $-3.6^{+1.0}_{-1.6}$ & -  & -  & $< -4.4$ & - \\  
log(TiO) & -  & -  & $< -6.2$ & $< -6.8$ & -  & -  & $-4.4^{+1.3}_{-3.9}$ & - \\  
log(VO) & -  & -  & $-4.8^{+0.5}_{-0.6}$ & $-5.8^{+0.9}_{-2.5}$ & -  & -  & $-4.2^{+1.0}_{-1.1}$ & - \\  
log(FeH) & -  & -  & $-6.1^{+1.3}_{-2.7}$ & $-7.6^{+2.0}_{-2.4}$ & -  & -  & $< -4.9$ & - \\  
log(e-) & -  & -  & $< -3.5$ & $-3.4^{+0.9}_{-4.8}$ & -  & -  & $-6.0^{+2.6}_{-3.4}$ & - \\  
log(Z) & $-2.1^{+1.1}_{-1.1}$ & $0.9^{+0.5}_{-0.5}$ & $-1.0^{+0.9}_{-0.9}$ & $-0.3^{+1.1}_{-1.1}$ & $-0.4^{+0.7}_{-0.4}$ & $-0.3^{+0.7}_{-0.4}$ & $-0.3^{+0.9}_{-0.9}$ & $-0.4^{+0.7}_{-0.5}$\\  
C/O & $0.7^{+0.4}_{-0.4}$ & $0.5^{+0.1}_{-0.1}$ & $0.6^{+0.5}_{-0.5}$ & $0.6^{+0.3}_{-0.3}$ & $1.0^{+0.0}_{-0.0}$ & $1.0^{+0.0}_{-0.0}$ & $0.3^{+0.4}_{-0.4}$ & $0.6^{+0.4}_{-0.4}$\\  
ln(E) & 187.1 & 199.2 & 189.8 & 201.6 & 184.7 & 198.0 & 173.4 & 166.3\\ 
\cutinhead{WASP-121\,b\tablenotemark{y}} \\
log(H$_2$O) & $< -8.3$ & $< -8.1$ & $-4.2^{+0.4}_{-1.4}$ & $< -4.6$ & -  & -  & $-2.7^{+0.5}_{-0.7}$ & - \\  
log(CH$_4$) & $< -7.2$ & $< -6.6$ & $< -5.4$ & $< -6.8$ & -  & -  & $< -6.0$ & - \\  
log(CO) & $< -5.0$ & $-6.2^{+2.3}_{-3.5}$ & $< -4.2$ & $< -3.6$ & -  & -  & $< -4.5$ & - \\  
log(CO$_2$) & $< -6.2$ & $< -6.7$ & $< -4.5$ & $< -6.4$ & -  & -  & $< -4.8$ & - \\  
log(TiO) & -  & -  & $< -7.7$ & $-4.5^{+0.3}_{-0.5}$ & -  & -  & $< -6.3$ & - \\  
log(VO) & -  & -  & $-6.9^{+0.5}_{-2.6}$ & $-4.8^{+0.3}_{-0.4}$ & -  & -  & $-7.9^{+1.5}_{-2.4}$ & - \\  
log(FeH) & -  & -  & $< -8.5$ & $< -7.4$ & -  & -  & $< -7.1$ & - \\  
log(e-) & -  & -  & $-2.8^{+0.5}_{-0.7}$ & $-4.2^{+0.7}_{-0.7}$ & -  & -  & $-5.2^{+1.9}_{-2.6}$ & - \\  
log(Z) & $-2.4^{+1.0}_{-1.0}$ & $-2.0^{+1.2}_{-1.2}$ & $-0.9^{+0.8}_{-0.8}$ & $-0.9^{+0.6}_{-0.6}$ & $0.7^{+0.9}_{-0.7}$ & $1.7^{+0.2}_{-0.3}$ & $0.3^{+0.5}_{-0.5}$ & $0.5^{+1.0}_{-0.9}$\\  
C/O & $0.8^{+0.4}_{-0.4}$ & $0.9^{+0.4}_{-0.4}$ & $0.5^{+0.4}_{-0.4}$ & $0.4^{+0.4}_{-0.4}$ & $1.6^{+0.3}_{-0.2}$ & $1.3^{+0.1}_{-0.1}$ & $0.2^{+0.2}_{-0.2}$ & $0.4^{+0.4}_{-0.2}$\\  
ln(E) & 198.1 & 306.7 & 201.4 & 313.8 & 183.0 & 289.5 & 193.4 & 189.6\\ 
\enddata
\tablenotetext{a}{CoRoT-1\,b: For comparison, the HST blackbody fit obtains ln(E) = 112.5, the HST+Spitzer blackbody fit obtains ln(E) = 125.5 and the featureless transit fit obtains ln(E) = 96.4.}
\tablenotetext{b}{HAT-P-2\,b: For comparison, the HST blackbody fit obtains ln(E) = 215.5 and the HST+Spitzer blackbody fit obtains ln(E) = 243.3.}
\tablenotetext{c}{HAT-P-7\,b: For comparison, the HST blackbody fit obtains ln(E) = 207.2, the HST+Spitzer blackbody fit obtains ln(E) = 225.3 and the featureless transit fit obtains ln(E) = 156.7.}
\tablenotetext{d}{HAT-P-32\,b: For comparison, the HST blackbody fit obtains ln(E) = 174.1, the HST+Spitzer blackbody fit obtains ln(E) = 188.8 and the featureless transit fit obtains ln(E) = 157.2.}
\tablenotetext{e}{HAT-P-41\,b: For comparison, the HST blackbody fit obtains ln(E) = 185.1, the HST+Spitzer blackbody fit obtains ln(E) = 194.7 and the featureless transit fit obtains ln(E) = 182.5.}
\tablenotetext{f}{HAT-P-70\,b: For comparison, the HST blackbody fit obtains ln(E) = 191.6.}
\tablenotetext{g}{HD\,189733\,b: For comparison, the HST blackbody fit obtains ln(E) = 146.4, the HST+Spitzer blackbody fit obtains ln(E) = 137.0 and the featureless transit fit obtains ln(E) = 184.6.}
\tablenotetext{h}{HD\,209458\,b: For comparison, the HST blackbody fit obtains ln(E) = 9.3, the HST+Spitzer blackbody fit obtains ln(E) = -400.0 and the featureless transit fit obtains ln(E) = 190.0.}
\tablenotetext{i}{KELT-1\,b: For comparison, the HST blackbody fit obtains ln(E) = 177.7, the HST+Spitzer blackbody fit obtains ln(E) = 192.2 and the featureless transit fit obtains ln(E) = 190.5.}
\tablenotetext{j}{KELT-7\,b: For comparison, the HST blackbody fit obtains ln(E) = 182.7, the HST+Spitzer blackbody fit obtains ln(E) = 144.1 and the featureless transit fit obtains ln(E) = 185.7.}
\tablenotetext{k}{KELT-9\,b: For comparison, the HST blackbody fit obtains ln(E) = 8.9 and the HST+Spitzer blackbody fit obtains ln(E) = 13.6.}
\tablenotetext{l}{Kepler-13\,A\,b: For comparison, the HST blackbody fit obtains ln(E) = 101.3 and the HST+Spitzer blackbody fit obtains ln(E) = 115.0.}
\tablenotetext{m}{TrES-3\,b: For comparison, the HST blackbody fit obtains ln(E) = 106.9 and the HST+Spitzer blackbody fit obtains ln(E) = 118.7.}
\tablenotetext{n}{WASP-4\,b: For comparison, the HST black-body fit obtains ln(E) = 119.0 and the HST+Spitzer black-body fit obtains ln(E) = 124.3.}
\tablenotetext{o}{WASP-12\,b: For comparison, the HST black-body fit obtains ln(E) = 119.1, the HST+Spitzer black-body fit obtains ln(E) = 117.5 and the featureless transit fit obtains ln(E) = 168.9.}
\tablenotetext{p}{WASP-18\,b: For comparison, the HST blackbody fit obtains ln(E) = 146.9, the HST+Spitzer blackbody fit obtains ln(E) = 172.7 and the featureless transit fit obtains ln(E) = 209.7.}
\tablenotetext{q}{WASP-19\,b: For comparison, the HST blackbody fit obtains ln(E) = 171.1, the HST+Spitzer blackbody fit obtains ln(E) = 189.2 and the featureless transit fit obtains ln(E) = 164.8.}
\tablenotetext{r}{WASP-33\,b: For comparison, the HST blackbody fit obtains ln(E) = -178.0 and the HST+Spitzer blackbody fit obtains ln(E) = -169.4.}
\tablenotetext{s}{WASP-43\,b: For comparison, the HST blackbody fit obtains ln(E) = 183.1, the HST+Spitzer blackbody fit obtains ln(E) = 163.9 and the featureless transit fit obtains ln(E) = 196.8.}
\tablenotetext{t}{WASP-74\,b: For comparison, the HST blackbody fit obtains ln(E) = 200.6, the HST+Spitzer blackbody fit obtains ln(E) = 192.5 and the featureless transit fit obtains ln(E) = 195.7.}
\tablenotetext{u}{WASP-76\,b: For comparison, the HST blackbody fit obtains ln(E) = 180.8, the HST+Spitzer blackbody fit obtains ln(E) = 195.0 and the featureless transit fit obtains ln(E) = 175.0.}
\tablenotetext{v}{WASP-77\,A\,b: For comparison, the HST blackbody fit obtains ln(E) = 157.0 and the HST+Spitzer blackbody fit obtains ln(E) = 65.7.}
\tablenotetext{w}{WASP-79\,b: For comparison, the HST blackbody fit obtains ln(E) = 124.2, the HST+Spitzer blackbody fit obtains ln(E) = 137.9 and the featureless transit fit obtains ln(E) = 172.9.}
\tablenotetext{x}{WASP-103\,b: For comparison, the HST blackbody fit obtains ln(E) = 174.0, the HST+Spitzer blackbody fit obtains ln(E) = 185.3 and the featureless transit fit obtains ln(E) = 161.8.}
\tablenotetext{y}{WASP-121\,b: For comparison, the HST blackbody fit obtains ln(E) = 91.1, the HST+Spitzer blackbody fit obtains ln(E) = 140.0 and the featureless transit fit obtains ln(E) = 173.2.}
\tablecomments{Table D1 is published in its entirety in the electronic 
edition of the {\it Astrophysical Journal}.}
\end{deluxetable}

%% file: sec_CO1.tex
The first transit of CoRoT-1\,b was observed in 2008 \citep{2008_barge_corot1b}, unveiling a 1.5 R$_J$ inflated planet. Later observations from the ground indicated a significant emission signal corresponding to a temperature of approximately 2300 K and potentially a poor heat redistribution between the day- and night-sides \citep{2009_alonzo_corot1b_eclipse,2009_Gillon_Corot1b}. The planet transit from HST WFC3 \cite[PN: 12181, PI:][]{deming_12181_prop} was previously studied but no evidence for molecular absorption was found due to the high uncertainties on the data \citep{2014_ranjan_5hj_staring}. The same proposal included eclipse observations, but no analysis of these has since been published. However, a follow-up study in 2010 analyzed Spitzer eclipse observations \citep{Deming_2010_co1b} and found potential evidence for high-altitude absorbers or an isothermal region in the atmosphere. We analyse the transmission and emission HST visits using our standardized pipeline and performed the retrieval analyses (see Materials and Methods section). For the eclipse, we considered the Spitzer observations reduced in \cite{Deming_2010_co1b}, which are taken without modifications.

On the day-side, the HST and HST+Spitzer are consistent with a thermal inversion. The posterior distributions indicate spectral signatures for H$_2$O, VO and H$^-$ opacities, which is consistent with the findings in \cite{Deming_2010_co1b}. The retrieved temperatures of about 2000 K could explain the presence of the detected absorbers. For the free runs including the full list of opacities, the chemistry is consistent with a solar metallicity and a sub-solar C/O ratio. We note that, since no C-bearing species are detected, the C/O ratio from the free runs is likely biased. When equilibrium chemistry is used, a thermal inversion is also recovered, with different thermal profiles depending whether Spitzer is included or not. The equilibrium retrievals tend to converge towards an atmosphere with solar or enriched metallicity and a C/O ratio of about 1. Our analysis provides decisive evidence in favour of the free model that includes H$_2$O and the optical absorbers VO and H$^-$.

For the transit observation, the spectrum shows a downward slope towards the red end of the spectrum.  In the free run, it is best fit with a cloudy atmosphere and a potential contribution from VO. This result is consistent with the detected absorption of VO on the day-side and the recovered temperature of around 1800 K. In this free scenario, the atmosphere might be consistent with a solar metallicity and C/O ratio, but we note that the uncertainties on those derived parameters are large, also allowing for sub-solar to super-solar C/O ratios. The equilibrium run leads to similar estimates of the metallicity and the C/O ratio, with better constraints, but the recovered temperature is un-physical, with a mean higher than at the day-side.

%% file: sec_H2.tex
HAT-P-2\,b was discovered in 2007 by \cite{Bakos_2007_hp2}. It is a massive hot Jupiter (9.1 M$_{jup}$) that orbits its host star in an highly eccentric orbit (e = 0.52) in about 5.6 days. Due to its large density, the planet is believed to require the presence of a large core. This large mass, combined with the highly eccentric orbit, raises many questions regarding the physics of this planet and its formation. For instance, along the entire orbit, the planet's equilibrium temperature varies from 1240K to 2150K \cite{Bakos_2007_hp2}. Studying the Rossiter-McLaughlin effect,\cite{Winn_2007_hp2, Loeillet_2008_hp2} found that stellar spin axis and orbital axis of the planet should be aligned, thus implying that the planet did not evolved through scattering or Kozai migration.

While being a very interesting planet, the atmosphere of HAT-P-2\,b was not studied with many instruments. A phase-curve observation with Spitzer at 3.6\,$\mu$m, 4.5\,$\mu$m, 5.6\,$\mu$m and 8\,$\mu$m was presented in \cite{Lewis_2013_hp2_spz}, highlighting a very complex atmosphere due to the particular orbital configuration of this planet. The study also suggested the planet might experience a temporary day side thermal inversion near periapse. In a follow-up work, \cite{Lewis_2014_hp2} performed a complementary analysis with GCM models to evaluate the impact of the eccentricity on the chemistry and the thermal structure of this planet, highlighting that dis-equilibrium processes on this planet might be important.  

Recently, a partial phase-curve was acquired with HST using the Grism G141 \cite[PN: 16194, PI:][]{Desert_2020_hp2Prop}. We fitted the publicly available data using our standardised \emph{Iraclis} pipeline. This data was complemented with the Spitzer observations, taken as-is from \cite{Lewis_2013_hp2_spz}.

We performed our the retrieval of this planet with \emph{Alfnoor}. All our retrieval models provide the same interpretation of the data and the HAT-P-2\,b eclipse spectrum does not show clear signatures for any molecule. The atmosphere is best interpreted by an isothermal profile, with eventually some marginal water feature. However, comparisons of the Bayesian evidence indicate that the detection is tentative at best. The spectrum is well explained by a simple blackbody fit and we are not able to confirm the thermally inverted profile predicted by \cite{Lewis_2013_hp2_spz}. Due to the fact that the HST and Spitzer observations were not carried simultaneously, we note that the atmospheric conditions might have changed between the different observations, in particular considering the additional variability introduced by the highly eccentric orbit.

%% file: sec_H7.tex
HAT-P-7\,b is an inflated hot Jupiter of 1.4 R$_J$ \citep{pal_hatp7}, which was first studied in emission during the commissioning program of Kepler when the satellite detected the eclipse as part of an optical phase curve \citep{borucki_hatp7}. These measurements indicated that HAT-P-7\,b could have a day-side temperature of around 2650 K which confirmed predictions from \citep{pal_hatp7, Fortney_2008}. This optical eclipse measurement was combined with Spitzer photometry over 3.5 - 8 $\mu m$ to infer the presence of a thermal inversion \citep{Christiansen_hatp7}, suggested by the high flux ratio in the 4.5 $\mu m$ channel of Spitzer compared to the 3.6\,$\mu m$ channel. In their paper, chemical equilibrium models associated these emission features with CO, H$_2$O and CH$_4$. A thermal inversion was also reported to provide the best fit to this data by the atmospheric models of \cite{spiegel_hatp7,madhu_hatp7} but all three studies noted that models without a thermal inversion could also well explain the data though only with an extremely high abundance of CH$_4$. Further Kepler phase curves identified an offset in the day-side hot-spot \citep{esteves_hatp7,esteves} as well as changes in its location \citep{armstrong_hatp7}, highlighting the complex dynamics of hot Jupiter atmospheres. However, while Spitzer phase curves at 3.5 \& 4.5 $\mu m$ were also best fitted with a thermal inversion on the day-side and relatively inefficient day–night re-circulation, \cite{wong_hatp7} did not find evidence of a hot-spot offset. Two eclipses were then obtained using HST WFC3 G141 which, when combined with previous observations, were found to be best fit with a thermal inversion due to optical absorbers but at a low significance when compared to a baseline black-body fit \citep{mansfield_hatp7}.

We fitted the two scanning mode eclipse observations of HAT-P-7\,b \cite[PN: 14792, PI:][]{bean_14792_prop}. We note that three staring mode eclipse observations were also taken \cite[PN: 12181, PI:][]{deming_12181_prop} but we ignore these given the higher precision of the scanning mode observations. Additionally, a staring mode transmission spectrum had been taken \cite[PN: 12181, PI:][]{deming_12181_prop} but there was no post-egress orbit and, due to significant systematics, our fit of the data presents large uncertainties. Models of HAT-P-7\,b suggest the terminator region should have patchy clouds with water being well mixed throughout the atmosphere \citep{helling}.

This planet, which was extensively observed with Spitzer, was not included in the population study from G20. We therefore considered the observations from \cite{Christiansen_hatp7}, which cover the wavelength range from 3.6 to 8 $\mu$m. In their paper, on top of the standard MCMC technique, they used the 'rosary bead' residual permutation method \citep{Winn_2008_rotary} to quantify potential remaining systematic errors. Their results highlighted differences in the 8\,$\mu$m point, so we use the 8\,$\mu$m channel from the rosary technique, as suggested in \cite{Christiansen_hatp7}. 

We performed our standardised retrieval study of this planet. Analysing the full HST-only run does not clearly favour a molecule and leads to an isothermal atmosphere. When Spitzer is included, however, additional constraints can be extracted from the 4 Spitzer photometric channels. The retrievals now include some evidence for sub-solar abundance of H$_2$O, high abundance of CO$_2$ and the presence of FeH. The associated temperature profile to best fit the HST+Spitzer data-set contains a localised thermal inversion. While the recovered temperature reaches 2500 K, we do not find evidence for H$^-$ opacities. Comparing the Bayesian evidence of the free models, it is not possible to validate the detection of molecular species from the HST only data-set, thus demonstrating that the HST spectrum is in this case, uninformative and consistent with a black-body. It is only when Spitzer is included that we find strong evidence in favour of the models with molecular opacities. The derived metallicity for the free models is only constrained when HST+Spitzer data are considered, favouring a super-solar metallicity. Due to the detection of large abundances of CO$_2$ in the model that includes Spitzer data, the derived C/O ratio takes a value close to 0.5. When testing the equilibrium retrievals, the thermal profile is decreasing with altitude if Spitzer is included or possesses a localised thermal inversion if only HST is considered. In both runs, the additional constraints of this chemical model do not allow to clarify the metallicity of this atmosphere, but the associated C/O ratio is restricted to high values. Looking at the spectra, the solution obtained by the equilibrium chemistry retrievals obtain a much worse fit of the observed Spitzer data, which is also confirmed by the lower Bayesian evidence.

At the terminator, we do not recover the presence of any molecular species, the spectrum being consistent with flat. The planet could therefore have high-altitude clouds at the terminator.

%% file: sec_H32.tex
The planet HAT-P-32\,b was first reported in 2011 \citep{Hartman_HATP32b_em} and presents a particularly large radius: 1.98 R$_J$ \citep{Wang_2019_hp}. The planet eclipse was then observed with HST WFC3 \citep{Nikolov_HATP32b_spectrum_em} and Spitzer \citep{Zhao_2014_hp32}. Both analyses suggested that thermal inversions could be present in this planet. The terminator region was also observed with HST WFC3 and it is consistent with a significant water feature at 1.4\,$\mu$m \citep{damiano_h32, tsiaras_30planets}. This data was analysed together with HST STIS and Spitzer IRAC data, which showed a thick cloud layer and a super-solar metallicity \citep{alam_h32}.

Since the transmission spectrum \cite[PN: 14260, PI:][]{Deming_14260_prop} was part of the T18 study and was already reduced with \emph{Iraclis}, we took the spectrum ``as is'', directly from this paper. For the eclipse \cite[PN: 14767, PI:][]{sing_14767_prop}, we re-analysed the raw images using \emph{Iraclis} following our standardised reduction technique. We include the Spitzer data from \cite{Zhao_2014_hp32} in our spectral retrieval analysis.

For the eclipse, we show the fitted spectra, the temperature profiles and the posteriors in Figure Set D1. In the full runs, the recovered temperature profile is slightly inverted and is similar between the HST and HST+Spitzer runs. In addition, both runs are consistent with H$^-$, which could be causing the observed thermal inversion. We note, however, that the recovered temperatures for this atmosphere are below the predictions for H$_2$ dissociation in \cite{Parmentier_2018_w121photodiss}. Such results are surprising but given the small differences in Bayesian evidence between with the reduced model, we believe this detection to be marginal. The addition of the Spitzer photometric points lead to some hints for VO and CH$_4$, but these detections also remain weak, with large tails in the posterior distributions. When comparing the full runs to the reduced runs, the inclusion of the Spitzer points favours the models with optical absorbers, but this is not the case when only HST is considered. For the free runs, we find that the full runs prefer a sub-solar metallicity case, which contrast with the results from equilibrium chemistry retrievals. In terms of C/O ratio, solar values are also allowed with large uncertainties for the full runs, while equilibrium chemistry retrievals require a super-solar C/O ratio that is unlikely from a planetary formation perspective. The equilibrium chemistry runs also feature a decreasing temperature structure, as opposed to the full free runs. In any case, looking at the Bayesian evidence obtained by the reference black-body fits, we conclude on the likely presence of molecular absorption in this planet.

For the transmission, which is shown in Figure Set D2, our analysis recovers a large abundance of water vapour and high-altitude opaque clouds, consistent with the previous analyses of the planet. We also find marginal evidence for FeH absorption, which is surprising as it is not detected on the day-side. This might be linked to circulation processes that prevent the molecule being seen on the day-side or it could come from an unfortunate fitting of the scattered datapoints between 1.1\,$\mu$m and 1.3\,$\mu$m. Constraints on the other near-optical absorbers are however relatively strong, with abundances restricted to less than 10$^{-5}$ for all of them. In terms of elemental ratios, both free and equilibrium chemistry are consistent with a large range of metallicities (from sub to super-solar). However, the C/O ratio for the free run is sub-solar, which is inconsistent with the equilibrium chemistry scenario.

%% file: sec_H41.tex
In August 2012 the HATnet survey reported the identification of three new inflated, transiting hot Jupiters orbiting bright F-type hosts. One of those systems hosts HAT-P-41\,b, which is in fact part of a binary system, with a K-dwarf companion \citep{Hartman_2012,Evans_2016}. The HST WFC3 transmission spectrum was studied in T18, finding the presence of water vapor and high-altitude clouds. In 2020, additional transmission data was obtained using the G280 grism (200-400 nm) onboard HST/WFC3 and analysed in conjunction with the Spitzer photometry (3.6\,$\mu$m and 4.5\,$\mu$m). Their best fit of the transmission spectrum, obtained using a grid of self-consistent models, confirmed the presence of clouds and water and contained evidence for molecular absorption from VO/TiO and CO$_2$ \citep{Wakeford_2020}.

As part of our work, we obtained the HST transmission spectrum from T18 and reduced the G141 observation of the eclipse from the raw data using our \emph{Iraclis} pipeline \cite[PN: 14767, PI:][]{sing_14767_prop}. The Spitzer data was obtained from the G20 population study of Spitzer eclipses. We then performed our retrieval analysis of the transit and eclipse spectra with \emph{TauREx3} via the \emph{Alfnoor} pipeline.

For the day-side, our full scenarios agreed on a thermal inversion in both HST and HST+Spitzer cases. The solutions display evidence for optical absorbers such as VO or FeH, but there remain large tails indicating the degeneracies between those two molecules. The full runs are, however, not statistically significant as they lead to similar Bayesian evidence to the reduced cases. In the reduced cases, the retrievals do not find evidence for any molecule and seem consistent with emission from continuous CIA and Rayleigh absorption with a decreasing thermal profile. The metallicity for the free cases is solar to sub-solar and as expected from the poor features in this spectrum, the C/O ratio of the planet remains unconstrained. The equilibrium chemistry runs are consistent with a decreasing to isothermal thermal structure, more similar to the free reduced case. For this planet, equilibrium chemistry provides a very good fit of the HST+Spitzer data-set, with a ln(E) comparable to the analogous free runs. Assuming equilibrium chemistry, the atmosphere is found to have a solar-like metallicity but a relatively high C/O ratio.

When analysing the terminator of the planet, we confirm the results from T18, finding a strong H$_2$O feature and evidence for high-altitude clouds. In the WFC3 data, VO/TiO or CO2 are not detected, suggesting that the additional constraints found in \cite{Wakeford_2020} come from the addition of the G280 grism and the Spitzer data. The metallicity of the planet is consistent with solar abundances, while the C/O ratio is difficult to constrain, ranging from sub-solar in the free run to solar with large uncertainties in the equilibrium case.

%% file: sec_H70.tex
HAT-P-70\,b is a very large ultra-hot Jupiter (1.87 R$_{J}$) orbiting a A-type star that was recently detected by TESS and the HATNet program. While radial velocity measurements of the star were performed in the discovery paper \cite{Zhou_2019_hp70}, only a upper limit on the mass of the planet (M$_p$ $<$ 6.78 M$_J$) could be obtained.

The planet was observed again in 2020 at high-resolution with the HARPS-N spectrograph \cite{Belllo_2021_hp70}. These observations detected a number of atomic and ionic species (Ca II, Cr I, Cr II, Fe I, Fe II, H I, Mg I, Na I and V I) and constrained further the orbital configuration of the system. The planet is in a highly misaligned orbit.

In July 2021, the eclipse of HAT-P-70\,b was observed with HST as part of the proposal \cite[PN: 16307, PI:][]{Fu_16307_prop}. We analyzed the raw images from this proposal using our \emph{Iraclis} pipeline. Due to the retirement of Spitzer, this planet is the only one in our sample that was not observed with this telescope. We therefore include the results from the HST only full case in the population study. Additionally, as of today, the planet's mass is still not known precisely. We tested different values, up to 6.78 M$_J$, to verify the impact of those uncertainties on our retrievals. Due to the geometry in eclipse, we did not find major differences in our test and we therefore assume a mass of 1M$_J$ for this study. 

From our retrievals, the full model recovers an inverted thermal profile with H$^-$ emission. This detection not decisive as the reduced run is not consistent with this picture and obtain a $\Delta$ln(E) $=$ 3.2. For the full model, the metallicity is sub-solar and the C/O ratio is consistent with solar. While the equilibrium chemistry run also presents a thermal inversion, this retrieval is consistent with a C/O ratio of 1.

%% file: sec_HD189.tex
HD\,189733\,b is one of the most studied exoplanets and was the first planet to exhibit evidence for molecular absorption \citep{tinetti_water}. The water detection for the Spitzer-IRAC data was subsequently confirmed by transit observations with HST-WFC3 \citep{mccull_hd189}, while HST-NICMOS data showed evidence for methane \citep{swain_hd189_nicmos_tr}. 

In eclipse \cite[PN: 12881, PI:][]{McCullough_12881_prop}, the HST-WFC3 spectrum was analysed in \cite{Crouzet_HD189_spectrum_em}, which highlighted a marginal evidence for water. In their analysis, they combined with the Spitzer data from \cite{Charbonneau_2008}, which provides photometric data up to 24\,$\mu$m, a rarity in the exoplanet world. Additional eclipse observations were also obtained with HST-NICMOS \citep{swain_2008_hd189_nicmos}, Spitzer-IRS \citep{Todorov_2014_hd189, grillmair2} and other instruments \citep{Swain_2010_hd189_gr}. For the Spitzer-IRS dataset, the re-analysis presented in \cite{grillmair2} led to differences in the recovered spectrum. Overall, the observed fluxes were consistent in terms of shape but the flux ratios appeared to be 20\% lower in the later study. Sources of these discrepancies could come from the additional data considered (stellar variability), or differences in the reduction techniques (ramp models). Either way, this example highlights the difficulties of analysing combined data-sets from different sources or taken at different epochs.

As the HST transit data \cite[PN: 12881, PI:][]{McCullough_12881_prop} was previously analysed with \emph{Iraclis} in T18, we take the data directly from this paper. For the eclipse, we perform our standard reduction of the visit and obtain the emission spectrum. We note that our recovered emission spectrum of the G141 Grism is consistent with the one from \cite{Crouzet_HD189_spectrum_em}. For the Spitzer data, we use the latest reduction from \cite{Charbonneau_2008} consisting of the 3.6\,$\mu$m, 4.5\,$\mu$m, 5.8\,$\mu$m, 8\,$\mu$m, 16\,$\mu$m and 24\,$\mu$m photometric channels. 

When performing the retrieval analysis of the emission spectrum, we find that the HST spectrum is almost uninformative. On the HST spectrum only, the free and equilibrium runs lead to an isothermal temperature structure and the solutions found do not lead to higher evidence than the simpler black-body fits. There is some slight indication that CH$_4$ might be there due to the modulation at 1.4\,$\mu$m and the increased absorption around 1.6\,$\mu$m, where CH$_4$ possesses absorbing properties. When the Spitzer data is included however, the free retrievals detect H$_2$O in low abundance, around 10$^{-5}$ along with CO$_2$. There might also be some hints of FeH, however, the later molecule is not required as shown by the almost similar Bayesian evidence between the reduced and full runs. The derived metallicity in the full HST+Spitzer run is slightly super-solar, which matches the findings from the equivalent equilibrium run. However, the C/O ratio is found to be 0.5 due to CO$_2$ being the only species detected with those observations. In the equilibrium case, the thermal profile displays a localised inversion and the C/O ratio is much more constrained to super-solar values. 

Our retrieval analysis of the transmission spectrum gives similar results to the one in T18. The free retrieval indicated that the terminator is cloudy with a large water absorption feature at 1.4\,$\mu$m, thus matching the findings from the eclipse retrievals that include Spitzer. The rest of the molecules are not detected but strong upper limits on the optical absorbers and H$^-$ can be extracted. The equilibrium retrieval suggests a deeper cloud cover but is consistent with a similar chemistry with solar metallicity and sub-solar to solar C/O ratio.

%% file: sec_HD209.tex
HD\,209458\,b is one of the historically most-studied exoplanets, together with HD\,189733\,b. In 1999, it was announced as the first transiting exoplanet \citep{charbonneau_2000,Henry_2000} and shortly became the prime target of many observing campaigns due to the detection of its atmosphere via absorption of Na \citep{charbonneau_2002}. Many molecules (H$_2$O, CO, CH$_4$, CO$_2$ and NH$_3$) have now been detected \citep{Barman_2007, Swain_2008_hd209, Snellen_2010, Deming_2013_hd209, Line_HD209_spectrum_em, tsiaras_30planets, MacDonald_HD209}. In particular, preliminary analyses of the day-side emission obtained with five photometric Spitzer channels indicated the possible presence of thermal inversions \citep{Burrows_2007,knutson_hd209,Crossfield_2012_HD209}. However, later analyses of HST G141 observations and re-analyses of the Spitzer data \citep{Line_HD209_spectrum_em,Diamond_Lowe_2014} were more consistent with the presence of water vapor, seen in absorption at 1.4\,$\mu$m and no thermal inversions. We re-analysed the G141 data-set \cite[PN: 13467, PI:][]{Bean_13467_prop} using \emph{Iraclis} and were able to recover an eclipse spectrum for all five visits. We note that \cite{Line_HD209_spectrum_em} also attempted to reduce the five observed eclipses, but did not manage to recover a spectrum from the 4th visit. We note that \cite{Line_HD209_spectrum_em} fitted all the visits together and did not rely on a normalization of the white light curve depths. Our pipeline does not offer this option at the moment, but future works could investigate if the observed differences are due to this point. For the Spitzer data-sets, HD\,209458\,b was not included as part of the population study from G20, so we used the six broadband points from 3.6\,$\mu$m to 25\,$\mu$m used in \cite{Diamond_Lowe_2014}, including the data-sets reduced in \cite{Crossfield_2012_HD209} and \cite{Swain_2008_hd209}.

For the terminator, the HST G141 transit observation \cite[PN: 12181, PI:][]{deming_12181_prop} was previously reduced with \emph{Iraclis} in T18, so we take the transmission spectrum from this study. 

We present the results of our retrievals on the eclipse spectrum of HD\,209458\,b in Figure Set D1 but also to illustrate our methodology in the Appendix B and Appendix C. We can immediately notice that our spectrum, as compared to the one obtained in \cite{Line_HD209_spectrum_em} possesses a similar shape of the water feature between 1.2\,$\mu$m and 1.3\,$\mu$m, but with an apparent vertical offset. In all our free runs, except the full run with HST only, we however recover a similar solution that contains water vapour in sub-solar abundances. Due to the particularly high signal-to-noise for this spectrum, the recovered abundances are very tightly-constrained. When Spitzer is included, CH$_4$ and CO are also detected. For the full run with HST only, we find a very different solution that does not involve H$_2$O nor CH$_4$. Instead, the spectrum that was interpreted with an absorption at 1.4\,$\mu$m is now interpreted with an emission feature at 1.2\,$\mu$m. In this case, the temperature profile is inverted and the feature is fitted with a low amount of FeH and H$^-$ opacity. We highlight that this solution is most likely an artefact of the spectrum scatter but the difference is log(E) between the HST only runs is surprisingly high in favour of the FeH and H$^-$ solution, confirming that the shape of our WFC3 data is better fit with the inverted solution. A similarly inverted temperature solution also develops in the equilibrium chemistry retrieval, further confirming the odd shape of the water feature in our data. In all the retrievals for HD\,209458\,b, except the HST equilibrium chemistry run, we find sub-solar metallicity solutions that are most likely attributed to the low abundance of water we retrieve. We note that the full and equilibrium Spitzer solutions are very close, both in terms of retrieved thermal structure and elemental ratios, with a C/O of about 1. 

In transit, we observe water in the atmosphere of HD\,209458\,b, associated with a high-altitude cloud-cover. No other molecules are detected. In both the free and equilibrium runs, the atmosphere is consistent with solar metallicity. A low C/O ratio is found in the free run, while the equilibrium chemistry does not seem to provide constraints on this quantity.

%% file: sec_K1.tex
The first low-mass object discovered by the KELT-North survey, KELT-1\,b is a $~27M_{J}$, $1.12R_{J}$ planet with a very short period circular orbit of 29 hours. In 2012, \cite{2012_siverd_kelt1} presented spectroscopy, photometry and radial velocity data and obtained an equilibrium temperature of $T_{eq} \approx 2400$ K, due to the significant amount of stellar irradiation received by this planet. Its extreme temperature and significant inflation make KELT-1\,b a valuable case-study for short-period atmospheric characterisation. In 2014, \cite{Beatty_2014} observed KELT-1\,b's secondary eclipse using Spitzer, obtaining eclipse depths which are compatible with the presence of a strong sub-stellar hot-spot, suggesting poor or moderate heat redistribution for this atmosphere. Subsequently their investigations favour an atmosphere without a TiO inversion layer, where a mechanism of ``day-to-night TiO cold trap'' is proposed. This study was followed up with ground-based spectrophotometry in 2017, with \cite{Beatty_2017} presenting a H-band emission spectrum obtained with the LUCI1 multiobject spectrograph on the Large Binocular Telescope. Modeling of the atmospheric emission using the obtained average day-side brightness temperature of 3250 K suggested a monotonically decreasing temperature-pressure profile. The team highlighted these findings as unusual, since many other hot Jupiters of similar temperatures were believed to be in possession of either an isothermal or a thermally-inverted temperature structure. The differences were attributed to the higher surface-gravity of KELT-1\,b, which could contribute to the creation of TiO cold traps.

Using our standardised methodology, we re-analysed the WFC3 eclipse observations \cite[PN: 14664, PI:][]{beatty_14664_prop} of KELT-1\,b. To our knowledge, this data-set has not been analysed in a publication before. Since the Spitzer observations were not included in the systematic analysis of G20, we chose to include the original data-set from \cite{Beatty_2014}. For the transit, since the raw data \cite[PN: 14664, PI:][]{beatty_14664_prop} was already analysed with \emph{Iraclis} by T18, we took the spectrum obtained in this study.

For the day-side, the spectra, temperature profiles and posterior distributions of our standardised retrievals are presented in Figure Set D1. Broadly speaking, in both HST and HST+Spitzer runs, the atmosphere presents indications of a localised thermal inversion associated with VO, FeH and H$^-$. We note that some degeneracies exist with the posteriors presenting bi-modal behaviour and solutions that include either H$^-$, or FeH and VO. By comparing the Bayesian evidence with the reduced runs (with no metal hydrides/oxides), our retrievals demonstrate the need for these optical absorbers, especially when Spitzer is included. In the full scenarios, the atmosphere is most consistent with a sub-solar to solar metallicity but as for many other planets, the recovered C/O ratio is not well constrained and all values are allowed. When assuming equilibrium chemistry, we note that the thermal profiles decrease with altitude, contrasting with our full free runs. With this assumption, the atmosphere prefers a solar metallicity with high C/O ratio if Spitzer is not included, or a super-solar metallicity and solar C/O ratio when Spitzer is also considered. For this planet, chemical equilibrium is not a good assumption as demonstrated by the much lower Bayesian evidence obtained by these fits.

When analysing the transmission spectrum of KELT-1\,b , we do not find evidence for any molecular absorption, noting that clouds are most likely masking the presence of molecules in this atmosphere.

%% file: sec_K7.tex
As a planet with a high equilibrium temperature and low surface gravity \citep{2015_bieryla_kelt7}, KELT-7\,b is an excellent candidate for atmospheric characterisation. The transmission spectrum \cite[PN: 14767, PI:][]{sing_14767_prop} of KELT-7\,b was analysed by \cite{pluriel_aresIII} who found a rich transmission spectrum which is consistent with a cloud-free atmosphere and suggests the presence of H$_2$O and H-. The same study also analysed the WFC3 emission spectrum \cite[PN: 14767, PI:][]{sing_14767_prop}, which could be explained by a varying temperature-pressure profile, collision-induced absorption (CIA) and H-. \cite{Pluriel_2020} explored the effect of including Spitzer data from G20, finding little changes when including these observations in the emission retrievals. As the raw data was reduced using the \emph{Iraclis} pipeline in \cite{pluriel_aresIII}, we took the data as is and started our analysis from there. 

Our retrieval analysis differs from the one in \cite{Pluriel_2020} in a few aspects. First, our free retrieval includes two more molecules: CH$_4$ and CO$_2$. Here, we also considered the appropriate Spitzer instrument response functions, which were considered flat in \cite{Pluriel_2020}. The star was also modelled using an interpolated Phoenix spectrum at metallicity [Fe/H] = 0.139. 

From the free run, we recover the results from \cite{Pluriel_2020}, which found a thermally inverted temperature profile associated with emission of H$^-$, in both HST only and HST+Spitzer cases. In the HST+Spitzer case, we also find hints for additional absorption from TiO. When comparing the ln(E) between reduced and full runs, we find that the addition of optical absorbers is in fact justified only for the HST+Spitzer case. As discussed previously, for an atmosphere with a high level of thermal dissociation, such as KELT-7\,b, the derived metallicity is likely biased. From the equilibrium chemistry runs, we find metallicities that are solar to super-solar, but we also note that the recovered thermal structure is very different to the ones found by the free-chemistry runs. In terms of C/O ratio, the equilibrium chemistry retrieval favours high values.

For the transmission retrieval, we recover the same results as \cite{Pluriel_2020}. The terminator region is best characterised by H$_2$O and $H^-$ absorption. Again, the derived metallicity for the free run is likely to be biased by the dissociation of the tracers used for its computation. The equilibrium chemistry retrieval on the transit spectrum does not seem to well fit the observed data and does not lead to strong constraints.

%% file: sec_K9.tex
KELT-9\,b is the hottest exoplanet known so far. It orbits an A0V/B9V star (T = 10140 K) and reaches day-side temperature of about $\sim$5000 K, being itself hotter than many stars \citep{gaudi_k9}. Given the extreme temperatures, the majority of molecules were anticipated to be dissociated, which would leave only atomic species on the day-side and a featureless broadband spectrum \citep{kitzmann_k9}. Ground-based high resolution observations have detected a number of metals including iron, titanium and calcium \citep{hoeij_k9,hoeij_k9_2,yan_k9, cauley_k9, turner_k9, pino_k9}, which are consistent with this picture. Observations of KELT-9\,b phase-curve with TESS and Spitzer have revealed an asymmetric transit \citep{ahlers_k9} induced by the fast rotation of its host star. The rotation leads to a non-uniform structure in the star, which has a larger and brighter equator than the poles, whereas KELT-9\,b orbits with a 87$^{\circ}$ spin-orbit angle. The same studies indicate that the planet possesses a low day-night temperature contrast with indications for H$_2$ dissociation and recombination \citep{wong_k9,mansfield_k9}. As KELT-9\,b is subject to an intense irradiation from the star and has a large extended hydrogen envelope reaching the Roche-Lobe limit, it is experiencing extreme atmospheric escape \citep{yan_k9}. As of today, there is no transmission data for this planet. The HST eclipse observation \cite[PN: 15820, PI:][]{Lorenzo_2019_proposals} was previously analysed by the same authors in \cite{Changeat_2021_k9} and using the same methodology, so we do not reproduce the reduction of the spectrum from this planet. We note that the physical characteristics of this system leads to a particularly precise spectrum in a single observation.

The only difference with the analysis in \cite{Changeat_2021_k9} is the addition of CO$_2$, which is required for comparing with the rest of our population. This does not change the results and we recover a strongly-inverted temperature profile, with the presence of molecules on the day-side. This was noted in \cite{Changeat_2021_k9} as a rather surprising finding since the associated temperatures are high enough to dissociate them in a solar composition and equilibrium chemistry. This therefore suggested the presence of dis-equilibrium processes, non-solar chemistry or biases from other sources. When analysing the equilibrium scenarios, we note that, similarly to the reduced cases, they do not perform well and lead to much worse fits of the observed data (see the Bayesian evidence).
For this planet, the estimates of the metallicity and the C/O ratio from our free chemistry are highly inaccurate due to the expected dissociation of the main molecules and our lack of constraints on the elements in atomic/ionic forms. Since the planet is likely experiencing disequilibrium chemistry, the equilibrium runs, which return a a slightly sub-solar metallicity and a C/O ratio of about 1, are also most likely biased and should be considered with caution.

%% file: sec_Ke13.tex
The exoplanet Kepler-13A\,b \citep{Shporer_2011_kepler13AB_discov} orbits a rapidly rotating A-type star in a triple star system. The planet's day-side was studied in great detail in two previous studies \citep{Shporer_Kepler13_em, Beatty_Kepler13_spectrum_em}, which indicated a very high temperature but did not find thermal inversions. Due to the presence of the two stars Kepler-13\,A and Kepler-13\,BC, the extraction process for this planet involves many steps that are not automated in our pipeline \emph{Iraclis}. We therefore choose not to re-analyse the HST emission spectrum obtained in \cite{Beatty_Kepler13_spectrum_em}, which carefully removed the contribution from the two stars to extract the spectrum. The data was taken as part of the proposal PN:13308 led by Ming Zhao \citep{zhao_13308_prop}. We also added the Spitzer photometric points from \cite{Shporer_Kepler13_em} as-is and started our analysis from these spectra. We note that this atmosphere was not studied in transit spectroscopy with HST. 

The spectra, temperature profiles and posteriors from our atmospheric retrievals can be found in Figure Set D1. As in the previous studies in the literature, we recover a decreasing temperature gradient with altitude and a well-constrained abundance of water vapour in all scenarios. This is driven by the strong and well defined 1.4\,$\mu$m water feature in the HST spectrum. Given the temperature of this atmosphere, one might expect the presence of metal oxide and hydrides, but none were detected in our analysis. The addition of the Spitzer data does not impact the results of our retrievals, except by providing further limits on the CO$_2$ abundance. For the free and equilibrium runs, the metallicity is similar and consistent with solar values. The C/O ratio in the free runs is difficult to constrain due to the non detection of carbon-bearing species. When considering the equilibrium chemistry retrievals, the C/O ratio is found to be roughly solar with large uncertainties.

%% file: sec_TR3.tex
TrES-3\,b is a hot Jupiter of about 2 M$_{jup}$ \citep{O_Donovan_2007_tr3}. At the time of its discovery in 2007, it had one of the shortest orbital periods of the known planets and was deemed a good candidate for atmospheric follow-up observations. The planet was then studied using photometric measurements, leading to some of the first exoplanet detections of emission via secondary eclipse \citep{Mooij_2009_tr3, Croll_2010}. An eclipse spectrum was then captured by the HST-WFC3 camera and studied as part of a population analysis of five planets \citep{2014_ranjan_5hj_staring}. According to their findings, the TrES-3\,b spectrum is consistent with a simple black-body emission but the authors point out that combining these observations with the Spitzer data from \cite{Fressin_2010_tr3} is inconsistent with the black-body fit. In their study, they find that an atmosphere with a solar to low metallicity best represents the combined data-set. Similar results are also found in \cite{Line_2014_syst}, which is also able to rule out a high C/O ratio solution, thanks to a statistical retrieval analysis.

We reduced the eclipse observation from the raw HST images \cite[PN: 12181, PI:][]{deming_12181_prop} using our \emph{Iraclis} reduction pipeline. The eclipse spectrum we obtain differs strongly from the one in \cite{2014_ranjan_5hj_staring}, here displaying an emission feature at 1.4\,$\mu$m. In their study, we note, however, that the fit of their white-light curve does not present data-points covering the baseline or the ingress/egress of the transit. For our analysis, we add the Spitzer photometric observations recovered in \cite{Fressin_2010_tr3} and perform our standardised retrieval runs.

Due to the higher emission obtained at 1.4\,$\mu$m in our observed spectrum, the run that does not include the Spitzer data is well-fit with an emission of water. The recovered thermal profile is therefore inverted and no other molecules are necessary to explain this data-set. When the Spitzer data is added, however, the water is rejected due to the low emission in the Spitzer band-pass. The spectrum is best fit with small amounts of FeH and VO. The thermal profile in this case is more complex, with a temperature decrease with altitude deep in the atmosphere and a thermal inversion at lower pressures. The metallicity and C/O ratio derived from the free retrieval with Spitzer is not well constrained as water and carbon-bearing species are not recovered. In the equilibrium chemistry runs, the solutions also differ depending on whether Spitzer is included or not. In the HST only case, the thermal profile is inverted to best fit the 1.4\,$\mu$m feature and associated with a solar metallicity and C/O ratio, most similar to the results of the corresponding free chemistry. When Spitzer is included, the profile is reversed and the spectrum is best fit with a flat spectrum in the WFC3 wavelength range, associated with a super-solar metallicity and a high C/O ratio.

%% file: sec_W4.tex
WASP-4\,b is an inflated hot Jupiter \citep{Wilson_2008} with an equilibrium temperature of around 1800 K. The planet day-side was observed with HST during a single visit, taken in staring mode \cite[PN: 12181, PI:][]{deming_12181_prop}. The data, while having a low signal-to-noise, favoured an atmosphere with an isothermal or a weakly-inverted thermal profile and a poor water content \citep{2014_ranjan_5hj_staring}. Those results confirmed previous findings using Spitzer \citep{Beerer_2010}. In this study, we re-analysed the HST data using our \emph{Iraclis} pipeline. As the Spitzer data was not part of the G20 study, we used the data from \cite{Beerer_2010}. We note that an additional staring mode transit spectrum was taken as part of the same HST proposal, but an adequate fit to the data could not be obtained.

In our \emph{TauREx3} retrievals, we show that the HST spectrum does not contain a lot of information. Indeed, the pure black-body fits perform similarly to the full and reduced runs in the HST-only scenarios. When optical absorbers are considered, the retrievals indicate the possible presence of TiO, but large tails can be observed. The detection is not significant since the Bayesian evidence of the reduced runs is similar to the full cases. The thermal profile for this planet, found in our free runs, is either isothermal or locally-inverted. When running the equilibrium chemistry retrievals, it is again difficult to conclude, with the HST only case favouring a weak thermal inversion and the HST+Spitzer case favouring a decrease of temperature with altitude. In those two cases the metallicity is solar with large uncertainties, while the C/O ratio is super-solar.

%% file: sec_W12.tex
WASP-12\,b has previously been studied with transmission and emission spectroscopy using WFC3 G141 in staring mode \citep{swain_wasp12}. The study found an emission spectrum consistent with a black-body and only marginal evidence for molecular features in the transmission data-set. Phase curve observations with the Spitzer Space Telescope at 3.6\,$\mu m$ and 4.5\,$\mu m$ indicated large emission amplitudes, showing evidence for poor day-to–night heat transport \citep{cowan_wasp12_phase}.  Photometric eclipse observations with Spitzer over 3.6 - 8\,$\mu m$ \citep{Campo_2011} suggested a weak thermal inversion \citep{madhu_wasp12_spitzer} and a super-solar C/O ratio. These studies, however, did not account for the presence of a binary companion system \citep{Bergfors_2012, Bechter_2014}. When re-analysed to correct for this, the previous observations of WASP-12\,b were inconsistent with an isothermal atmosphere due to eclipse depth variations in the Spitzer band-passes \citep{Crossfield_2012, Stevenson_Wasp12b_spectrum_em}, and required the presence of a carbon-rich atmosphere (HCN and C$_2$H$_2$) on the day-side. On the terminator, analysis detected the possible presence of H$_2$O or CH$_4$/HCN and metal-bearing molecules \citep{Stevenson_2014}. Later studies confirmed the possibility of super-solar C/O ratio, detecting H$_2$O in the terminator \citep{kreidberg_wasp12} thanks to the addition of the G102 Grism, but highlighted the impact of priors and model choices \citep{kreidberg_wasp12,Oreshenko_2017_W12} in analysing this particularly difficult data-set.  

STIS transit observations were also taken and, when combined with the WFC3 data, showed no evidence for TiO \citep{Sing_wasp12}. However, as the WFC3 observations were in staring mode, the error bars on the data were large ($\sim$120-200\,ppm). The G141 transmission spectrum was subsequently obtained in scanning mode and, when combined with transit observations from the G102 grism, displayed a broad water feature (log(H$_2$O) = -2.7$^{+1.0}_{-1.1}$) \cite{kreidberg_wasp12}. The spectrum showed no signs of optical absorbers with a 3$\sigma$ upper limit of log(TiO) = -3.69. 

As for the other planets, we reduced the HST G141 eclipse observation \cite[PN: 12230, PI:][]{Swain_12230_prop} using our \emph{Iraclis} pipeline. We applied the dilution factor to correct for the un-resolved companion of this system \citep{Bergfors_2012, Bechter_2014}. For the Spitzer data, we obtained the corrected reductions from \cite{Stevenson_Wasp12b_spectrum_em}, which we included as-is. The transmission spectrum \cite[PN: 13467, PI:][]{Bean_13467_prop} with HST was already analysed in T18, so we took the spectrum from this study. \\

The emission of WASP-12\,b at HST wavelengths is most consistent with a black-body spectrum. This is shown by the similar Bayesian evidence obtained in our reduced and full retrievals as compared with the simpler black-body fit. When Spitzer is considered, the full retrieval recovers CO$_2$ and possibly FeH, with a localised thermal inversion. The FeH detection is however not robust as shown by the similar Bayesian evidence obtained in the reduced run that includes the Spitzer points. In the reduced run, CO$_2$ is still detected, which confirms the need for this molecule to explain the observed combined spectrum. The derived metallicity for the full run is about solar with a C/O ratio of 0.5, most likely biased by the CO$_2$ detection. In the equilibrium chemistry runs, the HST-only case is consistent with the results from the free chemistry retrievals. When Spitzer is added, the C/O ratio is found to be closer to 1 with a decreasing-with-altitude thermal profile. We note that the free chemistry runs do not provide a significantly better fit than the chemical equilibrium. 

At the terminator, the transit of WASP-12\,b is clearly indicative of high altitude clouds and the presence of water vapour, which was not detected on the day-side. Our results are similar to that of T18, and we derive an atmosphere with solar metallicity but large uncertainties on the C/O ratio. The free retrievals do not require the presence of carbon-bearing species.

%% file: sec_W18.tex
WASP-18\,b \citep{hellier_w18} has been thoroughly studied since its discovery in 2008. Spitzer, Hubble WFC3 and ground-based eclipses have been taken \citep{nymeyer_w18_spit,w18_ground_ec, sheppard_w18,arcangeli_wasp18_em, manjavacas_wasp103, iro_w18, gandhi_w18} as well as phase curves with Hubble WFC3 and TESS \citep{arcangeli_w18_phase,shporer_w18}. These have revealed a low albedo, poor redistribution of energy to the night-side and evidence for an inverted day-side temperature-pressure profile. In particular, the analysis from \cite{sheppard_w18} considered a similar data-set to us and detected a strong thermal inversion, associated with the presence of H$_2$O and CO.

The HST transmission spectrum was taken, along with two eclipses, as part of a phase curve \cite[PN: 13467, PI:][]{Bean_13467_prop}. As part of the same proposal, three additional eclipses were also obtained. For consistency, we re-analysed the raw data using the \emph{Iraclis} pipeline and our standardised methodology. We also added the Spitzer observations from \cite{nymeyer_w18_spit} for the four photometric channels from 3.6\,$\mu$m to 8\,$\mu$m. The transmission spectrum \cite[PN: 13467, PI:][]{Bean_13467_prop} was included in the T18 study, which also employed the \emph{Iraclis} reduction pipeline, so we conserved their spectrum.

In eclipse, we find that the the atmosphere of WASP-18\,b is well fit by a localised thermal inversion, with emission from H$_2$O and e$^-$.  For this planet, the results are consistent between the HST and HST+Spitzer runs. As opposed to the study from \cite{sheppard_w18}, we do not find evidence for CO, differences which might be due to their model not including H$^-$ opacities or differences in our reduction techniques. We find that the atmosphere has a solar to slightly super-solar metallicity, which is also confirmed by the equilibrium chemistry retrievals. For the C/O ratio, this parameter is not constrained by our free runs due to the lack of detection of carbon-bearing species, but when assuming equilibrium chemistry, we find this atmosphere consistent with a C/O ratio of about 1.

For the transit, our retrieval analysis reveals that a very flat spectrum provides a good fit to the data. This is most likely due to the presence of high-altitude clouds, opaque at those wavelengths. As noted in T18, we are not able to extract further constraints on the possible abundances of any molecular species.

%% file: sec_W19.tex
WASP-19\,b has been the subject of a number of investigations, from both the ground and from space. Work by \cite{anderson_wasp19_spitzer} analysed four Spitzer eclipses, taken across 3.6-8\,$\mu m$, and constructed a spectral energy distribution of the planet’s day-side atmosphere. They found no stratosphere, supporting the hypothesis that hot Jupiters orbiting active stars have suppressed thermal inversions \citep{knutson_therm_inv}. Analysis of the TESS optical phase curve showed moderately efficient day-night heat transport, with a day-side temperature of 2240 K and a day to night contrast of around 1000 K \citep{wong_wasp19_tess}. This study also utilised a host of ground-based observations by \cite{anderson_wasp19_ground,burton_wasp19,abe_wasp19,bean_wasp19}. However, they did not utilise the HST WFC3-G141 observations of the WASP-19\,b eclipse. WASP-19\,b has also been studied via transmission spectroscopy. The retrievals of the STIS-G430L, G750L, WFC-G141 and Spitzer-IRAC observations suggest the presence of water at log(H$_2$O) $\approx$ 4 but show no evidence for optical absorbers \citep{sing_pop,barstow_10_planets,pinhas}. Those results do not match the ground-based transits that were acquired with the European Southern Observatory’s Very Large Telescope (VLT), using the low-resolution FORS2 spectrograph which covers the entire visible-wavelength domain (0.43–1.04\,$\mu m$). When analysing this data, \cite{sedahati_tio_wasp19} detected the presence of TiO to a confidence level of 7.7$\sigma$.

In this study, we reduced the WFC3 data of the WASP-19\,b eclipse \cite[PN: 13431, PI:][]{Huitson_13431_prop} using \emph{Iraclis} and obtained the eclipse spectrum. The Spitzer data is available in G20 for the 3.6\,$\mu$m and the 4.5\,$\mu$m channels, but since \cite{anderson_wasp19_spitzer} also reduced the additional 5.3\,$\mu$m and 8\,$\mu$m channels, we choose to use the later Spitzer data in our retrieval analysis. As the transit observations of WASP-19\,b were not included in the T18 population study with \emph{Iraclis}, we also reduce this data-set \cite[PN: 12181 and 13431, PI:][]{deming_12181_prop, Huitson_13431_prop} before running our standardised retrieval analysis.

At the day-side, we find decreasing temperature profiles for all scenarios, with the data being well fit by absorption of water. In all our models, we do not find evidence for optical absorbers. This is confirmed by the ln(E), which is essentially the same in both reduced and full runs. The free chemistry runs are consistent with the equilibrium chemistry scenario and we find a solar to super-solar metallicity best fits this observed spectrum. The C/O ratio is, as expected, poorly constrained, but full runs seem to favour a low C/O ratio, which is also confirmed by the HST+Spitzer equilibrium chemistry run. 

At the terminator, the transmission spectrum is well fit by water vapour. In the free run, this is the only molecule that is clearly identified. Clouds could also be present with a wide range of possible altitudes, as shown by the posterior distribution. The metallicity inferred in both free and equilibrium runs is solar. For the C/O ratio, the free run prefers a sub-solar case due to the lack of detection of carbon-bearing species, while the equilibrium run does not provide strong constraints, again demonstrating the difficulty of constraining these parameters from HST only.

%% file: sec_W33.tex
WASP-33\,b is the first planet discovered to orbit a $\delta$-Scuti variable star \citep{Cameron_2010_w33, Herrero_2011_w33}. At the time of discovery, it was the hottest known exoplanet, with temperatures above 3000 K. It now belongs to the category of the extremely-hot Jupiters, with KELT-9\,b and WASP-189\,b. As such its hot day-side is believed to be deprived of water vapor and contain metal oxide and hydrides. Due to the complex pulsations of the star, the analysis of WASP-33\,b involves complex de-convolution of the stellar signal. Early studies of its atmosphere indicated the likely presence of a thermal inversion associated with TiO emission \citep{Deming_2012_w33,von_essen_2014_w33,Haynes_Wasp33b_spectrum_em}, which greatly contributed to the debate on the importance of TiO and VO in hot Jupiters \citep{Fortney_2008,Gandhi_2019_top}. Later follow-ups at high resolution from the ground confirmed the detection of TiO \citep{Nugroho_W33b}, while observations of the terminator during transit were also consistent with the presence of optical absorbers, this time AlO \citep{von_Essen_2019_w33}. While the picture appeared clear, recent studies and re-analysis of WASP-33\,b's atmosphere, however, shed some doubts on the robustness of the TiO and VO detections \citep{Herman_2020_w33, Serindag_2021_w33}. In this work, we consider the HST-WFC3 data obtained during the eclipse of WASP-33\,b \cite[PN: 12495, PI:][]{deming_12495_prop}. While the spectrum was extracted in \cite{Haynes_Wasp33b_spectrum_em} with another pipeline, we do not perform an \emph{Iraclis} reduction from the raw data due to the complexity of accounting for the stellar pulsations. Instead, we take the reduced spectra as-is from \cite{Haynes_Wasp33b_spectrum_em} and start our retrieval analysis there. For this planet, there are no WFC3 transit observations.

The results of our retrieval analysis are shown in Figure Set D1. From comparing the Bayesian evidence of the free models, it is evidence that this reduced spectrum requires a model including optical absorbers. In the full models, the solutions are the same, independently of whether Spitzer is considered or not. The models converge to a strong thermal inversion associated with emission features of water vapour, TiO and H$^-$. In \cite{Haynes_Wasp33b_spectrum_em}, their investigations also highlighted the evidence for TiO absorption in the same data-sets, but in their models they did not fit for this molecule and choose to fix its abundance to solar values. Here we find compatible results but also highlight the likely presence of H$^-$, an opacity that was not considered in the original study. Our solutions possess a sub-solar metallicity and unconstrained C/O ratios, but we highlight that our method to determine elemental abundances is likely inaccurate for this type of atmosphere as many of the considered tracers dissociate into atomic/ionic species, similarly to KELT-9\,b. In the equilibrium chemistry run, the atmosphere is found slightly sub-solar but with a well constrained C/O ratio of about 1.

%% file: sec_W43.tex
WASP-43\,b is one of the first \citep{hellier_w43} and most studied hot Jupiters. With a day-side temperature of about 1800 K, this 1 $R_{jup}$ planet of about 2 M$_{jup}$ has been at the center of many studies, thanks to the large number of observations in transit, eclipse and phase-curves. In particular, it is one of the rare planets that has been observed in phase-curve with HST \citep{Stevenson_2014}. This study demonstrated the use of HST in phase-curve studies for the first time and detected the presence of water vapour. A follow-up study included the Spitzer phase curve to the analysis \citep{Stevenson_2017}, extending the constraints to carbon-bearing species. Those studies also highlighted an offset in the hot-spot of the planet as well as a large day-night contrast. The night-side was found to be surprisingly cool, which suggested that global opaque clouds might cover the deeper layers of this side of the planet. We note however, that the use of a more complex method to analyse the same data \citep{Feng_2016_inomogeneou}, highlighted that the previous constraints on CH$_4$ might be biases due to the 1D assumption in the earlier studies. Using the data from the same phase curve, water vapour was also detected in the terminator region \citep{kreidberg_w43} but a later, more complete study \citep{Chubb_2020_w43}, also indicated that an optical absorber, AlO, might be present in this region.

For those reasons, WASP-43\,b became the go-to planet for the theoretical work on the global circulation \citep{Mendonca_2018_w43,venot_2020_w43, Carone_2020_w43} in exoplanet atmospheres and for cloud/haze modelling \citep{2020_helling_clouds_w43, helling2021_clouds}. Similarly, due to the availability of consistent HST and Spitzer phase-curves, the planet is often used as benchmark for the development of retrieval techniques exploiting the 3D aspects of atmospheres \citep{Taylor_2020, irwin_2020_w43, Changeat_2020_phasecurve1, changeat_2021_phasecurve2, Feng_2020_2D}. However, later contradictory re-analyses of the Spitzer phase curve demonstrated the difficulties in recovering robust estimates of atmospheric emission with Spitzer \citep{2018_mendonca_w43_spz, 2019_morello_w43, May_2020_w43, Bell_2020_w43}. This further demonstrates the need for cautious approaches when combining HST and Spitzer data-sets. As for the thermal profile it was believed to be decreasing with altitude, up until \cite{changeat_2021_phasecurve2}, which displayed model-dependent behaviour and concluded that thermal inversions could not be rejected for this planet.

In this paper, we re-analyse from scratch the eclipse observations \cite[PN: 13467, PI:][]{Bean_13467_prop} of WASP-43\,b. For the 3.6\,$\mu$m and 4.5\,$\mu$m Spitzer channels, we use the values from G20. For the transmission, the data \cite[PN: 13467, PI:][]{Bean_13467_prop} was first studied in \cite{kreidberg_w43}, but it was since  re-analysed in T18, which used the \emph{Iraclis} pipeline. We therefore proceed from the retrieval step directly, taking the spectrum from T18. 

Analysing the eclipse spectrum, we find that the HST+Spitzer retrieval contains a decreasing with altitude thermal profile. The 1.4\,$\mu$m feature is, in this case, associated with the presence of water vapour, as found in previous studies. The models also require the presence of CO$_2$, which absorbs in the 4.5\,$\mu$m Spitzer band. Now, considering HST only, the full run, including optical absorbers, converges to a different solution. This second solution displays a thermal inversion and low abundance of FeH, seen in emission. While this is most likely due to the data-point being scatter between 1.2\,$\mu$m and 1.3\,$\mu$m, this result highlights why the interpretation of eclipse spectra with HST alone is difficult. This re-enforces the discussion regarding model degeneracies made in Appendix C, which explains why a spectral feature at 1.4\,$\mu$m can be well interpreted by both, absorption of H$_2$O or emission of refractory species. For this planet, comparisons of the Bayesian evidence in the HST only case indicate that the simpler reduced model, with absorption of H$_2$O, in fact, provides an equivalent fit to the data. Regarding the derived metallicity and C/O ratio, the retrieval on the HST+Spitzer data is consistent with a super-solar metallicity and a C/O ratio about solar. The C/O ratio value is most likely due to CO$_2$ being the only detected species, but the the metallicity is also confirmed by the HST+Spitzer equilibrium chemistry retrieval. The HST-only run, is best represented by an isothermal thermal profile with large uncertainties on the metallicity and C/O ratio.

When considering the transit data, we find some evidence for water vapour but the features appear relatively muted, most likely due to opaque clouds. As highlighted in previous studies \citep{kreidberg_w43, Stevenson_2017}, the temperature found at the terminator is much lower than the one from the day-side. Our findings from the free chemistry are consistent with a solar to super-solar metallicity but an unconstrained C/O ratio, which matches the results found by our equilibrium chemistry model.

%% file: sec_W74.tex
Orbiting a slightly evolved F9 star, WASP-74\,b is a 0.95 M$_J$ moderately inflated hot Jupiter \citep{hellier_w74}. While we did not find any analysis of the eclipse of this planet, the HST transmission spectrum was first analysed in T18, uncovering a moderate water feature. Later, ground-based photometry with multiple filters was obtained and, when combined with the HST spectrum, favoured models containing TiO and VO absorption \citep{mancini_w74}. This was however argued against in a follow-up study using the same data as well as further photometric observations from the ground and Spitzer IRAC 1/2. In this later study, no evidence for these optical absorbers was found, and the data favoured a strong Rayleigh slope \citep{luque_w74}. 

We take the transmission spectrum from T18 \cite[PN: 14767, PI:][]{sing_14767_prop} and reduce the single G141 observation of the eclipse \cite[PN: 14767, PI:][]{sing_14767_prop} using our \emph{Iraclis} pipeline. For the Spitzer observations, we use the photometric emission recovered as part of the G20 population study. 

Analysing the eclipse spectrum, we do not find evidence for optical absorbers. The free retrieval including HST and Spitzer observations reveals the presence of H$_2$O and CH$_4$ as the main absorbers with a decreasing thermal profile with altitude. It is interesting to note that the HST-only case leads to a slightly inverted thermal profile with H$^-$ opacity, however, the Bayesian evidence of this retrieval as compared to the reduced case does not change, indicating that the increase in complexity is not justified. Looking at the metallicity and C/O ratio, we find a solar metallicity and an unconstrained C/O ratio. The metallicity recovered from the equilibrium chemistry runs is also solar, and the models are overall consistent with the findings from free chemistry. When equilibrium chemistry is assumed, the thermal profile is also decreasing with altitude. When only HST is considered in the equilibrium retrievals, the C/O ratio remains unconstrained, but adding Spitzer restricts the solutions to high values.   

At the terminator, we find some evidence for H$_2$O and FeH, but with large tails in the posterior distributions. In both free and equilibrium models, high-altitude clouds are likely to explain the flat shape of the transmission spectrum. The metallicity is found to be around solar values with large uncertainties, while the C/O ratio is unconstrained.

%% file: sec_W76.tex
WASP-76 b was first studied by HST WFC3 in transmission. These observations were analysed as part of the T18 population study \citep{tsiaras_30planets}, where their retrievals suggested a water-rich atmosphere (log(H$_2$O) = $-2.7\pm1.07$) with a 4.4$\sigma$ detection of both TiO and VO. However, as noted in the study, the abundance of TiO retrieved is likely to be nonphysical and driven by narrow spectral coverage. Retrieval analysis of this spectrum was also performed by \cite{fisher} who extracted a water abundance which was incompatible with the previous study (log(H$_2$O) = $-5.3\pm0.61$). \cite{fisher} did not include TiO and VO in their analysis, but instead used a non-grey cloud model to match the opacity at shorter wavelengths. In a more recent study, \cite{Edwards_2020_ares}, we accounted for a faint stellar companion to WASP-76 b \citep{bohn_w76, southworth_w76} and showed that the original transit spectrum was contaminated. A similar study was performed by \cite{Fu_2021_w76}. Once the contamination was removed, the transmission spectrum no longer showed substantial evidence for optical absorbers but retained its strong water feature. WASP-76 b has also been observed in emission with WFC3-G141 and our analysis of the corrected spectrum in the same study shows strong evidence for VO. However, the best-fit abundance of log(VO) $\approx$ -4 is likely unfeasible, especially as the evidence for TiO in the current data-set was marginal. In this initial study, only the WFC3 data was utilised. Here, we do not re-analyse the WFC3 data \cite[Transit PN: 14260; Eclipse PN: 14767, PI:][]{sing_14767_prop} and take it directly from \cite{Edwards_2020_ares}, but we also include the Spitzer points from G20. 
The only differences to the retrievals performed in \cite{Edwards_2020_ares} are that we model the star using a Phoenix spectrum, account for the proper instrument profiles for Spitzer and include CO$_2$ for the consistency in our approach. 

The analysis of the eclipse spectrum is consistent with a thermal inversion for all models. When optical absorbers are not included, the data is well fit with CH$_4$ only. When optical absorbers are included, the results are similar to \cite{Edwards_2020_ares} and we confirm that H$_2$O and TiO provide the best fit to the HST data, with decisive evidence in favour of the more complex model. When HST+Spitzer are included, however, a higher ln(E) is obtained for the reduced case. This solution, while simpler, can arguably be ruled out due to the high abundance of CH$_4$ that would be required in this case. The free retrievals obtain solar metallicity and sub-solar C/O ratios. When running our equilibrium chemistry retrievals, we also find an inverted thermal profile and the chemistry displays a solar to super-solar metallicity as well as a high C/O ratio.

Our interpretation of the transit spectrum of WASP-76\,b is the same as in \cite{Edwards_2020_ares}. The terminator region is most consistent with the presence of water vapour and high-altitude clouds. In the free run, this leads to solar metallicity and low C/O ratio but one can see in the equilibrium chemistry run that solutions with higher metallicities and a wide range of C/O ratios are also consistent with the observed spectrum.

%% file: sec_W77.tex
\begin{figure}
    \centering
    \includegraphics[width=0.45\textwidth]{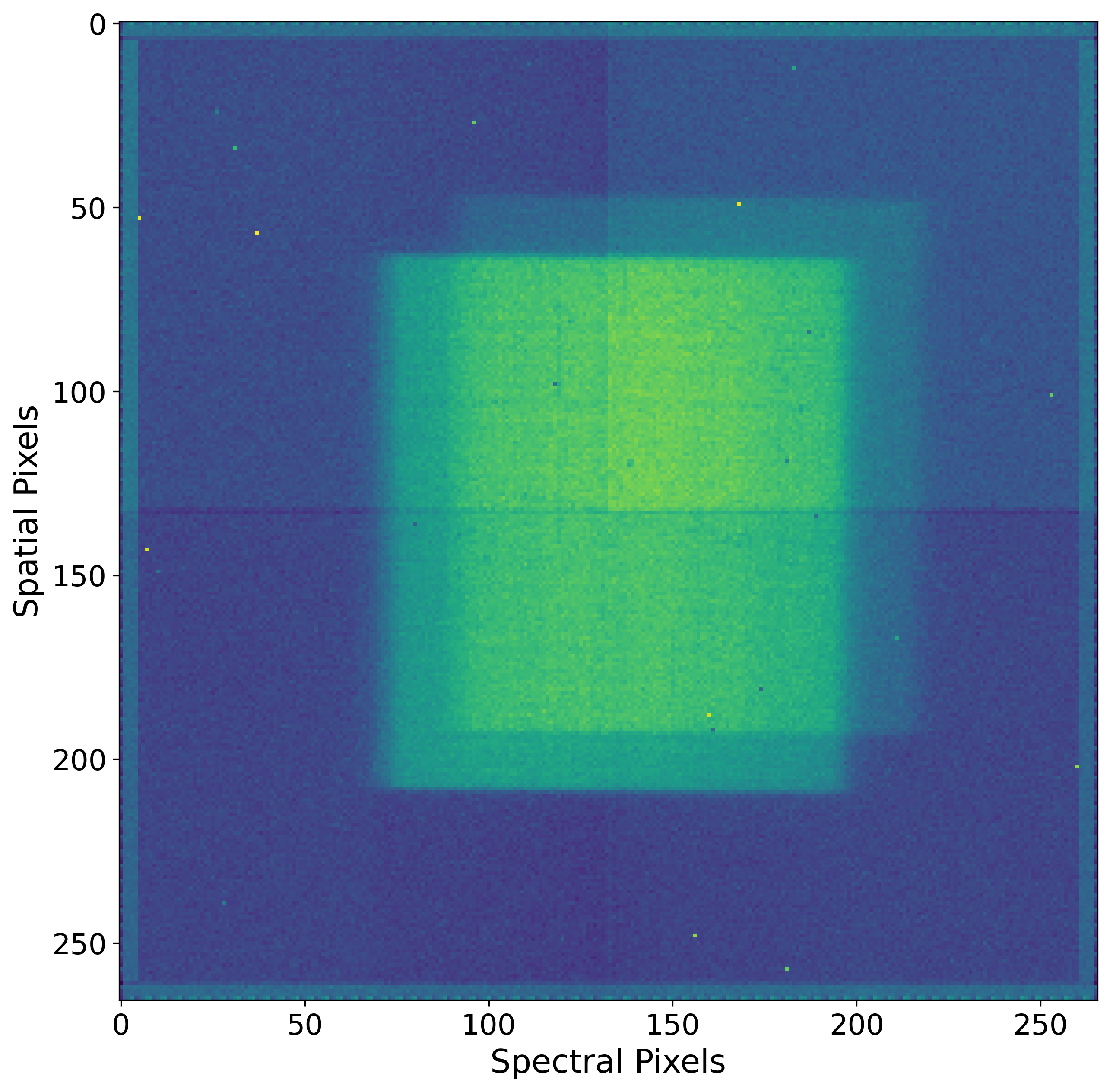}
    \caption{Example spatial scan from the second eclipse observation of WASP-77 A b. The contributions from WASP-77 A and WASP-77 B can clearly be seen to overlap.}
    \label{fig:w77_ex_image}
\end{figure}

WASP-77 A\,b is an inflated hot Jupiter in a wide binary system that was first discovered in 2012 \citep{Maxted_2013_w77}. The planet orbits WASP-77 A, a G8 V star. WASP-77 B, a fainter (2 mag) K-dwarf companion, is separated by 3". As such, the spatial scans from WASP-77 A and WASP-77 B overlap somewhat on the detector, as shown in Figure \ref{fig:w77_ex_image}, and cannot be separated. If not corrected for, the flux from WASP-77 B would evidently adversely affect the recovered emission spectrum for the planet. Hence, to overcome this, we utilised the specialised WFC3 simulator Wayne \citep{Varley_2017} to model the contribution of the secondary star.

We started by analysing the data \cite[PN: 16168, PI:][]{Mansfield_16168_prop} in the normal way, extracting the contaminated eclipse spectrum of WASP-77 A b from both visits. We then preceded to account for the contamination. In each of the two eclipses of WASP-77 A b, the relative positions of WASP-77 A and WASP-77 B were different. Therefore, we calculated the positions of WASP-77A and B for each visit from the direct image. We then extracted the high-resolution (10 angstrom bins) spectra of WASP-77A and B individually from the first non-destructive read of the out-of-eclipse observations during the second eclipse observation sequence as the star separation was better than the first. Using these spectra, we simulated high resolution spatially scanned WFC3 images for each star individually. Next, utilising the same setup used for the analysis of the real data, we extracted the stellar spectra of each star for each visit. For each bin of the eclipse spectrum, we then calculated the flux ratio between the two stars, using this to apply a correction to the eclipse spectrum of WASP-77 A b for each visit. Finally, we calculated the weighted average of the planetary emission.

As for the other planets, we included the Spitzer data. We obtained the photometric measurements in \cite{Maxted_2013_w77} and interpreted the spectrum using our standardized atmospheric retrievals. We note that \cite{Line_2021_w77} analysed observations of the same planet at high resolution and detected H$_2$O and CO. In their study, they concluded from constraints on the abundance of those two molecules that the planet should have a sub-solar metallicity and a solar C/O ratio. To our knowledge the HST observations of WASP-77\,A\,b have not been presented in previous works. We obtained the Spitzer data from the population study of \cite{Garhart_2020}.

Our atmospheric retrievals are consistent with the presence of water vapor and a decreasing with altitude thermal profile in both HST and HST+Spitzer runs. In the HST data, we note that additional modulations of the spectrum at 1.25\,$\mu$m could be associated with the molecule TiO. The detection is significant with an ln(E) $>$ 5 between the full and reduced models. When Spitzer is included, there are additional evidence for carbon-bearing species with the absorption of CH$_4$ and CO$_2$ being detected. In the full HST+Spitzer run, we recover a solar metallicity and an unconstrained C/O ratio. Chemical equilibrium runs are consistent with a sub-solar metallicity as in \cite{Line_2021_w77} but when Spitzer data is added, the C/O ratio is about 0.8. When assuming chemical equilibrium, we note that the 1.25\,$\mu$m spectral modulation that was attributed to TiO in the free runs is not well fit.

%% file: sec_W79.tex
WASP-79\,b \citep{smalley12} is a very large hot Jupiter, believed to have an evaporating atmosphere \citep{Bourrier2015}. It has been shown to have a polar orbit through the Rossiter McLaughlin effect \citep{addison_w79}. 

The transmission spectrum of WASP-79\,b \cite[PN: 14767, PI:][]{sing_14767_prop} has been previously analysed with \emph{Iraclis} as part of \cite{skaf_aresII}. A water rich atmosphere was found, with a $>5\sigma$ confidence level in models with H$_2$O and FeH included. The latter was included as the optical absorber of choice after initial models without it struggled with nonphysical values. 

\cite{sotzen_w79} also analysed the HST WFC3, combining it with observations from Magella/LDSS-3C R, TESS, and Spitzer, resulting in a transmission spectrum range of 0.6-5\,$\mu$m. Whilst the spectra extracted from the LDSS-3C R data indicated clouds initially, their retrieval code ATMO found that including FeH and H- as absorbers provided better data fits. \cite{rathcke2021} studied the most complete transmission range. This study included the STIS instrument on HST, with two transits observed through the G430L grating and another through the G750L grating, the HST WFC3 instrument, Spitzer 3.6\,$\mu$m and 4.5\,$\mu$m channels, and LDSS-3C R data. The study used three different retrieval tools: NEMESIS, POSEIDON, and ATMO's retrieval tool, ARC. Similar to the results found by the other two studies, their best fitting model spectrum is characterised by an absorption feature at 1.4\,$\mu$m of H$_2$O, as well as the relatively smooth continuum from 0.4 to 1.3\,$\mu$m indicative of H- bound free absorption. 

Our analysis of the transmission data \cite[PN: 16168, PI:][]{sing_14767_prop} from HST WFC3 is shown in Figure Set D2. Whilst the water feature is relatively well constrained in the posterior, as with the other studies mentioned, there is little evidence to support the presence of any carbon-based species. The transmission spectrum also shows detection of H$^-$ opacity, which was not included in \cite{skaf_aresII}. In their study, they did not include H$^-$, which explains the differences found here. However, their Bayesian Evidence (191.2 for their best model with FeH) is very similar to ours, indicating that from an observational point of view, it is difficult to conclude on whether FeH or H$^-$ (or both) is responsible for the absorption. We note that the equilibrium retrieval does not provide a good fit of the transmission spectrum.  

The emission data \cite[PN: 14767, PI:][]{sing_14767_prop} was reduced with \emph{Iraclis} in Bieger et al 2021 (in prep), and is shown in Figure Set D1 along with the fits from our standardised retrievals. The emission analysis from our retrievals supports what is found in the transmission--again, the most unconstrained parameters are carbon-based species. Inspection of the full retrieval indicates the detection of VO and FeH, associated with an increasing with altitude thermal profile. In the full model, the spectral modulation in the HST spectrum is fit with an emission feature from optical absorbers. On the opposite, if optical absorbers are not included, or if additional constraints from chemical equilibrium are considered, the HST spectrum is better fit with a decreasing temperature profile and the absorption of water vapour. This behaviour is independent from whether Spitzer is considered or not and is due to the degeneracies induced by the narrow spectral region of HST, as discussed in Appendix C. The full retrievals, however, provide a strong evidence in favor of optical absorbers as the difference in Bayesian Evidence is larger than 3 ($\Delta$ln(E)=4.5). 
 
These results will require further retrievals and analysis in order to understand the role of TiO, VO, FeH or H- in the atmosphere of WASP-79\,b. The JWST Early Release Science (ERS) program includes WASP-79b as a primary target; JWST will be observing it for 42 hours over four different modes, which will present more opportunity to study this planet in depth with more precise data. More detailed analysis of WASP-79\,b will be presented in Bieger et al 2021 (in prep).

%% file: sec_W103.tex
WASP-103\,b is an ultra-short period planet ($P$ = 22.2 hr) whose orbital distance is less than 20\% larger than its Roche radius, resulting in the possibility of tidal distortions and mass-loss via Roche-lobe overflow \citep{gillon_wasp103}.

WASP-103 b's HST WFC3 emission spectrum was found to be featureless down to a sensitivity of 175 ppm, showing a shallow slope toward the red \citep{cartier_wasp103}. Work by \cite{manjavacas_wasp103}, which performed a reanalysis of the same data-set, found that the emission spectrum of WASP-103\,b was comparable to that of an M-3 dwarf. \cite{delrez_wasp103} obtained several ground-based high-precision photometric eclipse observations which, when added to the HST data, could be fit with an isothermal black-body or with a low water abundance atmosphere with a thermal inversion. However, their Ks band observation showed an excess of emission compared to both these models. Recently, a phase-curve analysis of the planet was taken and reported in \cite{Kreidberg_w103}. The study also utilized the previous HST emission spectra and confirmed a seemingly featureless day-side. The phase-resolved spectra of WASP-103\,b were compared in this work to those of brown dwarfs and directly-imaged companions of similar temperature which show evidence for water absorption at 1.4\,$\mu m$, whereas WASP-103\,b showed no such feature. The result could be due to WASP-103\,b’s irradiation environment and its low surface gravity. A later study on the same data \citep{changeat_2021_w103} and employing a unified phase-curve retrieval technique obtained a more complex picture of the planet. It confirmed the presence of thermal inversion and dissociation processes on the day-side of the planet and found signature of FeH emission. The study also constrained water vapor across the entire atmosphere. Ground-based transmission observations found strong evidence for sodium and potassium \citep{lendl_wasp103}. \\

Two HST G141 phase curves of WASP-103\,b were obtained \cite[PN: 14050, PI:][]{kreidberg_w103_wfc3_w18_spit_prop} which each contained a single transit and eclipse. Additionally, two further eclipses were taken with the same instrument \cite[PN: 13660, PI:][]{zhao_13600_prop}. 

Our eclipse spectrum of WASP-103\,b, is not consistent with a pure black-body fit, thus contrasting with the previous studies. In our re-analysis we find that our models prefer a thermal inversion with the presence of optical absorber such as VO or FeH. When Spitzer is included, the model also includes H$^-$, which was predicted to be an important opacity source for this atmosphere \citep{Kreidberg_w103}. We note, however, that comparing the Bayesian evidence with the reduced models does not provide decisive evidence in favour of the full models, meaning that an atmosphere without these active molecules could also well fit this data-set. The metallicity associated with our full runs is centred around solar values but with large uncertainties. For the C/O ratio, since CO$_2$ is detected in the HST+Spitzer case, the derived value converges to 0.5, but this is most likely due to the lack of detection of other species. When considering the equilibrium chemistry case, a thermal inversion is also inferred from the data. This is associated with a solar metallicity and a C/O ratio of about 1.

Analysing the transit spectrum \cite[PN: 14050, PI:][]{kreidberg_w103_wfc3_w18_spit_prop}, we observed that the downward slope is well fit by VO. Other optical absorbers might be present (TiO, H$^-$) but the data does not allow verification of this. The solution found possesses a wide range of metallicities; sub-solar in nature. When applying equilibrium chemistry to this data-set, the thermal profile converges to unrealistic values for this atmosphere.

%% file: sec_W121.tex
Significant observational time has been spent on WASP-121\,b \citep{Delrez_Wasp121b_em}. In transmission, the analyses of ground-based observations, Hubble-STIS and Hubble-WFC3 have shown the presence of H$_2$O and optical absorption attributed to VO and/or FeH \citep{Evans_wasp121_t1, Evans_wasp121_t2}. The authors of these studies note that chemical equilibrium models with solar abundances cannot reproduce the spectrum seen, while free chemical retrievals can only do so by converging to high abundances of VO and FeH. WASP-121\,b has also been observed in eclipse and the presence of a thermal inversion provides the best fit to the data \citep{bourrier,daylan_tess_wasp121,Parmentier_2018_w121photodiss,Evans_wasp121_e3}. \cite{bourrier} attributed this to VO with a best-fit abundance of log(VO) = -6.03 while \cite{Evans_wasp121_e3} performed chemical equilibrium retrievals, finding evidence for a muted water feature due to dissociation and H$^-$ opacity.
 
In parallel, high resolution ground-based observations of the transit have put upper limits on the abundances of TiO and VO at the terminator with log(VMR) $<$ -7.3 and 7.9 respectively, suggesting these cannot be causing the inversion seen \citep{merritt_w121}. However, the study highlighted that these limits are largely degenerate with other atmospheric properties such as the scattering properties or the altitude of clouds on WASP-121\,b. Another study found a host of atomic metals, including V, which are predicted to exist if a planet is in equilibrium and has a significant quantity of VO \citep{hoeij_w121}. They too noted the absence of TiO which could support the hypothesis that Ti is depleted via a cold-trap.

Here, we take the transmission spectrum from T18 \cite[PN: 14468, PI:][]{evans_14468_prop} already analysed using the \emph{Iraclis} pipeline, and supplement it with the two new transit observations recently obtained as part of a phase curve proposal \cite[PN: 15134, PI:]{evans_15134_prop}. All of these were taken with the WFC3 G141 grism. In emission, we analyse the two available visits with the G102 grism \cite[PN: 15135, PI:][]{evans_15135_prop} and the five observations taken using G141, PN: 14767, PI: \cite{sing_14767_prop}; PN: 15134, PI: \cite{evans_15134_prop}. In our most comprehensive retrievals, we also added the Spitzer data-sets from G20.

For the planet day-side, we find that the solution depends on the data-set considered but since the reduced runs achieve a much lower ln(E) than the full runs, there is decisive evidence in favour of the presence of optical absorbers for this planet. The full runs are consistent with a thermal inversion, detecting VO and H$^-$. This result is independent of whether Spitzer is considered or not. If only HST G141 and G102 Grisms are included, we also find evidence for H$_2$O. When Spitzer is added, the H$_2$O detection disappears in favour of TiO, a molecule that might be consistent with the temperatures found for this atmosphere, but was thought from previous studies to not be present. Those results are inconsistent with the recent findings in \cite{Evans_wasp121_e3}, which detected H$^-$ and H$_2$O from the same data-set. At this point, it remains difficult to know whether the differences in our interpretation come from the retrieval setup or from the reduction pipelines. As described previously, different pipelines often lead to different results when combining the data from various instruments. In any case, all our retrievals seem consistent with a strong thermal inversion and H$^-$ emission. Whereas the partition between the sources (TiO, VO, FeH, H-) could be different, optical absorption is required by all scenarios. The associated metallicity of those free retrievals is sub-solar but the C/O ratios remain unconstrained. We also ran equilibrium chemistry retrievals, finding different thermal structures to the free runs. The models still include a thermal inversion, but the thermal profile is decreasing with altitude for the highest pressures. The metallicity in this case is solar to super-solar and associated with high C/O ratios. We note, however that the bayesian evidence obtained by the equilibrium chemistry models is much lower than with the free runs. 

Our analysis of the transmission spectrum shows the presence of water vapor and H$^-$. There is also evidence for absorption due to VO as seen in the posterior distribution. In addition to this, a large pressure range for opaque clouds is allowed, which might consequently have an impact on the degeneracies highlighted for the upper limits of VO and TiO in \cite{merritt_w121}. In general, our analysis of the transmission spectrum is consistent with the results from \cite{Evans_wasp121_t2}. The metallicity for the free retrieval of the transit spectrum is solar, while the C/O ratio is found to be sub-solar. When assuming equilibrium chemistry, a similar metallicity is found, but the C/O ratio is unconstrained.